\documentclass[usegraphicx, longauth]{aa}

\usepackage{natbib}
\bibpunct{(}{)}{;}{a}{}{,} 
\usepackage{subfigure}
\usepackage{subfig}
\usepackage{caption}
\usepackage{longtable}
\usepackage{graphicx}
\usepackage{graphics}
\usepackage{keyval}
\usepackage{trig}
\usepackage[american]{babel}
\usepackage{url}
\usepackage{color}
\usepackage{amsmath}
\usepackage{bm}
\usepackage{longtable}
\usepackage{amssymb}
\usepackage{textcomp}

\def\beq{\begin{equation}}
\def\eeq{\end{equation}}
\def\bey{\begin{eqnarray}}
\def\eey{\end{eqnarray}}

\def\masssun{\mathcal{M}_\odot}

\def\lsim{\mathrel{\raise.3ex\hbox{$<$\kern-.75em\lower1ex\hbox{$\sim$}}}}
\def\gsim{\mathrel{\raise.3ex\hbox{$  $\kern-.75em\lower1ex\hbox{$\sim$}}}}

\def\lsun{L_\odot}

\def\a0{A_{\mathrm{0}}}

\def\teff{T_\mathrm{eff}}
\def\fehzw{\mathrm{[Fe/H]_{ZW}}}
\def\fehuves{\mathrm{[Fe/H]_{UVES}}}
\def\dmyr{{\rm d\,Myr}^{-1}}

\def\apjl{ApJ}

\def\aap{A\&A}      
\def\apjs{ApJS}       

\newcommand\Eq[1]{Eq.~(\ref{#1})}
\newcommand\Fig[1]{Fig.~\ref{#1}}
\newcommand\Tab[1]{Table~\ref{#1}}
\newcommand\Sec[1]{Sec.~\ref{#1}}

\def\clustermet{-2.07 \pm 0.06}
\def\clustermetuves{-2.20 \pm 0.10}

\def\murrab{15.00 \pm 0.11}
\def\murrc{15.00 \pm 0.05}
\def\musx{14.97 \pm 0.11}
\def\mumvfeh{15.00 \pm 0.07}

\def\drrab{10.00 \pm 0.49}
\def\drrc{9.99 \pm 0.21}
\def\dsx{9.84 \pm 0.50}
\def\dmvfeh{10.00 \pm 0.30}

\def\metrrzw{-2.07 \pm 0.06}
\def\metrruves{-2.20 \pm 0.10}

\def\metmvfeh{-2.09 \pm 0.26}
\def\metmvfehuves{-2.24 \pm 0.42}

\makeatletter\def\CT{\def\@captype{figure}}\makeatother

\begin{document}

   \title{A census of variability in globular cluster M68 (NGC~4590)}

   \author{N. Kains\inst{\ref{stsci}, \ref{eso}}\and
          A. Arellano Ferro\inst{\ref{unam}}\and
          R. Figuera Jaimes\inst{\ref{eso}, \ref{standrews}}\and
          D. M. Bramich\inst{\ref{qeeri}}\and
	     J. Skottfelt\inst{\ref{nbi}, \ref{copenhagen}} \and
          U. G. J{\o}rgensen \inst{\ref{nbi}, \ref{copenhagen}} \and
          Y.~Tsapras\inst{\ref{lcogt}, \ref{queenmary}} \\
          and\\
	R.~A.~Street\inst{\ref{lcogt}}
\and P. Browne\inst{\ref{standrews}}
\and M. Dominik\inst{\ref{standrews}}\fnmsep\thanks{Royal Society University Research Fellow}
\and K. Horne\inst{\ref{standrews}}
\and M. Hundertmark\inst{\ref{standrews}}
\and S. Ipatov\inst{\ref{qnrf}}
\and C. Snodgrass\inst{\ref{mps}}
\and I.~A.~Steele\inst{\ref{ljmu}}\\
          (The LCOGT/RoboNet consortium)\\
 	and \\
          K.A. Alsubai \inst{\ref{qnrf}} \and
	V. Bozza \inst{\ref{salerno}, \ref{fisicanucleare}} \and
	S. Calchi Novati \inst{\ref{salerno}, \ref{iiass}} \and
	S. Ciceri \inst{\ref{mpia}} \and
	G. D'Ago \inst{\ref{salerno}, \ref{fisicanucleare}} \and
	P. Galianni \inst{\ref{standrews}} \and
	S.-H. Gu \inst{\ref{yunan}, \ref{chinasci}} \and
	K.~Harps{\o}e \inst{\ref{nbi}, \ref{copenhagen}} \and
	T.C. Hinse \inst{\ref{kasi},\ref{nbi}} \and
	D. Juncher  \inst{\ref{nbi}, \ref{copenhagen}} \and
	H. Korhonen \inst{\ref{finca}, \ref{nbi}, \ref{copenhagen}} \and
	L. Mancini \inst{\ref{mpia}} \and
	A. Popovas\inst{\ref{nbi}, \ref{copenhagen}} \and
	M. Rabus \inst{\ref{puc}, \ref{mpia}} \and
	S.~Rahvar \inst{\ref{sharif}, \ref{perimeter}} \and
	J. Southworth \inst{\ref{keele}} \and
	J. Surdej \inst{\ref{liege}} \and
	C. Vilela \inst{\ref{keele}} \and
	X.-B. Wang \inst{\ref{yunan}, \ref{chinasci}} \and
	O. Wertz \inst{\ref{liege}} \\ 
	(The MiNDSTEp Consortium)
          }
          
\institute{Space Telescope Science Institute, 3700 San Martin Drive, Baltimore, MD 21218, United States of America \label{stsci}\\
\email{nkains@stsci.edu}
\and European Southern Observatory, Karl-Schwarzschild Stra\ss e 2, 85748 Garching bei M\"{u}nchen, Germany\label{eso}\\
\and Instituto de Astronom\'{i}a, Universidad Nacional Aut\'{o}noma de Mexico\label{unam}\\
\and SUPA School of Physics \& Astronomy, University of St Andrews, North Haugh, St Andrews, KY16 9SS, United Kingdom \label{standrews}\\
\and Qatar Environment and Energy Research Institute, Qatar Foundation, Tornado Tower, Floor 19, P.O. Box 5825, Doha, Qatar \label{qeeri}\\
\and Niels Bohr Institute, University of Copenhagen, Juliane Maries vej 30, 2100 Copenhagen, Denmark \label{nbi}\\	
\and Centre for Star and Planet Formation, Geological Museum, {\O}ster Voldgade 5, 1350 Copenhagen, Denmark \label{copenhagen}\\	
\and Las Cumbres Observatory Global Telescope Network, 6740 Cortona Drive, Suite 102, Goleta, CA 93117, USA  \label{lcogt}\\	
\and School of Mathematical Sciences, Queen Mary, University of London, Mile End Road, London E1 4NS, UK \label{queenmary}\\
\and Qatar Foundation, P.O. Box 5825, Doha, Qatar \label{qnrf}\\	
\and Max Planck Institute for Solar System Research, Justus-von-Liebig-Weg 3, 37077 G\"{o}ttingen, Germany \label{mps}\\	
\and Astrophysics Research Institute, Liverpool John Moores University, Twelve Quays House, Egerton Wharf, Birkenhead, Wirral., CH41 1LD, UK \label{ljmu}\\
\and Dipartimento di Fisica ``E.R. Caianiello", Universit\`a degli Studi di Salerno, Via Giovanni Paolo II 132, I-84084 Fisciano, Italy \label{salerno}\\	
\and Istituto Nazionale di Fisica Nucleare, Sezione di Napoli, Italy \label{fisicanucleare}\\	
\and Istituto Internazionale per gli Alti Studi Scientifici (IIASS), Via Giuseppe Pellegrino, 19, 84019 Vietri Sul Mare Salerno, Italy \label{iiass}\\		
\and Max Planck Institute for Astronomy, K\"{o}nigstuhl 17, 69117 Heidelberg, Germany \label{mpia}\\	
\and Yunnan Observatories, Chinese Academy of Sciences, Kunming 650011, China \label{yunan}\\
\and Key Laboratory for the Structure and Evolution of Celestial Objects, Chinese Academy of Sciences, Kunming 650011, China \label{chinasci}\\
\and Korea Astronomy and Space Science Institute, Daejeon 305-348, Korea \label{kasi}\\
\and Finnish Centre for Astronomy with ESO (FINCA), University of Turku, V{\"a}is{\"a}l{\"a}ntie 20, FI-21500 Piikki{\"o}, Finland \label{finca}\\
\and Instituto de Astrof\'isica, Facultad de F\'isica, Pontificia Universidad Cat\'olica de Chile, Av. Vicu\~na Mackenna 4860, 7820436 Macul, Santiago, Chile \label{puc}\\
\and Department of Physics, Sharif University of Technology, P.~O.\ Box 11155--9161, Tehran, Iran \label{sharif}\\	
\and Perimeter Institute for Theoretical Physics, 31 Caroline St. N., Waterloo ON, N2L 2Y5, Canada \label{perimeter}\\
\and Astrophysics Group, Keele University, Staffordshire, ST5 5BG, United Kingdom \label{keele}\\	
\and Institut d'Astrophysique et de G\'{e}ophysique, Universit\'{e} de Li\`{e}ge, All\'{e}e du 6 Ao\^{u}t 17, Sart Tilman, B\^{a}t.\ B5c, 4000 Li\`{e}ge, Belgium \label{liege}\\	
}

   \date{Received ... ; accepted ...}


  \abstract
   {}
   {We analyse 20 nights of CCD observations in the $V$ and $I$ bands of the globular cluster M68 (NGC~4590), using these to detect variable objects. We also obtained electron-multiplying CCD (EMCCD) observations for this cluster in order to explore its core with unprecedented spatial resolution from the ground.}
   {We reduced our data using difference image analysis, in order to achieve the best possible photometry in the crowded field of the cluster. In doing so, we showed that when dealing with identical networked telescopes, a reference image from any telescope may be used to reduce data from any other telescope, which facilitates the analysis significantly. We then used our light curves to estimate the properties of the RR Lyrae (RRL) stars in M68 through Fourier decomposition and empirical relations. The variable star properties then allowed us to derive the cluster's metallicity and distance.
   }
   {M68 had 45 previously confirmed variables, including 42 RRL and 2 SX Phoenicis (SX Phe) stars. In this paper we determine new periods, and search for new variables, especially in the core of the cluster where our method performs particularly well. We detect an additional 4 SX Phe stars, and confirm the variability of another star, bringing the total number of confirmed variable stars in this cluster to 50. We also used archival data stretching back to 1951 in order to derive period changes for some of the single-mode RRL stars, and analyse the significant number of double-mode RRL stars in M68. Furthermore, we find evidence for double-mode pulsation in one of the SX Phe stars in this cluster. Using the different classes of variables, we derived values for the metallicity of the cluster of [Fe/H]=$\clustermet$ on the ZW scale, or $\clustermetuves$ on the UVES scale, and found true distance moduli $\mu_0=\murrab$ mag (using RR0 stars), $\murrc$ mag (using RR1 stars), $\musx$ mag (using SX Phe stars), and $\mumvfeh$ mag (using the $M_V-$[Fe/H] relation for RRL stars), corresponding to physical distances of $\drrab$, $\drrc$, $\dsx$, and $\dmvfeh$ kpc, respectively. Thanks to the first use of difference image analysis on time-series observations of M68, we are now confident that we have a complete census of the RRL stars in this cluster.
}
   {}

\maketitle


\section{Introduction}\label{sec:intro}

Globular clusters in the Milky Way are ideal environments to study the properties and evolution of old stellar populations, thanks to the relative homogeneity of the cluster contents. Over the last century, a sizeable observational effort has been devoted to studying globular clusters, in particular their horizontal branch (HB) stars, including RR Lyrae (RRL) variables. Increasingly precise photometry has allowed for the detailed study of pulsation properties of these stars, both from an observational point of view \citep[e.g.][]{kains12b, kains13b, arellano13, figuera13, kunder13b}, and from a theoretical point of view using stellar evolution \citep[e.g.][]{dotter07} and pulsation models  \citep[e.g.][]{bono03, feuchtinger98}. RRL and other types of variables can also be used to derive estimates of several properties for individual stars, and for the cluster as a whole.

In this paper we analyse time-series observations of M68 (NGC~4590, C1236-264 in the IAU nomenclature; $\alpha = 12^h39^m27.98^s, \delta = -26^{\circ}44'38.6^{\prime\prime}$ at J2000.0), one of the most metal-poor globular clusters with [Fe/H] $\sim$ -2.2 at a distance of $\sim 10.3$ kpc. This is a particularly interesting globular cluster, with suggestions that it might be undergoing core collapse, as well as showing signs of rotation \citep{lane09}. It has also been suggested that M68 is one of a number of metal-poor clusters that were accreted into the Milky Way from a satellite galaxy, based chiefly on their co-planar alignment in the outer halo \citep{yoon02}. Indeed metal-poor clusters are of particular importance to our understanding of the origin of globular clusters in our Galaxy, as they are essential to explaining the Oosterhoff dichotomy. This phenomenon was postulated by \cite{oosterhoff39}, who noticed that globular clusters fell into two distinct groups, Oosterhoff types I and II, traced by the mean period of their RRL stars, and the relative numbers of RR0 to RR1 stars. Since then, many studies have confirmed the existence of the Oosterhoff dichotomy as a statistically significant phenomenon \citep[e.g.][and references therein]{sollima14}, with very few clusters falling within the ``Oosterhoff gap" between the two groups. However, two metal-rich clusters are now generally thought to be part of a new Oosterhoff III type of clusters \citep[e.g.][]{pritzl01, pritzl02}, and, interestingly, \cite{catelan09} notes that globular clusters in satellite dwarf spheroidal (dSph) galaxies of the Milky Way have been observed to fall mostly \textit{within} the Oosterhoff gap, which would seem to go against theories stating that the Galactic halo was formed from accretion of dwarf galaxies \citep[e.g.][]{zinn93a, zinn93b}. The strength of such arguments rests partly on our ability to obtain complete censuses of RRL stars in globular clusters. This has only really been achievable since difference image analysis (DIA) made it possible to obtain precise photometry even in the crowded cores of globular clusters \citep[e.g.][]{alard99, alard00, bramich08, albrow09, bramich13}. 

Here we use time-series photometry to detect known and new variable stars in M68, which we then analyse to derive their properties, and the properties of their host cluster. In particular, we were able to detect a number of SX Phoenicis (SX Phe) stars thanks to improvements in observational and reduction methods since the last comprehensive time-series studies of this cluster were published by \cite{walker94}, hereafter W94, and \cite{clement93}, hereafter C93; this is also the first study of this cluster making use of DIA, meaning that we can now be confident that all RRL stars in M68 are known. 

We also use electron-multiplying CCD (EMCCD) data to study the core of M68 with unprecedented resolution from the ground. EMCCD observations, along with methods that make use of them, such as Lucky Imaging (LI), are becoming a powerful tool to obtain images with a resolution close to the diffraction limit from the ground. We first demonstrated the power of EMCCD studies for globular cluster cores in a pilot study of NGC~6981 \citep{skottfelt13}, where we were able to detect two variables in the core of the cluster that were previously unknown due to their proximity to a bright star.

In \Sec{sec:observations}, we describe our observations and reduction of the images. In \Sec{sec:variables}, we summarise previous studies of variability in this cluster, and outline the methods we employed to recover known variables and to detect new ones. We also discuss period changes in several of the RRL stars in this cluster. We use Fourier decomposition in \Sec{sec:fourdec} in order to derive physical parameters for the RRL stars, using empirical relations from the literature. The double-mode pulsators in M68 are discussed in \Sec{sec:doublemode}, and we use individual RRL properties to estimate cluster parameters in \Sec{sec:clusterprop}. Finally, we summarise our findings in \Sec{sec:conclusions}.

\section{Observations and reductions}\label{sec:observations}
\subsection{Observations}

We obtained Bessell $V$- and $I$-band data with the LCOGT/RoboNet 1m telescopes at the South African Astronomical Observatory (SAAO) in Sutherland, South Africa, and at Cerro Tololo, Chile. The telescopes and cameras are identical and can be treated as one instrument. The CCD cameras installed on the 1m telescopes are Kodak KAF-16803 models, with 4096 $\times$ 4096 pixels and a pixel scale of 0.23$^{\prime\prime}$ per pixel, giving a 15.7 $\times$ 15.7 arcmin$^2$ field of view (FOV). The images were binned to $2048 \times 2048$ pixels, meaning that the effective pixel scale of our images is 0.47$^{\prime\prime}$ per pixel. The CCD observations spanned 74 days, with the first night on March~8 and the last night on May~20, 2013. These observations are summarised in \Tab{tab:observations}. 

We also observed M68 using the EMCCD camera mounted on the Danish 1.54m telescope at La Silla, Chile. The camera is an Andor Technology iXon+ model 897 EMCCD, with 512$\times$512 16$\mu{\rm m}$ pixels and a pixel scale of 0.09$^{\prime\prime}$ per pixel, giving a FOV of 45$\times$45 arcsec$^2$. The small FOV means that only the core of M68 was imaged in the EMCCD observations. The filter on the camera is approximately equivalent to the SDSS $i' + z'$ filters \citep{bessell05}; more details on the filter are given in \cite{skottfelt13}. 72 good EMCCD observations were taken, spanning 2.5 months (May 1 to July 18 2013), each observation consisting of a data cube containing 4800 0.1s exposures; in general, 1 or 2 observations were taken on any one night.

\begin{table}
\begin{center}
  \begin{tabular}{ccccc}

     \hline
    Date		 &$N_V$ 	&$t_V$(s)	&$N_I$	&$t_I$(s) \\
  \hline  
    20130308		&$1$		&120		&1		&60	\\
    20130311		&$7$		&120		&9		&60	\\
    20130314		&$15$	&120		&16		&60	\\
    20130316		&$17$	&120		&16		&60	\\
    20130317		&$14$	&120		&16		&60	\\
    20130318		&$16$	&120		&11		&60	\\
    20130319		&$12$	&120		&7		&60	\\
    20130320		&$8$		&120		&8		&60	\\
    20130321		&$15$	&120		&16		&60	\\
    20130323		&$5$		&120		&6		&60	\\
    20130329		&$20$	&40		&20		&40	\\
    20130331		&$9$		&120		&10		&60	\\
    20130401		&$4$		&120		&2		&60	\\
    20130402		&$30$	&40-120	&28		&40-60\\
    20130403		&$7$		&120		&5		&60	\\
    20130404		&$19$	&40-120	&19		&40-60	\\
    20130427		&$9$		&40		&9		&40	\\
    20130428		&$10$	&40		&10		&40	\\
    20130429		&$10$	&40		&10		&40	\\
    20130520		&$8$		&120		&$-$		&$-$	\\

\hline
   Total		&236	&	&219	& \\
\hline \hline
  \end{tabular}
  \caption{Numbers of images and exposure times for the $V$ and $I$ band observations of M68. When varying exposure times were used, a range is given. \label{tab:observations}}
  \end{center}
\end{table}

\subsection{Difference image analysis}

\subsubsection{CCD observations}

We used the DIA software {\tt DanDIA}\footnote{{\tt DanDIA} is built from the {\tt DanIDL} library of {\tt IDL} routines available at {\tt http://www.danidl.co.uk}}\citep{bramich13, bramich08} following the recipes devised in our previous publications of time-series globular cluster observations (\citealt{kains12b, kains13b, figuera13, arellano14}), to reduce our observations. DIA is particularly adept at dealing with crowded fields like the cores of globular clusters, as described in detail in our previous papers \citep[e.g.][]{bramich11}; here we summarise the main steps of the reduction process. We note here that an interesting advantage of using data from networks of identical telescopes and setups such as the LCOGT/ RoboNet network is that one can use a reference image constructed from observations from one telescope for the other telescopes in the network.

After applying bias level and flatfield corrections to our raw images, we blurred our images with a Gaussian of appropriate $\sigma$, in order for all images to have a full-width half-maximum (FWHM) of 3.5 pixels; images that already have a FWHM $\ge 3.5$ pixels were not blurred. This is to avoid under-sampling, which can cause difficulties in the determination of the DIA kernel solution. Images were stacked from the best-seeing photometrically stable night in order to obtain a high signal-to-noise reference image in each filter; if an image had too many saturated stars, it was excluded from the reference. The resulting images in $V$ and $I$ are made up of 12 and 20 stacked images respectively, with combined exposure times of 480s (in $V$) and 800s (in $I$), and respective point-spread function (PSF) FWHM of 3.17 pixels (1.49$^{\prime\prime}$) and 3.09 pixels (1.45$^{\prime\prime}$). The reference frames were then used to measure source positions and reference fluxes in each filter. Following this, the images were registered with the reference frame, and the convolution of the reference with the kernel solution was subtracted from each image. This resulted in a set of difference images, from which we extracted difference fluxes for each source, allowing us to build light curves for all of the objects detected in the reference images. The light curves of the variable stars we detected in M68 are available for download at the CDS, in the format outlined in \Tab{tab:onlinedata}\footnote{The full light curves can be obtained from the CDS via anonymous ftp to cdsarc.u-strasbg.fr (130.79.128.5) or via {\tt http://cdsarc.u-strasbg.fr/viz-bin/qcat?J/A+A/}}.

\subsubsection{EMCCD observations}

Each EMCCD data cube was first pre-processed using the algorithms of \cite{harpsoe12}, which included bias correction, flatfielding and alignment of all exposures (corresponding to a tip-tilt correction). This procedure also yielded the point-spread function (PSF) width of each exposure. Each data cube was then subdivided into 10 groups of exposures of increasing PSF size. 

We then reduced the pre-processed EMCCD cubes using a modified version of the {\tt DanDIA} pipeline, and a different noise model was adopted to account for the difference between CCD and EMCCD observations, as detailed by \cite{harpsoe12}. Conventional LI techniques only keep the best-quality exposures within a data cube and therefore usually discard most of them; here we build the reference image from the best-seeing groups only, but the photometry is measured from all exposures within the data cubes. That is, once a reference image has been built, the full sets of exposures (including the ones with worse seeing) are stacked for each data cube; we do this to achieve the best possible signal-to-noise (S/N). The sharp reference image is then convolved with the kernel solution and subtracted from each of the stacked data cubes. 

Our EMCCD reference image has a PSF FWHM of 4.5 pixels, or 0.40$^{\prime\prime}$, and has a total exposure time of 302.4s (3024 $\times$ 0.1s).

\begin{table*}
\begin{center}
  \begin{tabular}{ccccccccccc}

     \hline
    \#		&Filter	&HJD 	&$M_{\rm std}$	  & $m_{\rm ins}$ 	&$\sigma_m$		&$f_{\rm ref}$ 	&$\sigma_{\rm ref}$	& $f_{\rm diff}$ &$\sigma_{\rm diff}$ &$p$  \\
	&	&($d$, UTC)	&(mag)	&(mag)	&(mag)	&(ADU s$^{-1}$)	&(ADU s$^{-1}$)	&(ADU s$^{-1}$)	&(ADU s$^{-1}$)	&	\\
  \hline  
   
V1 &V & 2456360.56581 &    15.994 &    18.128 &     0.008 &   873.400 &     1.095 &  -784.689 &    10.536 &    2.5083 \\
V1 &V & 2456363.70715 &    15.999 &    18.133 &     0.008 &   873.400 &     1.095 &  -935.364 &    11.470 &    2.9662 \\
\vdots &\vdots&\vdots&\vdots&\vdots&\vdots&\vdots&\vdots&\vdots&\vdots&\vdots \\
V1 &I & 2456360.68185 &    15.057 &    18.303 &     0.009 &   447.298 &     0.950 &    44.675 &     5.617 &    1.4841 \\
V1 &I & 2456363.71698 &    15.370 &    18.616 &     0.009 &   447.298 &     0.950 &  -130.860 &     4.495 &    1.4636 \\
\vdots &\vdots&\vdots&\vdots&\vdots&\vdots&\vdots&\vdots&\vdots&\vdots&\vdots \\

\hline \hline
  \end{tabular}
  \caption{Format for the time-series photometry of all confirmed variables in our $V-$ and $I-$ band CCD observations. The standard $M_{\rm std}$ and instrumental $m_{\rm ins}$ magnitudes listed in columns 4 and 5 respectively correspond to the variable star, filter and epoch of mid-exposure listed in columns 1-3, respectively. The uncertainty on $m_{\rm ins}$ and $M_{\rm std}$ is listed in column 6. For completeness, we also list the reference flux $f_{\rm ref}$ and the differential flux $f_{\rm diff}$ (columns 7 and 9 respectively), along with their uncertainties (columns 8 and 10), as well as the photometric scale factor $p$. Definitions of these quantities can be found in e.g. \cite{bramich11}, Eq. 2-3. This is a representative extract from the full table, which is available with the electronic version of the article. \label{tab:onlinedata}}
  \end{center}
\end{table*}

\subsection{Photometric calibration}

\subsubsection{Self-calibration}

For the CCD data, we carried out self-calibration of the light curves to correct for some of the systematics. Although systematics cannot be removed completely, substantial corrections can be made in the case of time-series photometry, as shown in our previous papers \citep[e.g.][]{kains13b}. 

\begin{figure}
  \centering
  \includegraphics[width=8cm, angle=0]{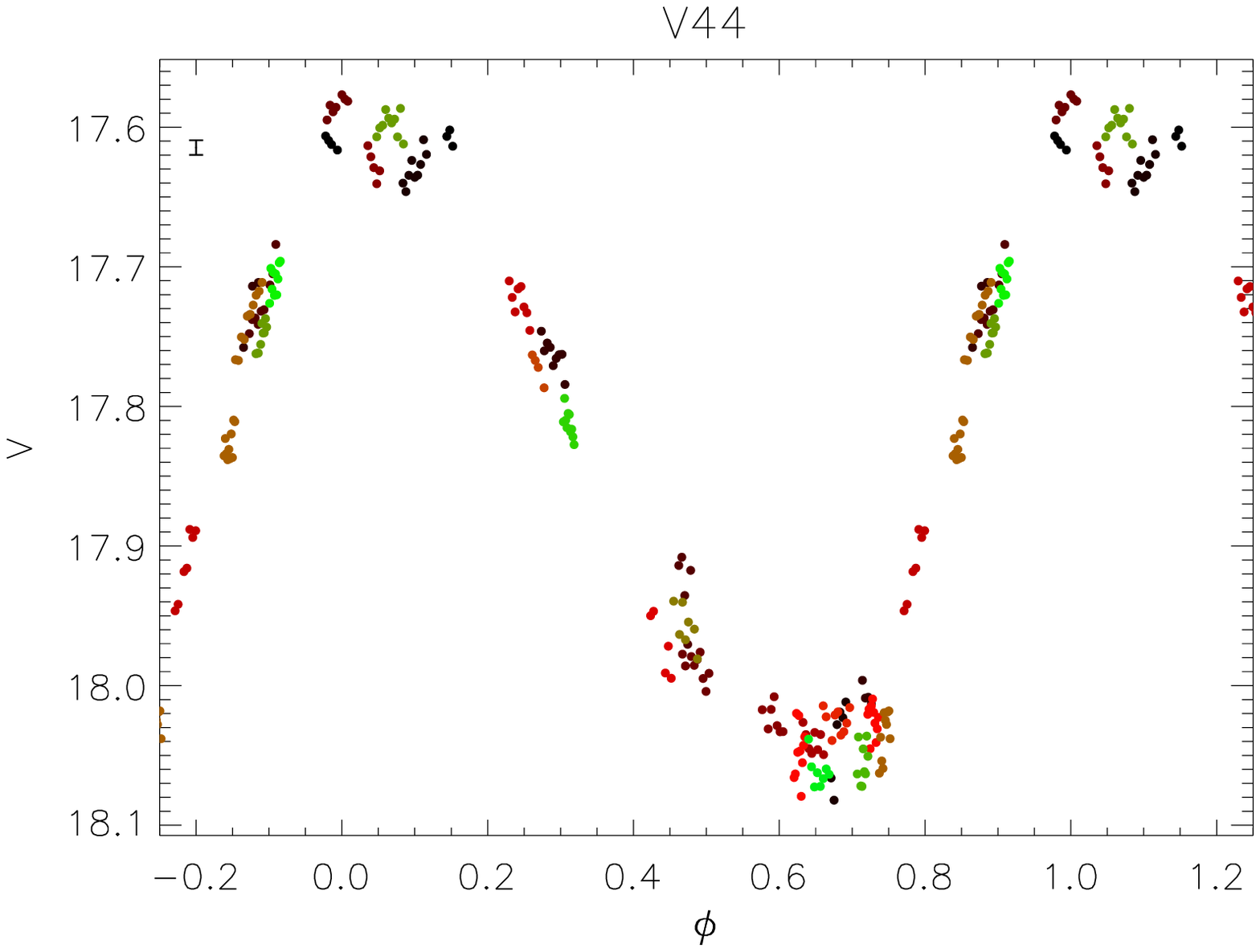}
  \includegraphics[width=8cm, angle=0]{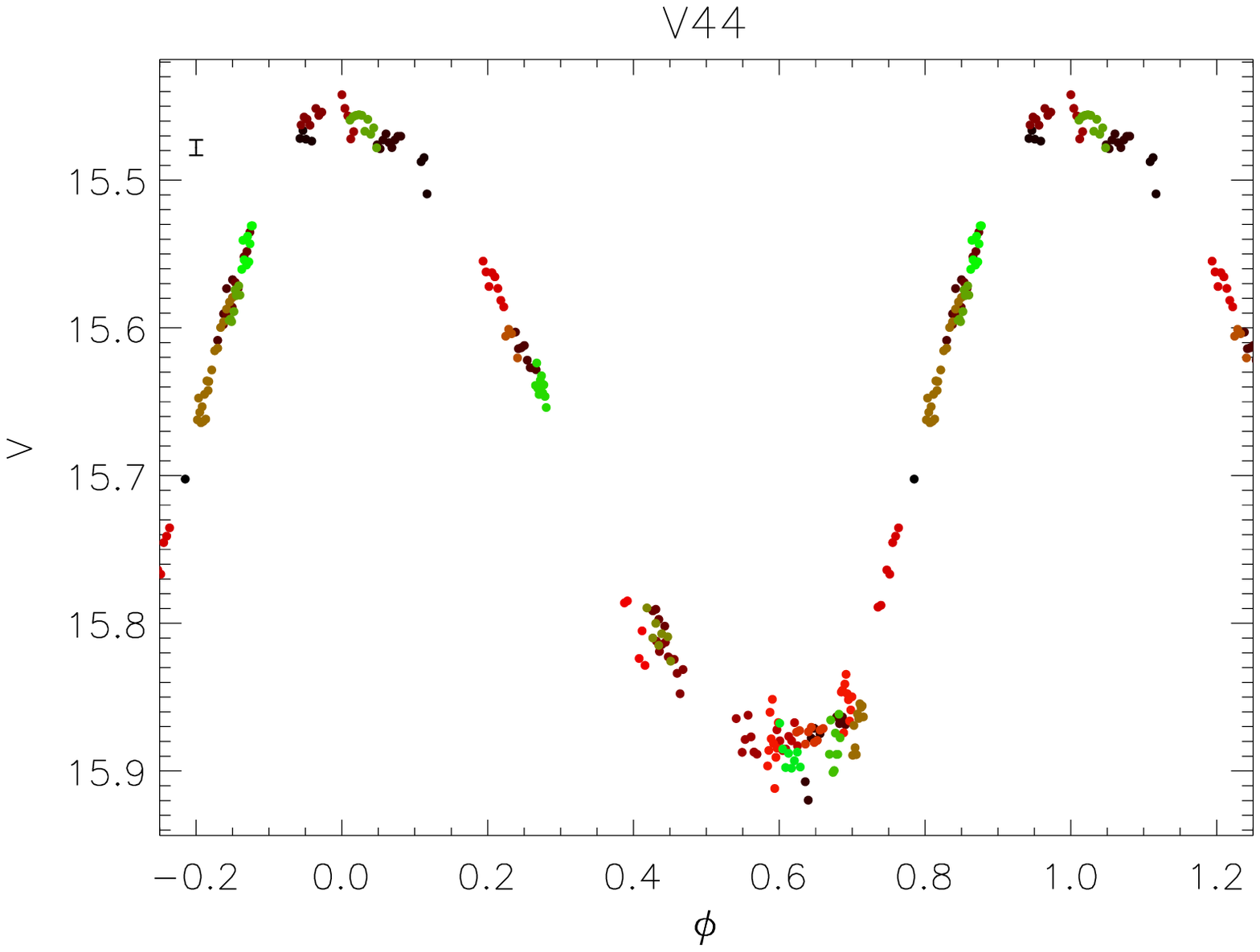}
  \caption{$V-$band light curves for V44 before (top) and after (bottom) self-calibration using the method of \cite{bramich12b}, showing significant improvement in light curve quality. \label{fig:calib}}
\end{figure}

We used the method of \cite{bramich12b} to derive magnitude offsets to be applied to each epoch of the photometry, which serve to correct for any errors in the fitted values of the photometric scale factors. The method involves setting up a (linear) photometric model for all of the available photometric measurements of all stars and solving for the best-fit parameter values by minimising $\chi^{2}$. In our case the model parameters consist of the star mean magnitudes and a magnitude offset for each image. The offsets we derive are of the order of a few percents, and lead to significant improvements in the light curves for this cluster. An illustration of this is shown in \Fig{fig:calib}.

\subsubsection{Photometric standards}

\begin{figure}
  \centering
  \includegraphics[width=8cm, angle=0]{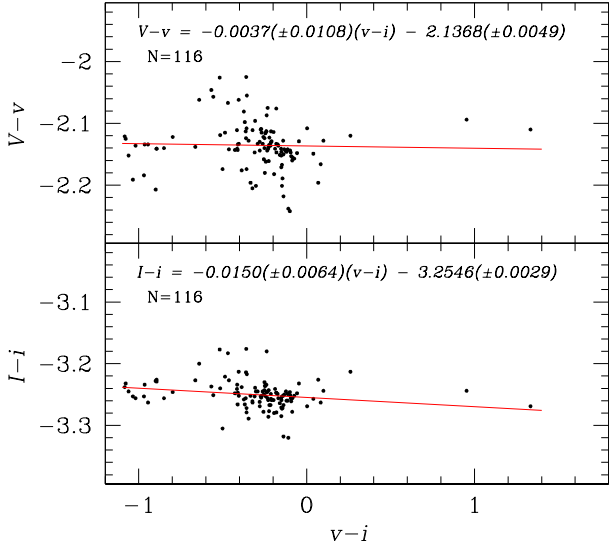}
  \caption{Relations used to convert from instrumental to standard Johnson-Kron-Cousins magnitudes for the $V$ (top) and $I$ (bottom) bands. \label{fig:transf}}

\end{figure}

We used secondary photometric standards in the FOV from \cite{stetson00} covering the full range of colours of our CMD to convert the instrumental magnitudes we obtained from the pipeline reduction of the CCD images to standard Johnson-Kron-Cousins magnitudes. This was done by fitting a linear relation to the difference between the standard and instrumental magnitudes, and the instrumental $v-i$ colour of each photometric standard in our images. The resulting transformation relations are shown in \Fig{fig:transf}. We compared our photometry with that of W94 by comparing mean magnitudes of RRL stars; this showed small differences of $< 1\%$ in both $V$ and $I$ (see \Tab{tab:variables} in \Sec{sec:variables} for our mean magnitudes).

\subsection{Astrometry}

We derived astrometry for our CCD reference images by using \textit{Gaia}\footnote{\tt http://star-www.dur.ac.uk/{\textasciitilde}pdraper/gaia/gaia.html} to match $\sim 300$ stars manually with the UCAC3 catalogue \citep{zacharias10}. For the EMCCD reference, we derived the transformation by matching 10 stars to HST/ WFC3 images \citep[e.g.][]{bellini11}. The coordinates we provide for all stars in this paper (Table \ref{tab:coordinates}) are taken from these astrometric fits. The rms of the fit residuals are 0.27 arcsec (0.57 pixel) for the CCD reference images and 0.09 arcsec (0.96 pixel) for the EMCCD reference.

\section{Variables in M68}\label{sec:variables}

The first 28 (V1-V28) variables in this cluster were identified by H. Shapley and M. Ritchie \citep{shapley19, shapley20}, using fifteen photographs obtained with the 60-inch reflector telescope at Mt Wilson Observatory. All of these are RRL stars, except for V27, which was identified in the 1920 paper as a long-period variable; \cite{greenstein47} used a spectrum of V27 taken at the McDonald Observatory in 1939 to work out its radial velocity, and compared this to the cluster's radial velocity to conclude that V27 is a long-period foreground variable star. \cite{rosino54} then used observations taken between 1951 and 1953 at Lojano Observatory to derive periods for 20 variables, and discovered three additional RRL stars (V29-V31). They also noted some irregularities in V3, V4, V29 and V30, whereby they could not derive precise periods for those four stars. \cite{vanagt59} found 7 more variables (V32-V38) analysing observations taken in 1950 with the Radcliffe 74-inch reflector telescope in South Africa. They also noticed many discrepancies between their derived periods and those published a few years earlier by \cite{rosino54}, as well as differences in light curve morphologies. \cite{terzan73} announced another four variables in M68 (V39-V42), including the first SX Phe star in this cluster. \cite{clement90} and C93 then studied 30 of the RRL stars, identifying 9 double-mode pulsators and detecting period changes since the earlier studies. \cite{brocato94} carried out the first CCD-era study of this cluster using the 1.5m ESO Danish Telescope, and soon after, W94 used CCD observations made in 1993 at the CTIO 0.9m telescope in Chile to discover an additional six variables (V43-V48), including another SX Phe star. Finally, \cite{sariya14} recently used observations from 2011 at the Sampurnanand telescope in Nainital, Northern India, to claim 9 new variable detections, including 5 RR1 stars, bringing the total number of published variables in this cluster to 57.

\subsection{Detection of variables}

\subsubsection{CCD observations}

\begin{figure}
  \centering
  \includegraphics[width=8cm, angle=0]{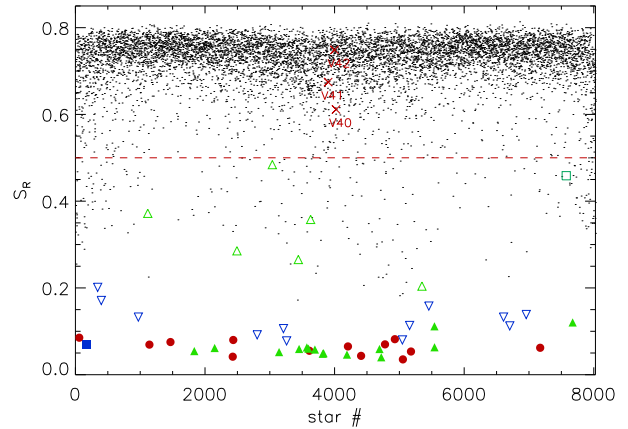}
  \caption{Plot of the distribution of the $S_R$ statistic as defined in the text, for our $V$-band light curves. A dashed line marks the threshold below which we searched for periodic variables. RR0, RR1 and RR01 variables are shown as red filled circles, green filled triangles and blue open inverted triangles respectively, SX Phe as open green triangles, the field Mira variable V27 (FI Hya) as a filled blue square, and the variable of unknown type V53 as an open green square. The three marked stars above the threshold are V40-V42, which we found to be non-variable. \label{fig:sq}}

\end{figure}

We searched for variables using two methods: firstly we inspected the difference images visually and checked light curves of any objects that had residuals on a significant number of images. This did not enable us to detect any new variables. We also constructed an image from the sum of the absolute values of all difference images and inspected light curves at pixel positions with significant peaks on this stacked image; as with the first method, this method recovered most known variables, but did not enable us to detect new variables. Finally, we also conducted a period search, for periods ranging from 0.02 to 2 days, on all light curves using the ``string length" method \citep[e.g.][]{dworetsky83}, and computed the ratio $S_R$ of the string length for the best-fit period to that for the worst (i.e. the string length for phasing with a random period). For light curves without periodic variations, $S_R$ is expected to be close to 1, although in practice, due to light curve scatter, the mean value is around 0.75; for true periodic light curves, $S_R \ll 1$. The distribution of $S_R$ is shown in \Fig{fig:sq}. We inspected all light curves that fell below an arbitrary threshold of $S_R=0.5$, chosen by visually inspecting light curves sorted with ascending $S_R$. 

\begin{figure}
  \centering
  \includegraphics[width=8cm, angle=0]{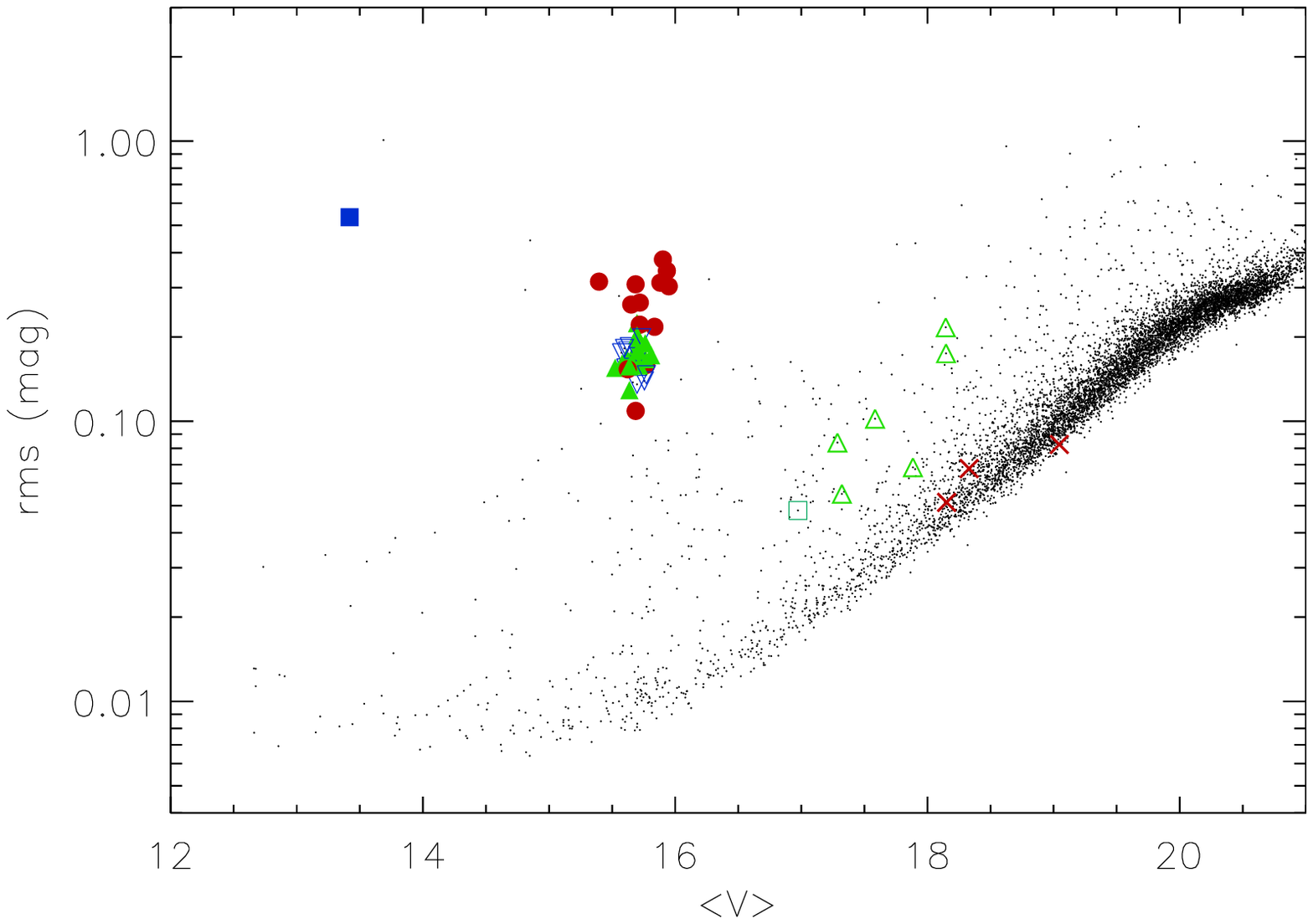}
  \includegraphics[width=8cm, angle=0]{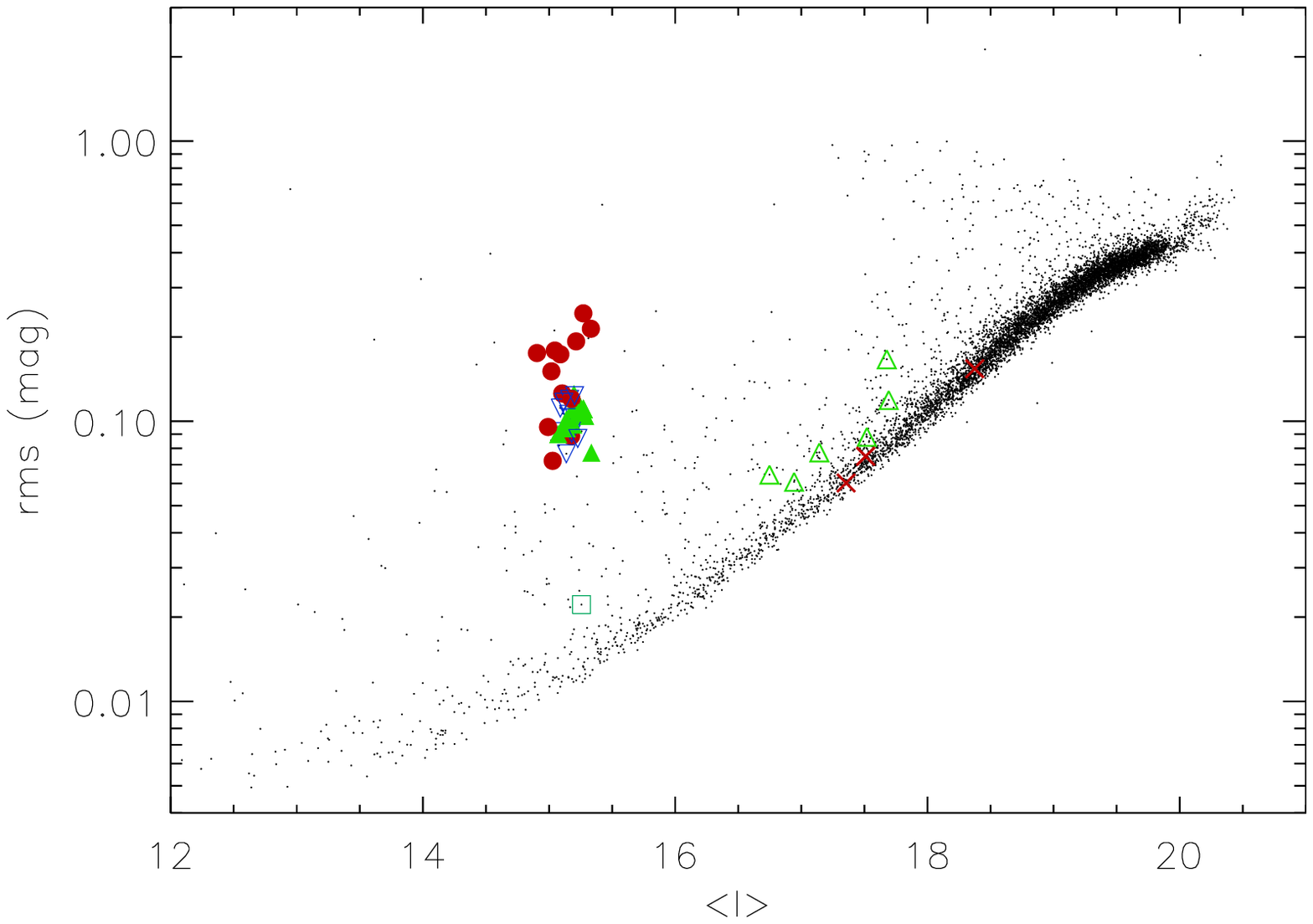}
  \caption{Root mean square magnitude deviation versus mean magnitude for all stars for which photometry was obtained. Plots are for the $V-$band (top) and $I-$band (bottom). Classified variables are marked with red filled circles (RR0), green filled triangles (RR1) and blue open inverted triangles (RR01), green open triangles (SX Phe), a blue square (for V27, the field Mira variable FI Hya), and a green open square (variable V53, of unknown type). Non-variable objects previously catalogued as variable in the literature are marked with red crosses. \label{fig:rms}}
\end{figure}

Using this method, we recovered all known variables except for V32, which is outside our field of view; we were also able to derive periods for all of them, except for V27, which is now known to be a foreground variable star with a period of $\sim322$ d \citep{pojmanski02}; for V27 we also do not have an $I-$band light curve because it is saturated in our reference image. We also discovered four new variables, all of them SX Phe stars. Furthermore, we find that V40-42 are not variable within the limits of the rms in our data, given in \Tab{tab:rms_nonvar}, in agreement with the findings of W94. We also find that none of the new variables recently claimed by \cite{sariya14} is variable within the rms scatter of our data (Table \ref{tab:rms_nonvar}, see also \Fig{fig:rms}), and we therefore continue the variable numbering system from its standing prior to the publication of their paper. 

We also confirm that many of the RRL stars are double-mode variables, as previously reported by \cite{clement90} and C93, and we determine pulsation periods for both modes, when possible. Double-mode RRL stars in this cluster are discussed in \Sec{sec:rrd}.

In order to obtain the best possible period estimate for each star, we used archival data from previous studies of variability in this cluster by \cite{rosino53}, \cite{rosino54}, C93, W94, and \cite{brocato94}. This gives us a baseline of up to 62 years for stars which were observed by \cite{rosino53}, and of over 20 years for the stars that were observed in the 1993-1994 studies. The data of \cite{rosino53}, \cite{rosino54}, and \cite{brocato94} were previously not available in electronic format, and we have uploaded the light curves to the CDS\footnote{The CDS data is available through the data link on the ADS records for these papers.} for interested readers. For some of the stars, it was not possible to phase-fold the data sets without also fitting for a linear period change; for some even this did not lead to well-phased data sets, suggesting that some other effect is at work, such as a non-linear period change. Those are discussed in \Sec{sec:individual}.

We also performed frequency analysis on all of the light curves in order to characterise the Blazhko effect \citep{blazhko1907}, which can cause scatter in the phased light curve due to modulation of amplitude, frequency, phase, or a combination of those. We discuss the results of this search in \Sec{sec:individual}.

Finally, we inspected the light curve of the standard star S28, which W94 found to be variable with an amplitude of $\sim 0.1$ mag. Our data show an increase in brightness of $\sim 0.1$ mag as well over the time span of our observations, confirming the variable nature of this star; we therefore assign it the variable number V53. Interestingly, however, we do not find significant variation in the $I-$band light curve of this star.

\begin{figure*}
  \centering
  \includegraphics[width=14cm, angle=0]{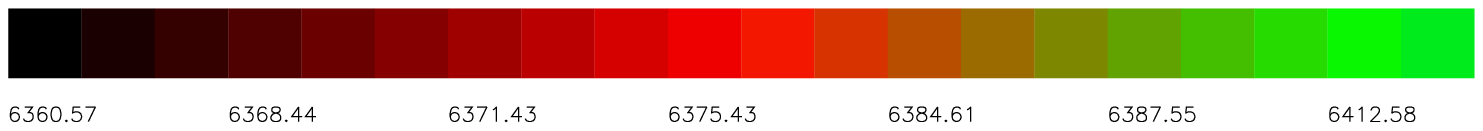}   \\
   \includegraphics[width=6cm, angle=0]{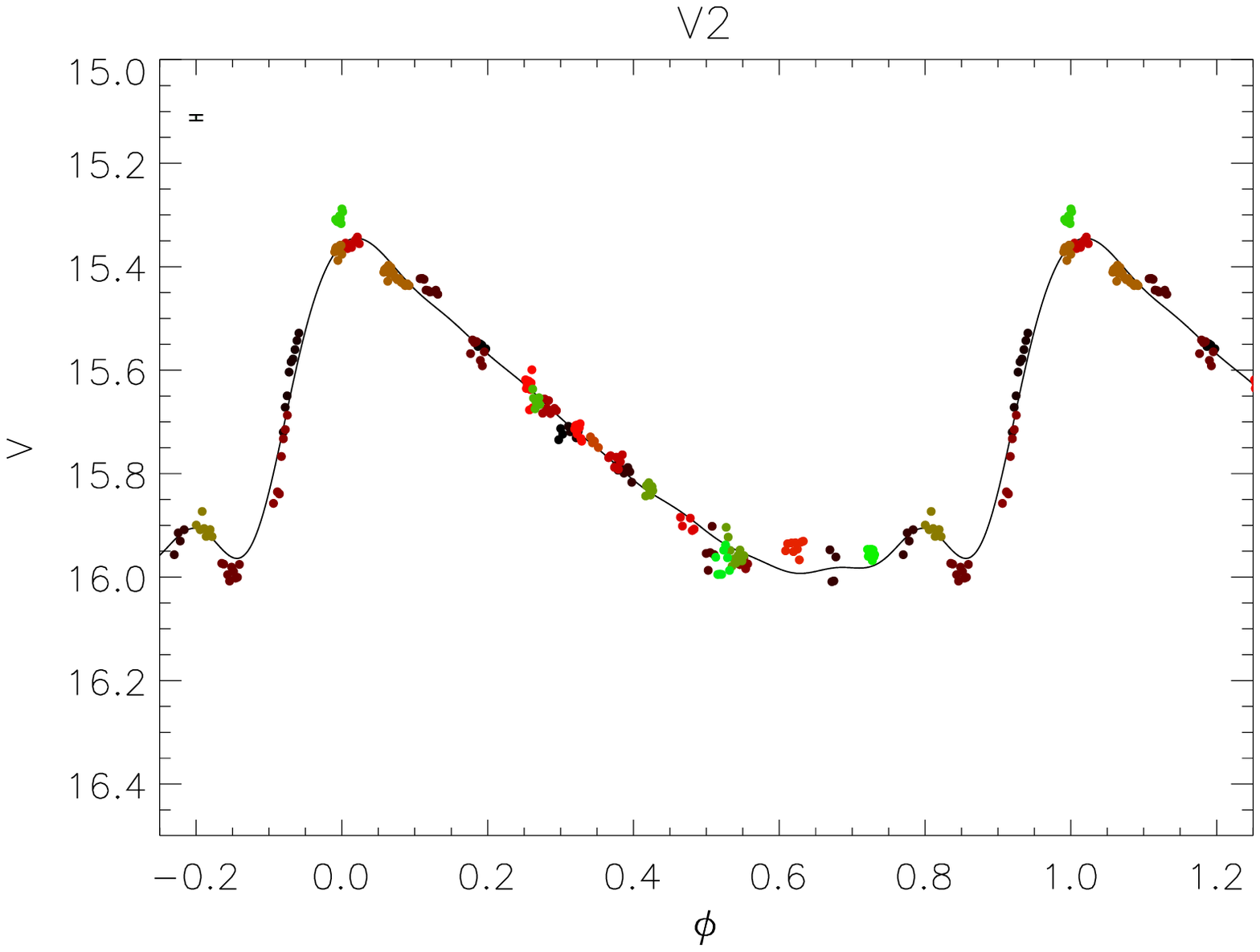}
  \includegraphics[width=6cm, angle=0]{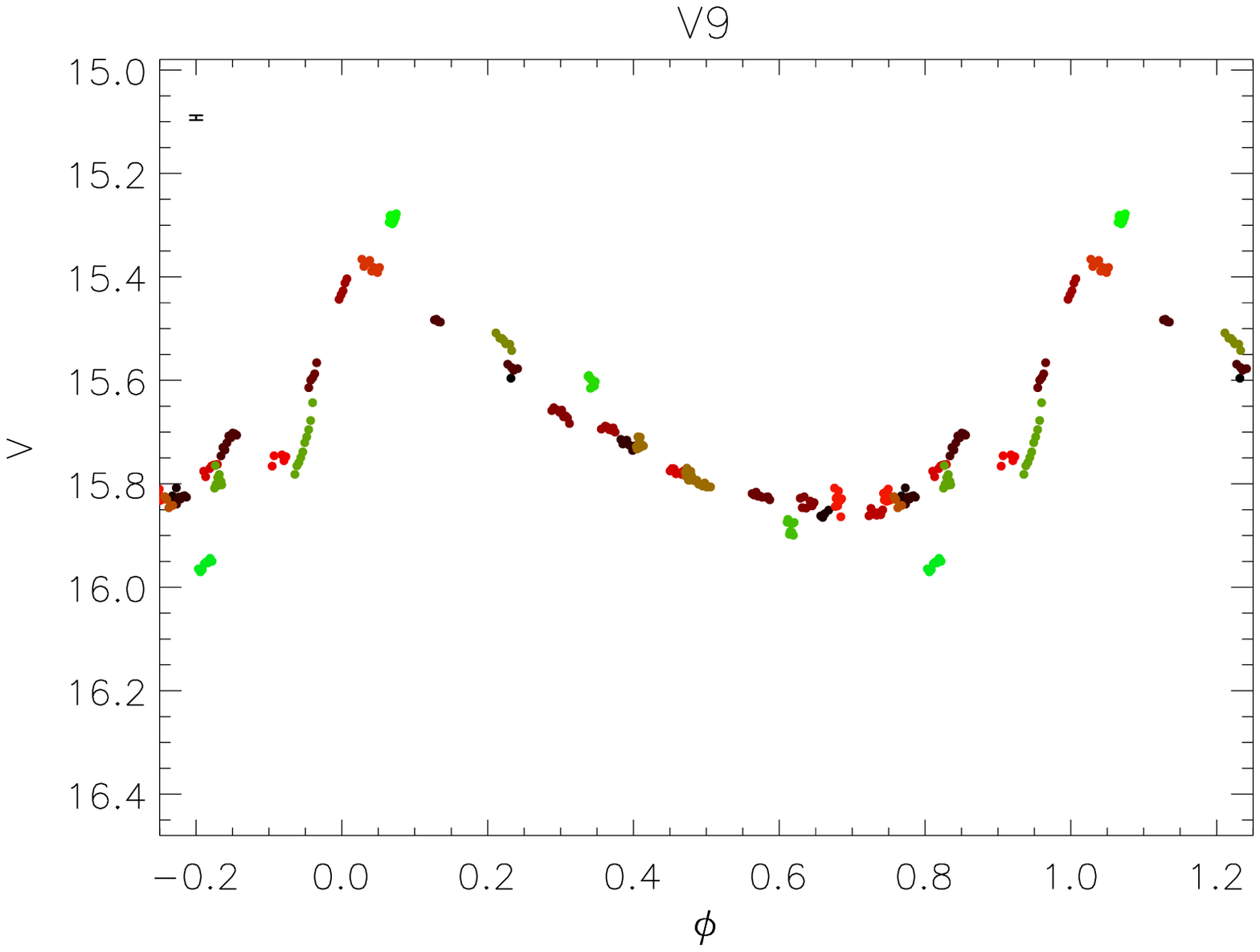}
  \includegraphics[width=6cm, angle=0]{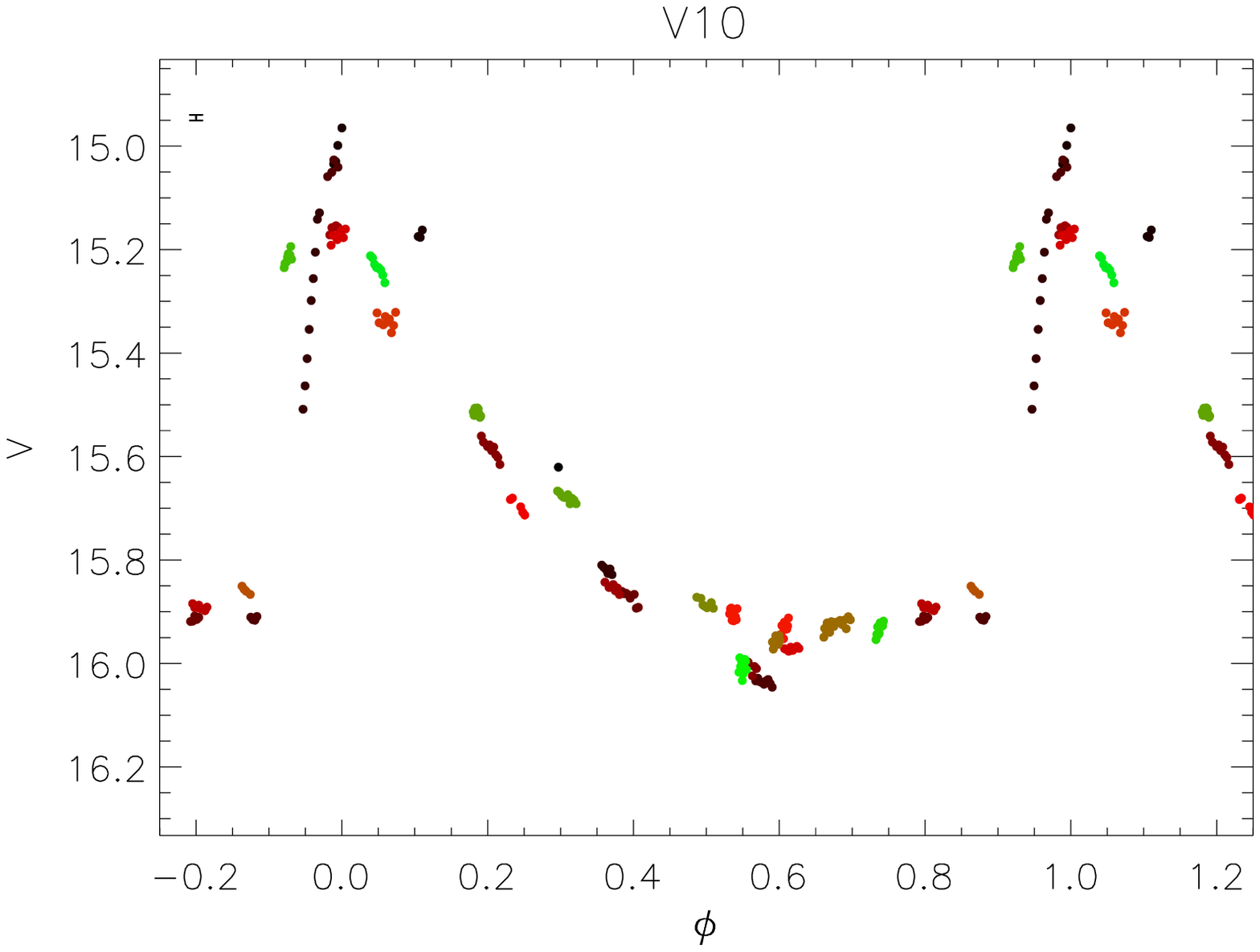}
  \includegraphics[width=6cm, angle=0]{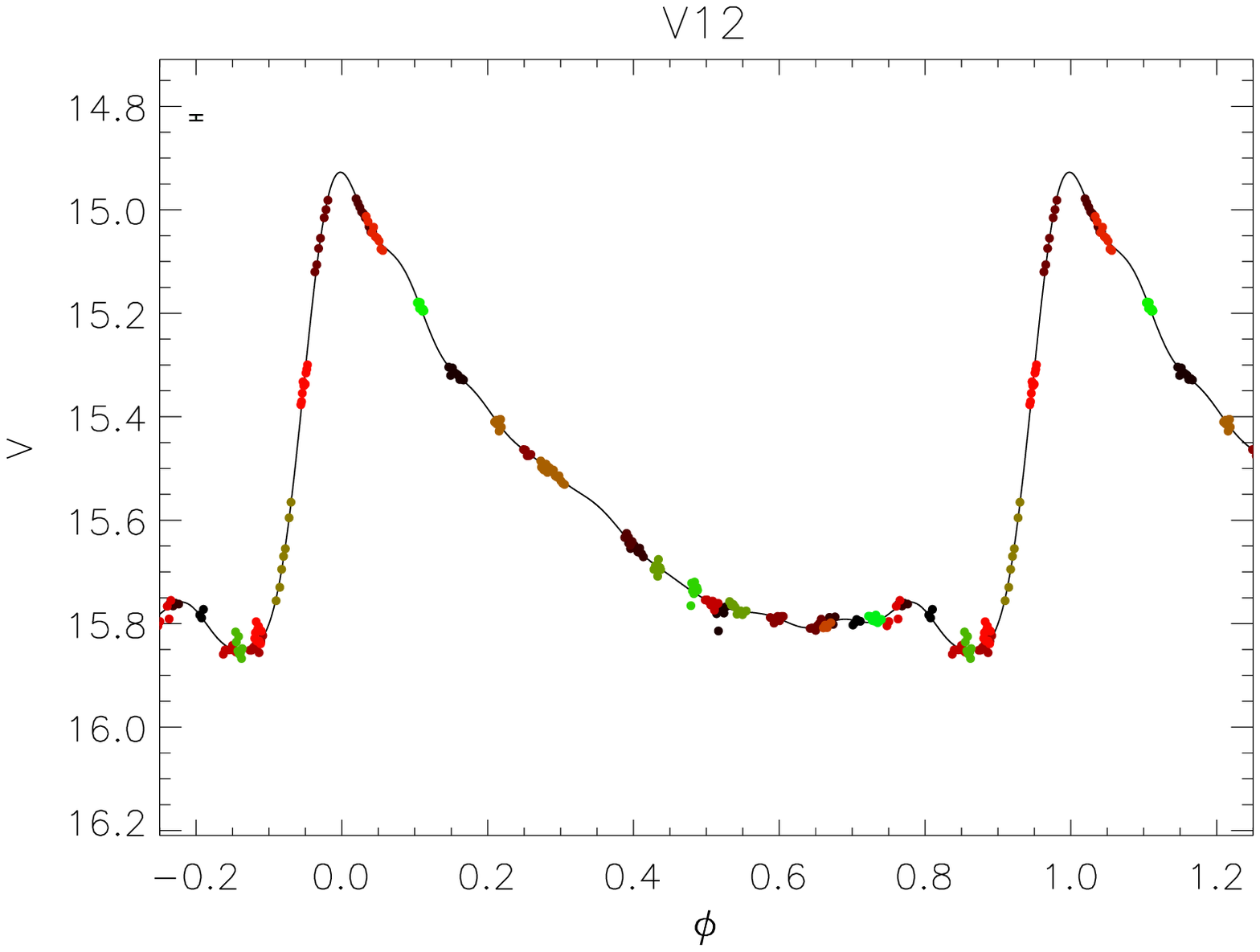}
   \includegraphics[width=6cm, angle=0]{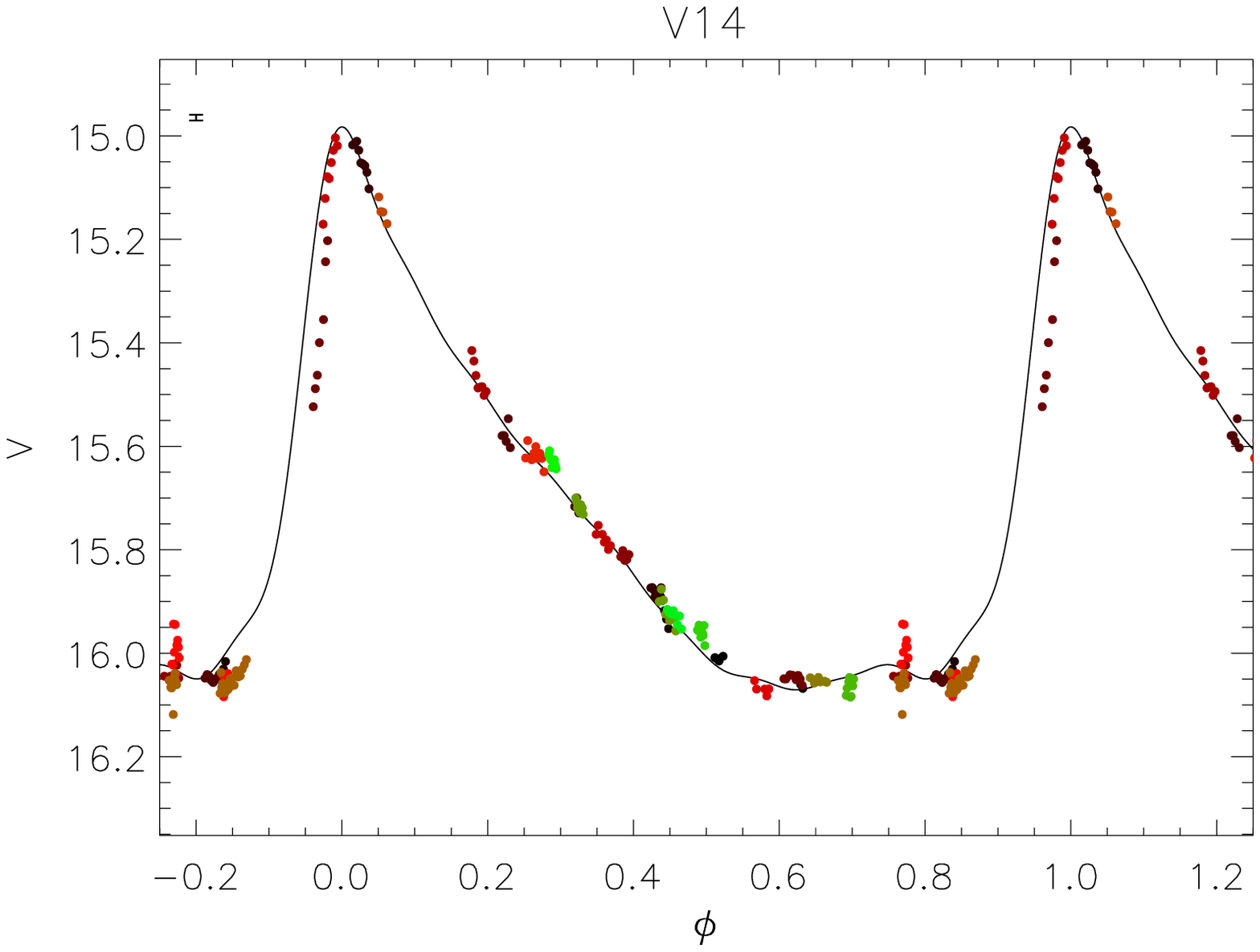}
  \includegraphics[width=6cm, angle=0]{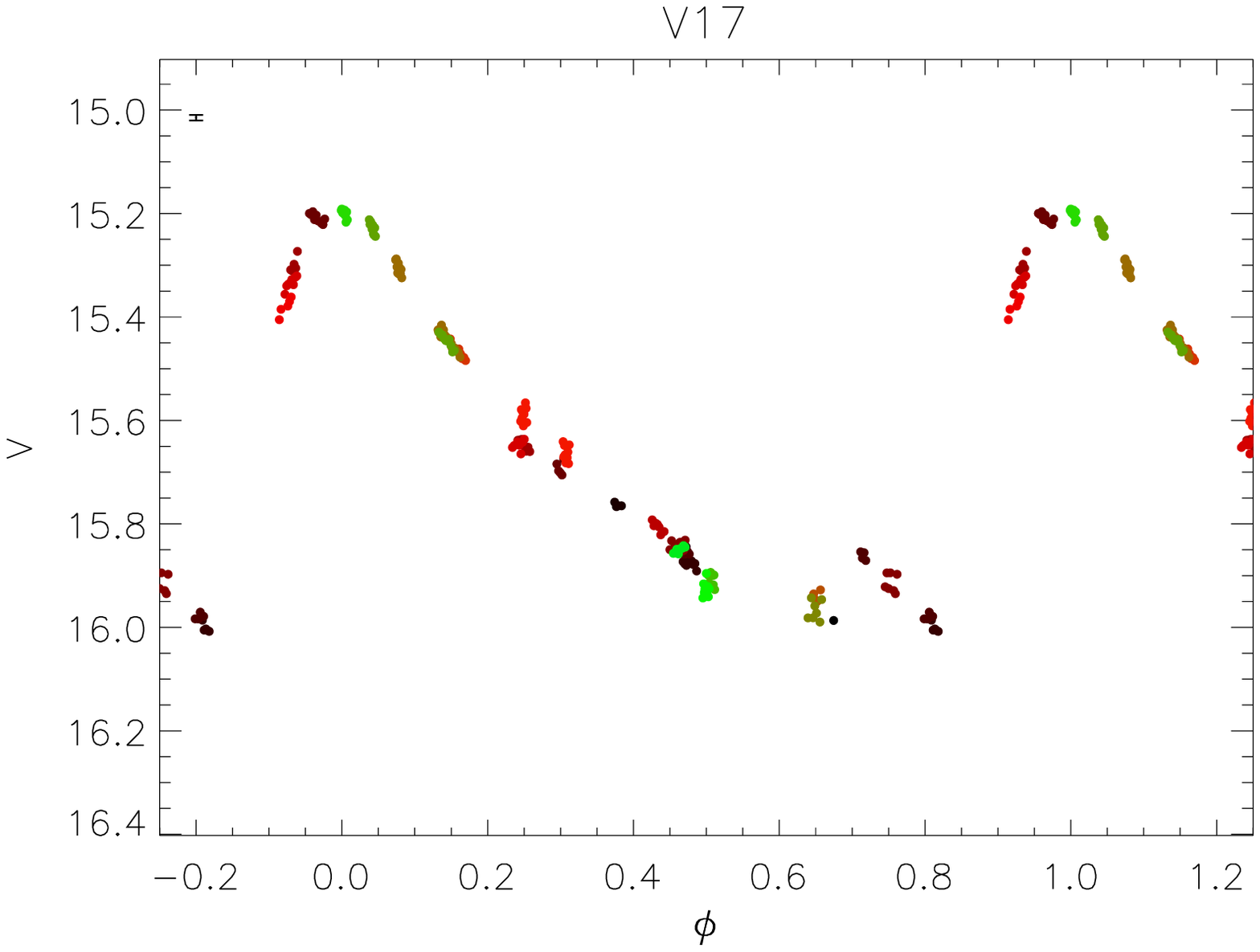}
  \includegraphics[width=6cm, angle=0]{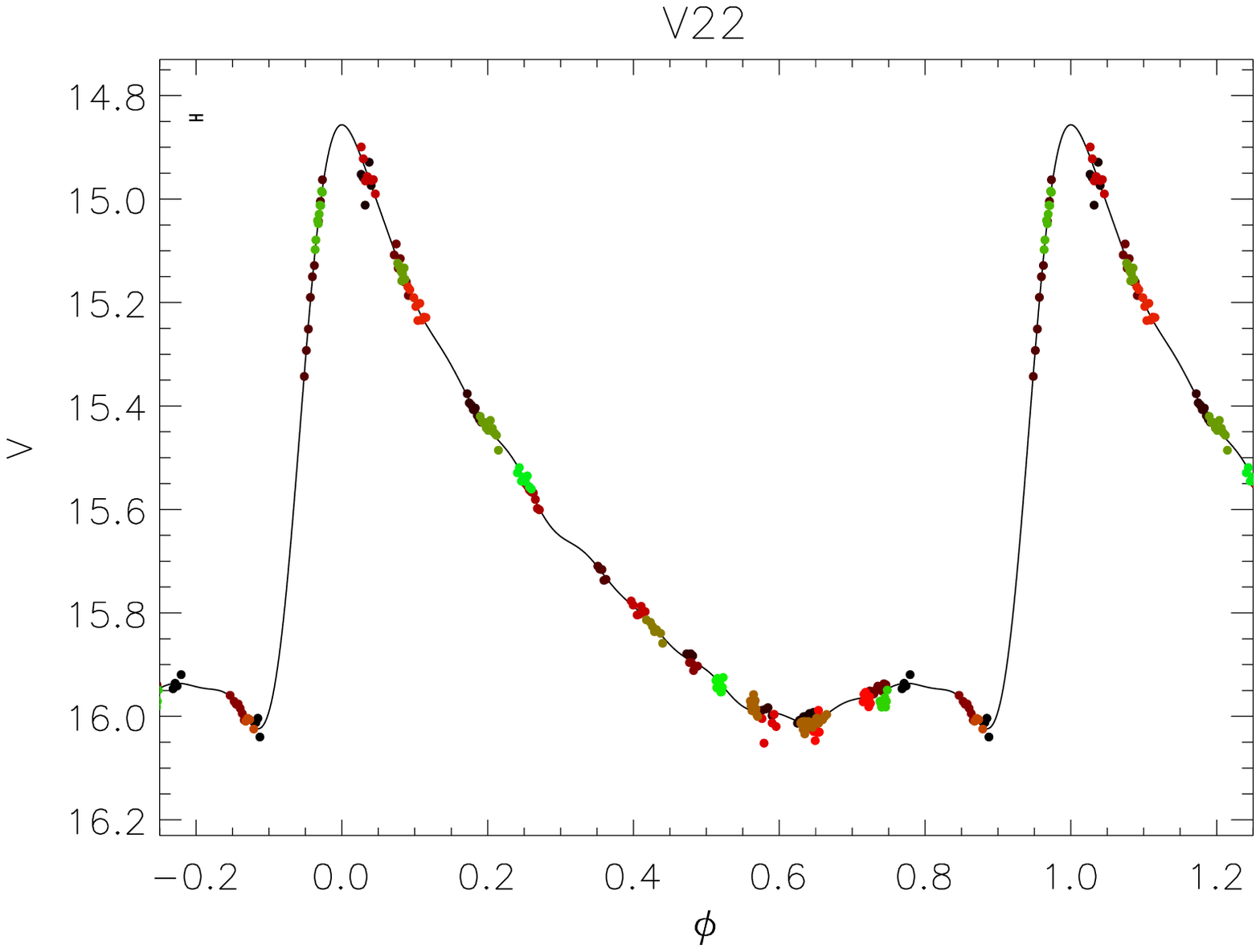}
  \includegraphics[width=6cm, angle=0]{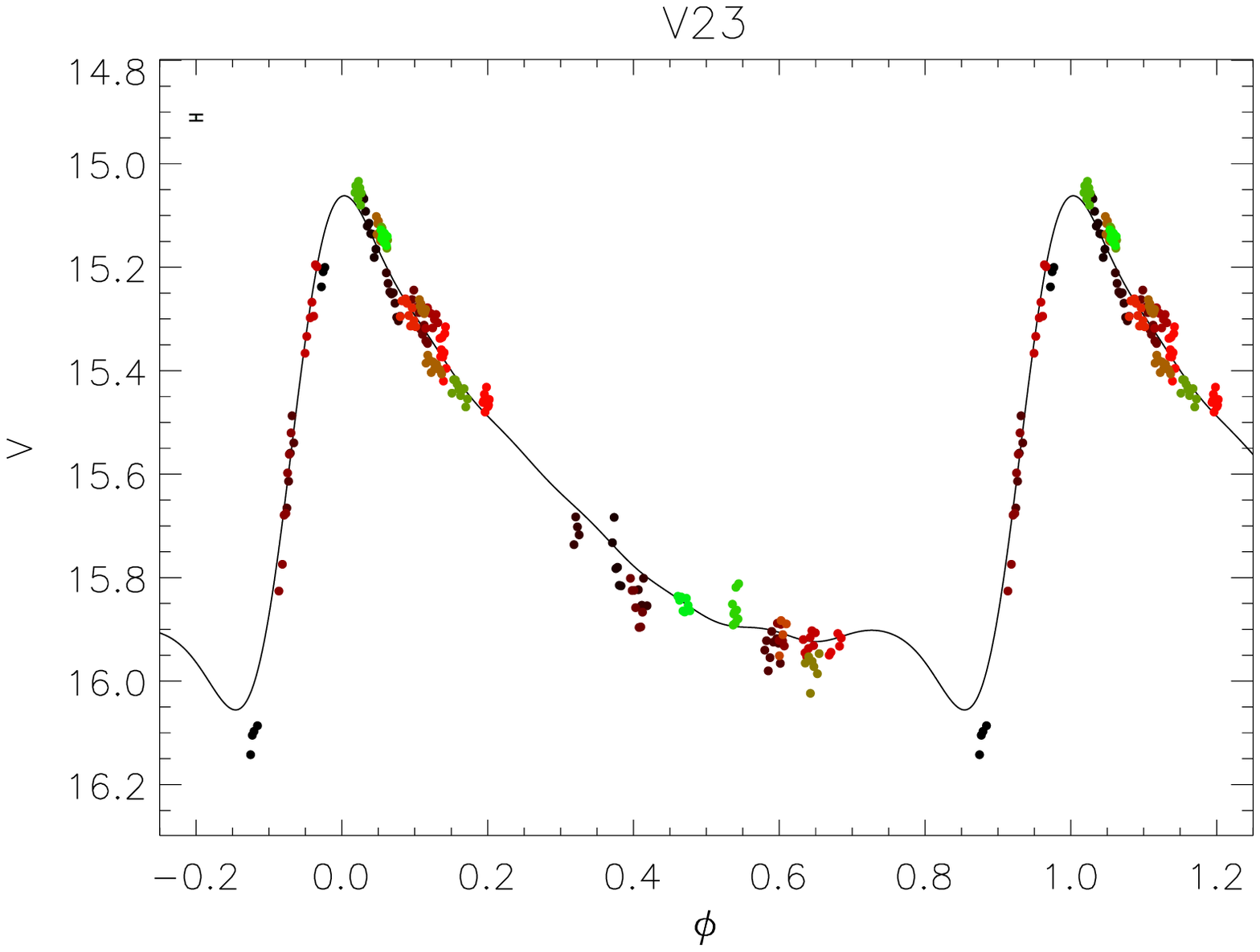}
  \includegraphics[width=6cm, angle=0]{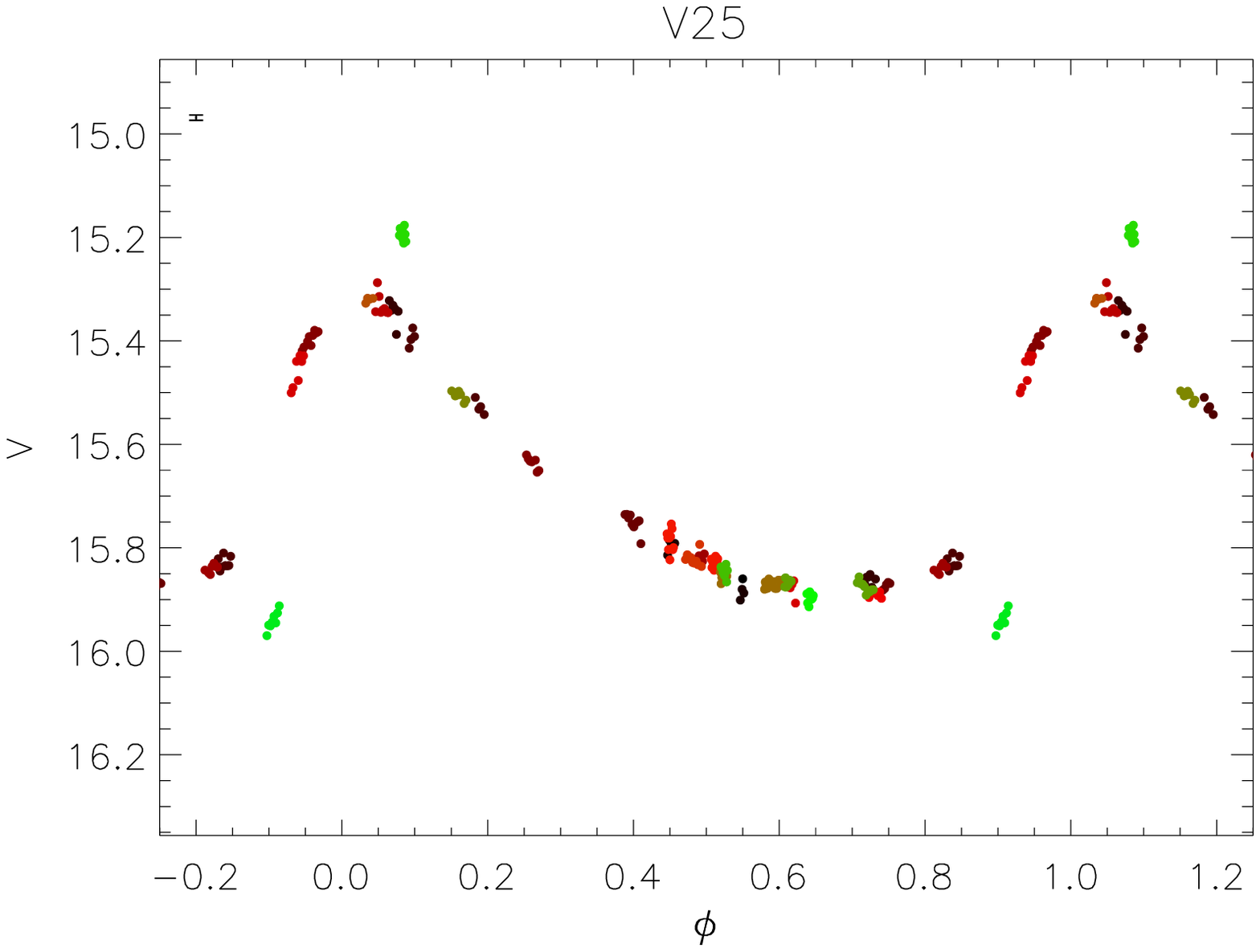}
  \includegraphics[width=6cm, angle=0]{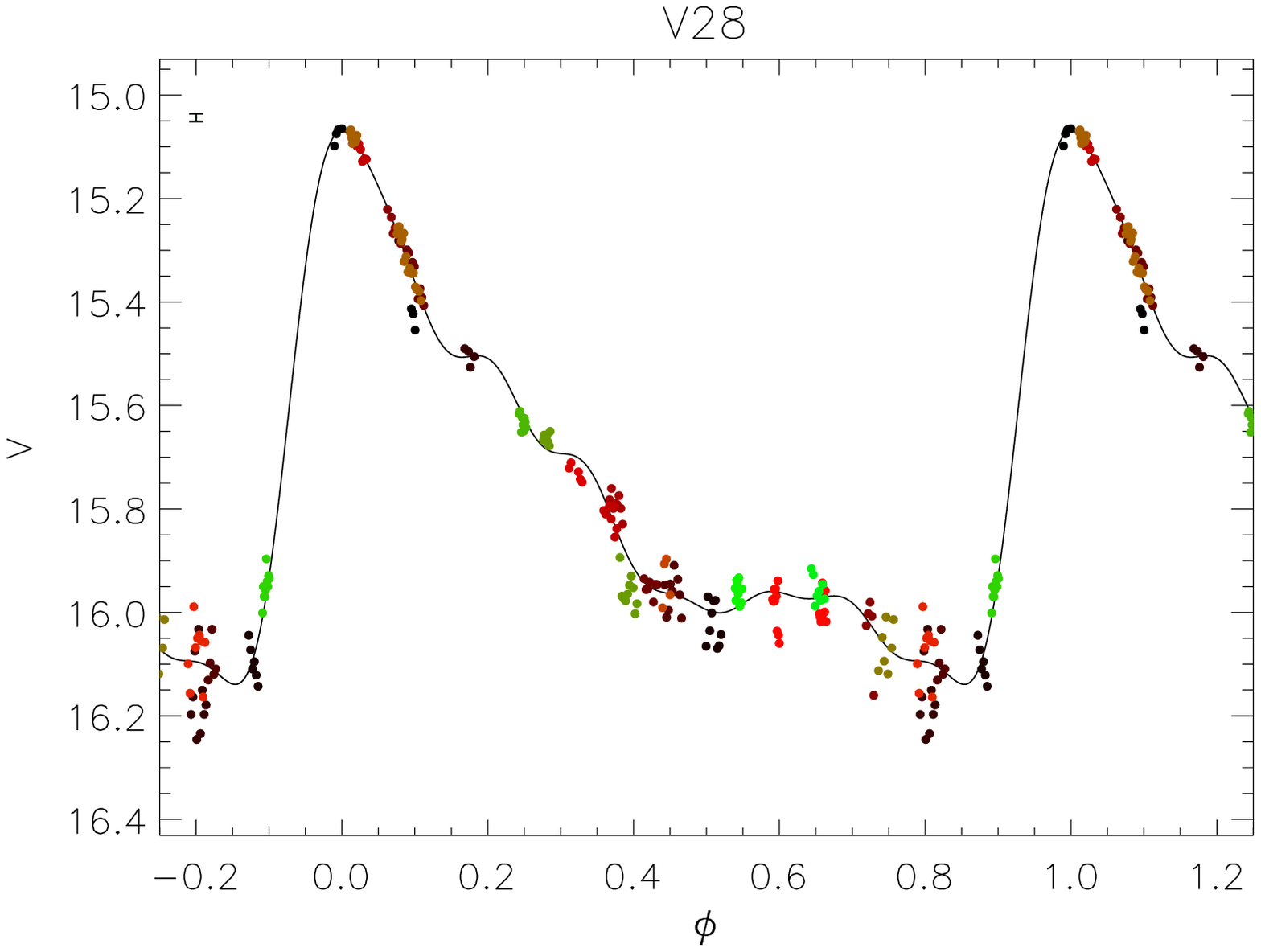}
  \includegraphics[width=6cm, angle=0]{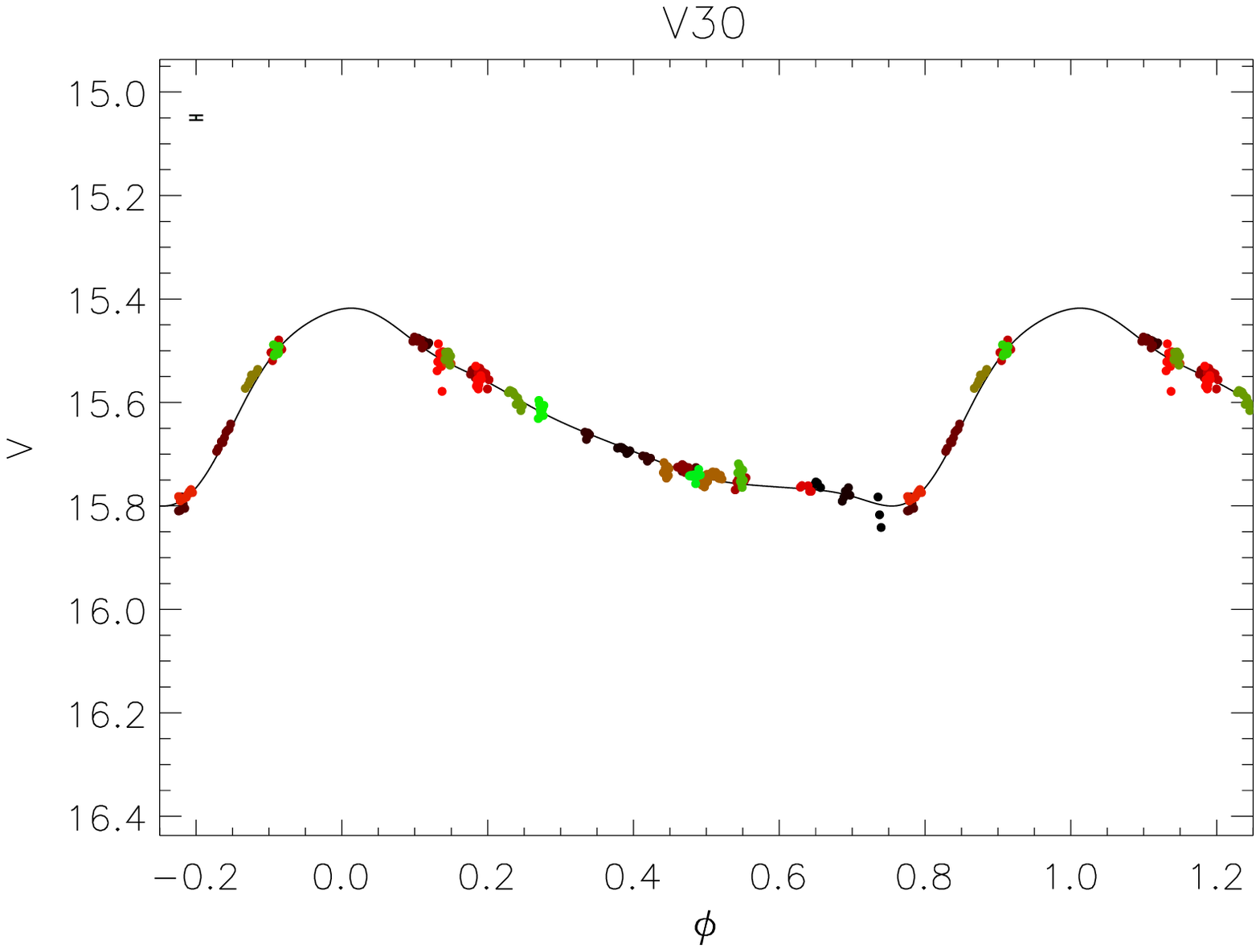}
  \includegraphics[width=6cm, angle=0]{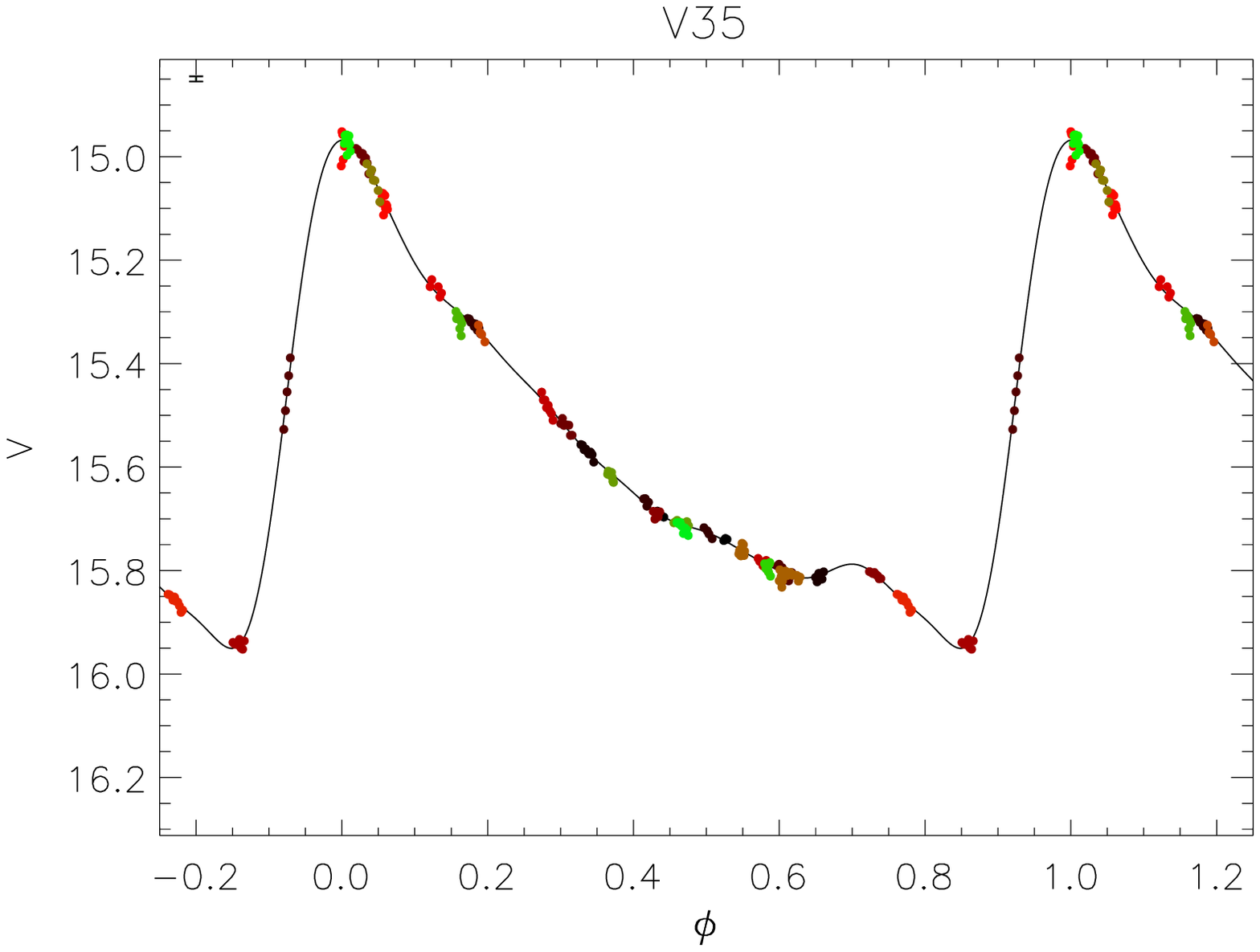}
  \includegraphics[width=6cm, angle=0]{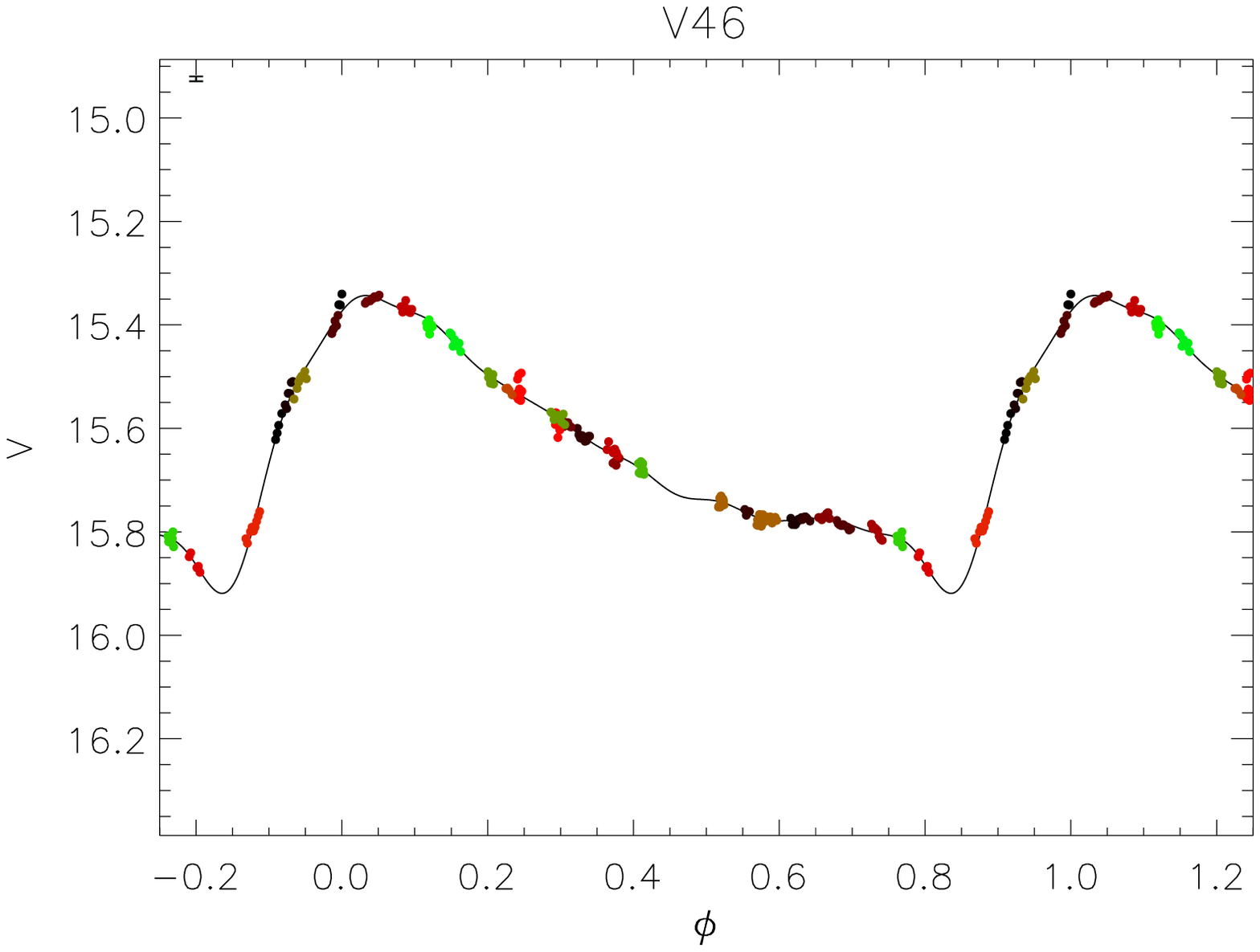}

\caption{Phased $V$-band light curves of the confirmed RR0 variables in M68. Data from different nights are plotted in different colours (electronic version only), with a colour bar provided for reference (top panel). On the light curves for which a good Fourier decomposition could be obtained, the fit is overplotted. The size of typical 1-$\sigma$ error bars is plotted in the top left corner. The magnitude scale is the same on all plots in order to facilitate comparison of variation amplitude. \label{fig:lc_RR0V}}

\end{figure*}

\begin{figure*}
  \centering
  \includegraphics[width=14cm, angle=0]{fig/colourcode.ps}   \\
  \includegraphics[width=4.5cm, angle=0]{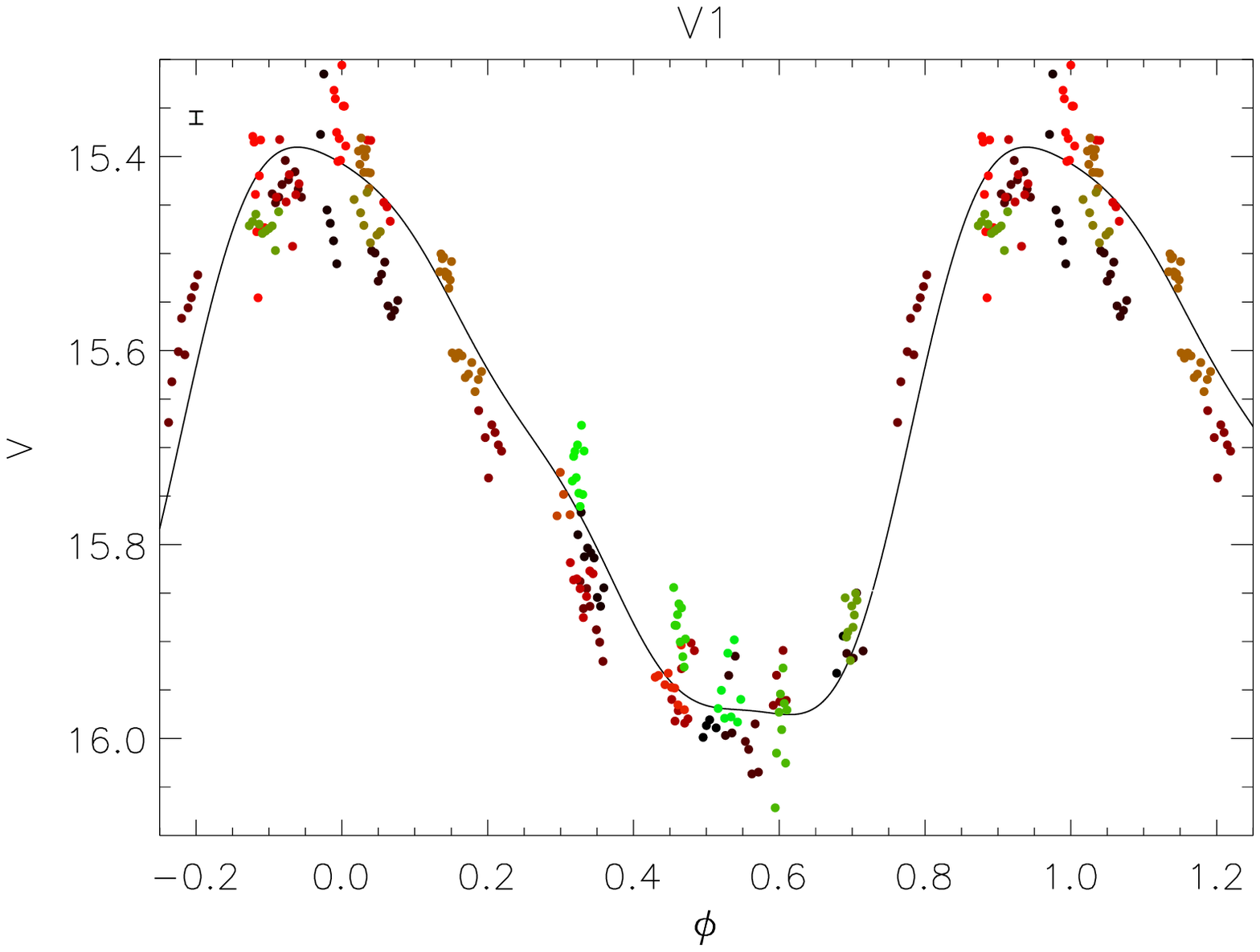}
  \includegraphics[width=4.5cm, angle=0]{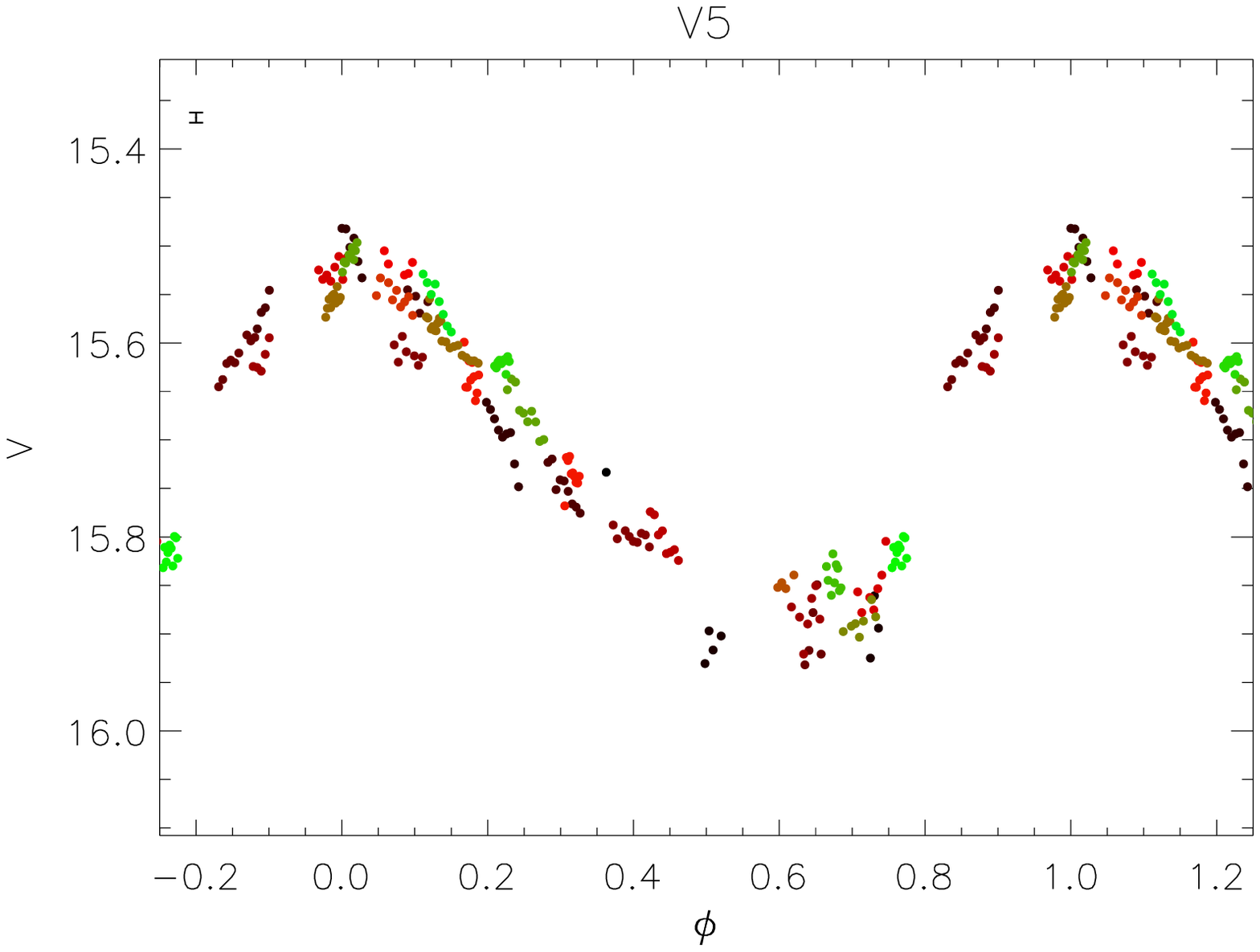}
  \includegraphics[width=4.5cm, angle=0]{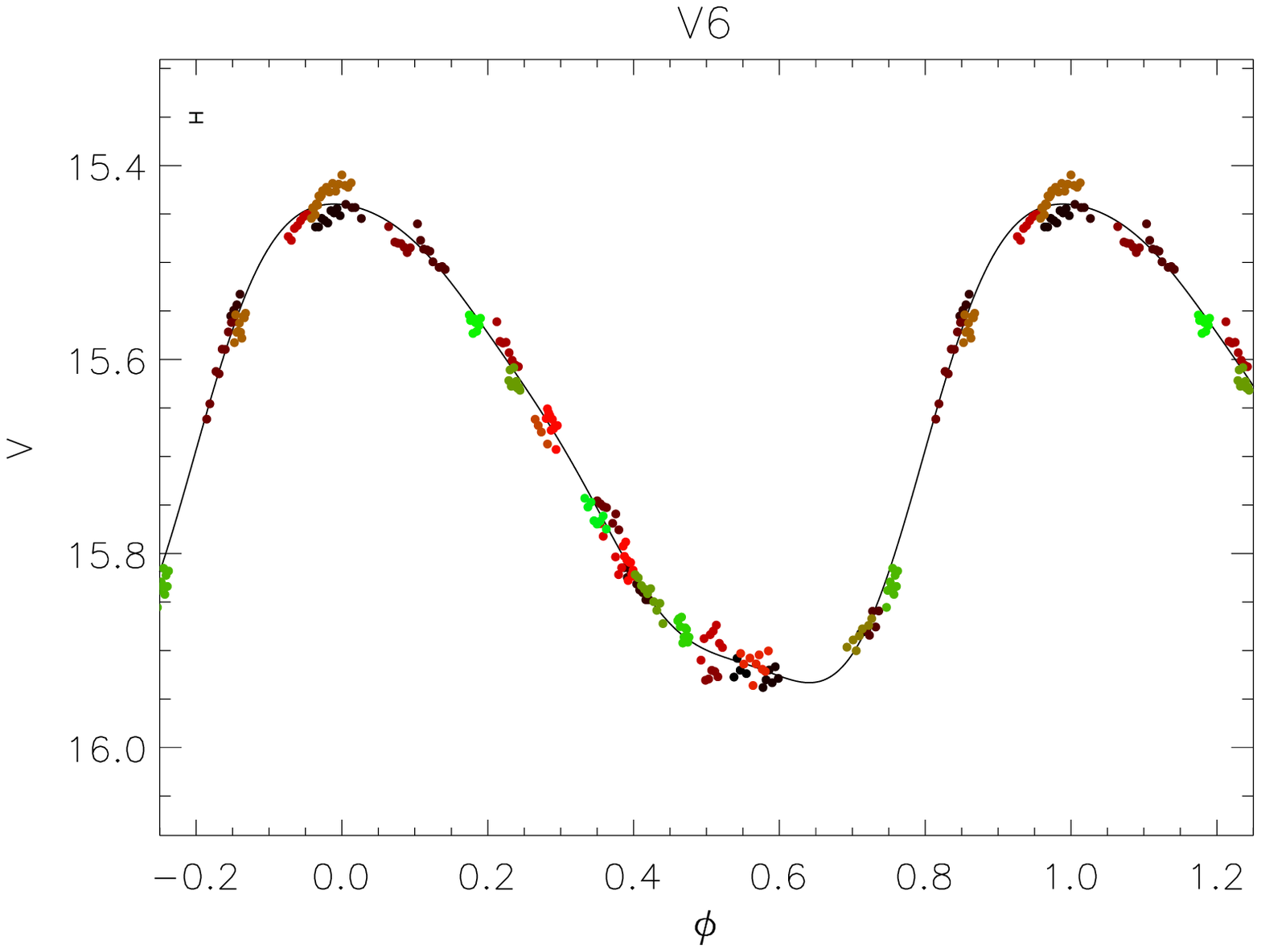}
  \includegraphics[width=4.5cm, angle=0]{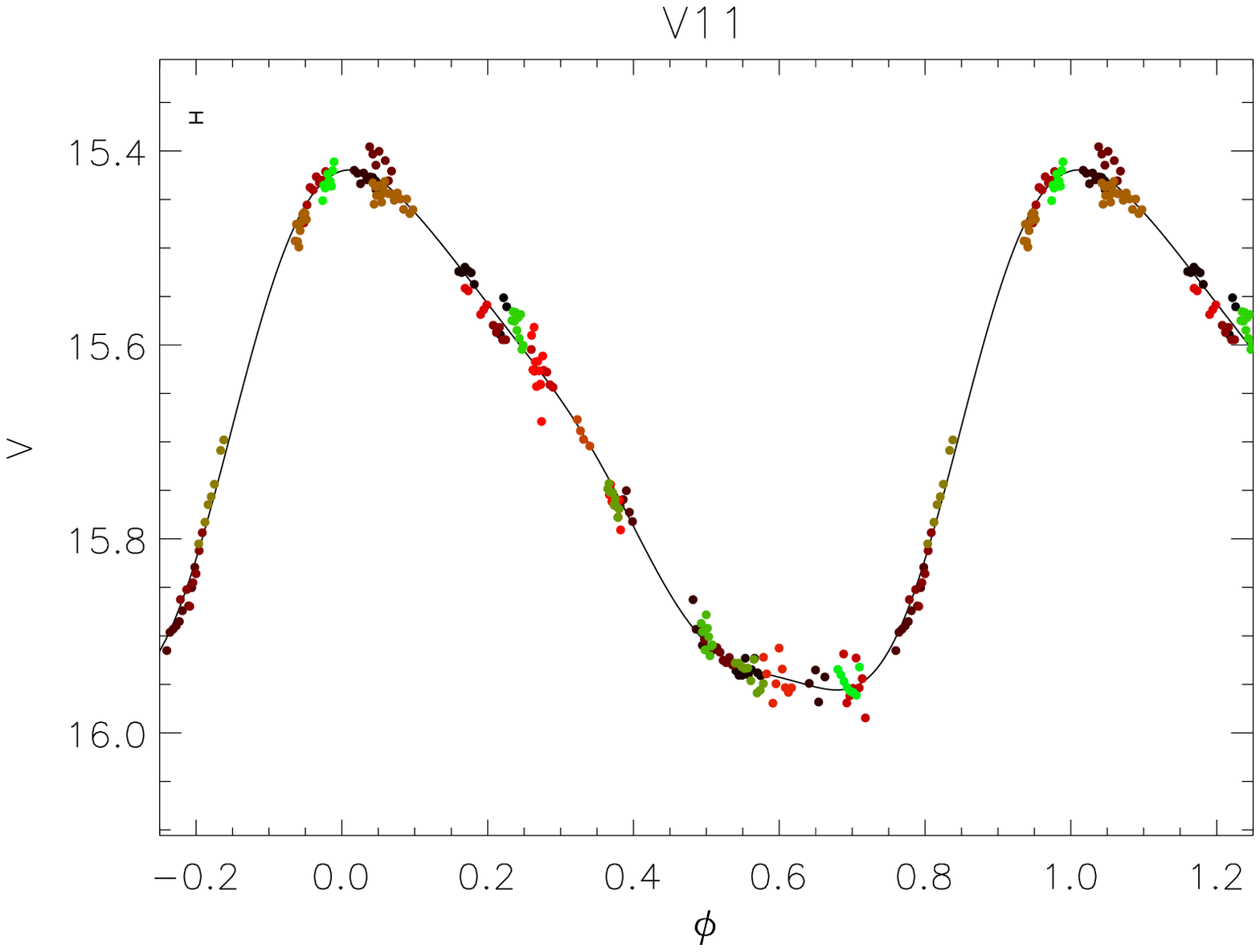}
  \includegraphics[width=4.5cm, angle=0]{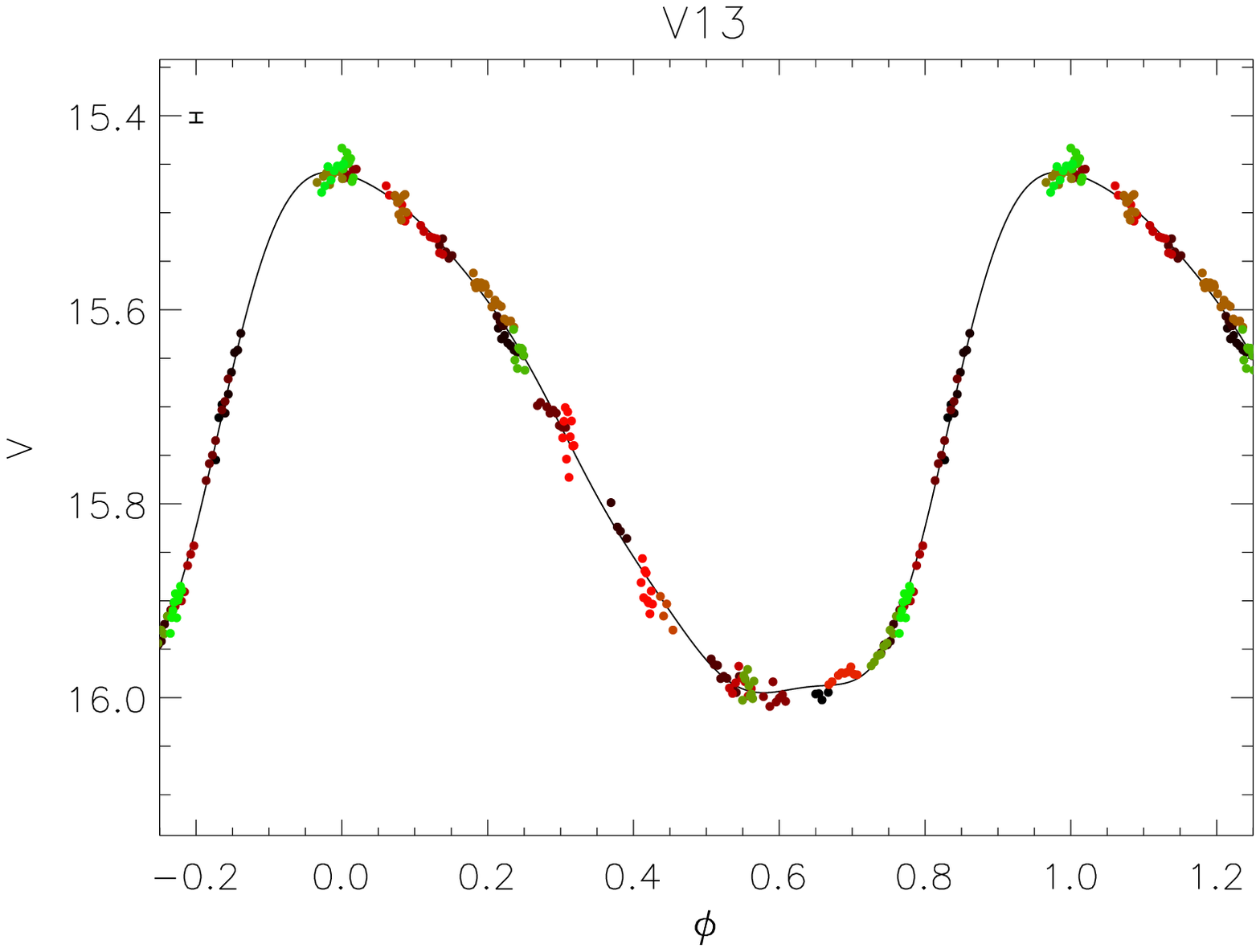}
  \includegraphics[width=4.5cm, angle=0]{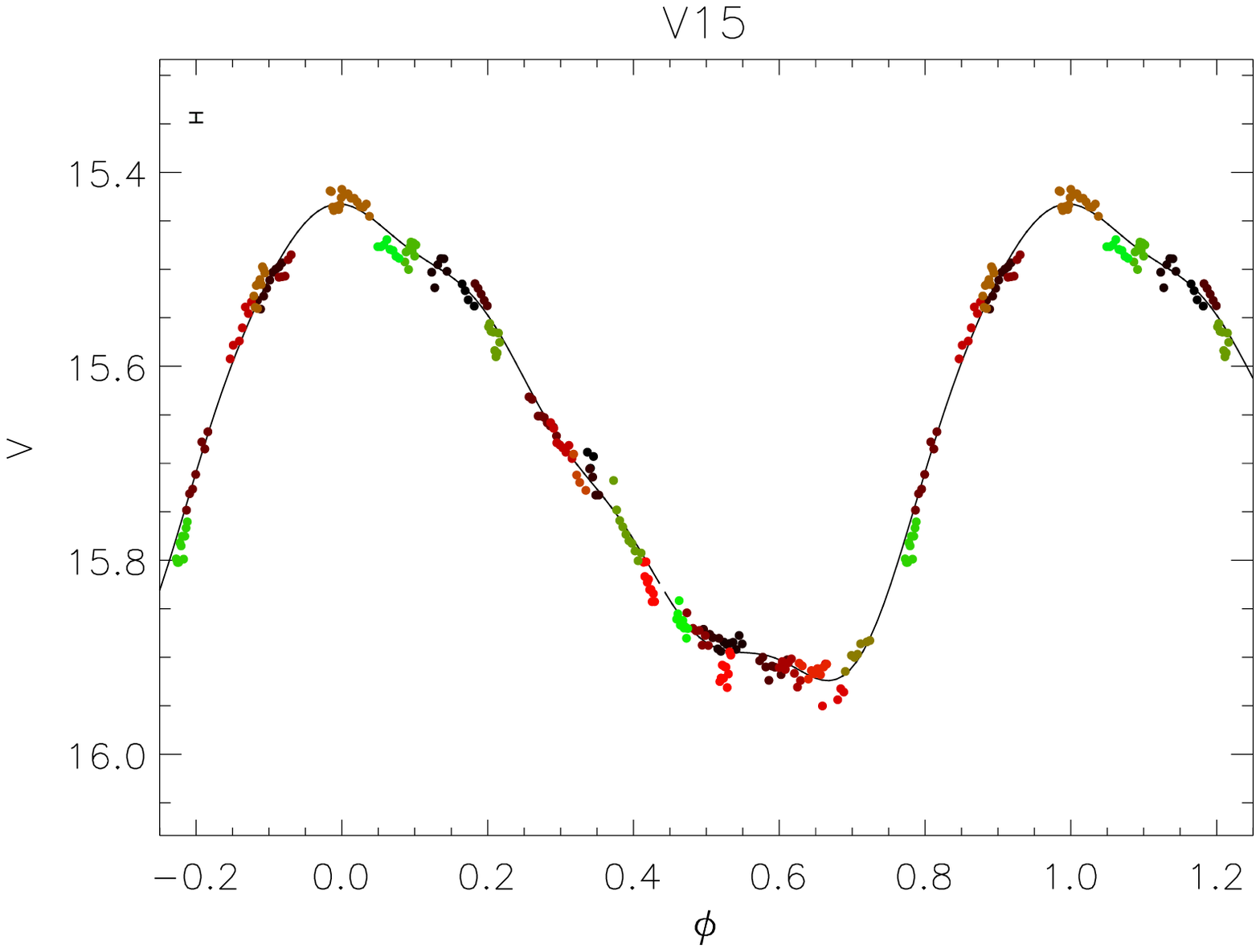}
  \includegraphics[width=4.5cm, angle=0]{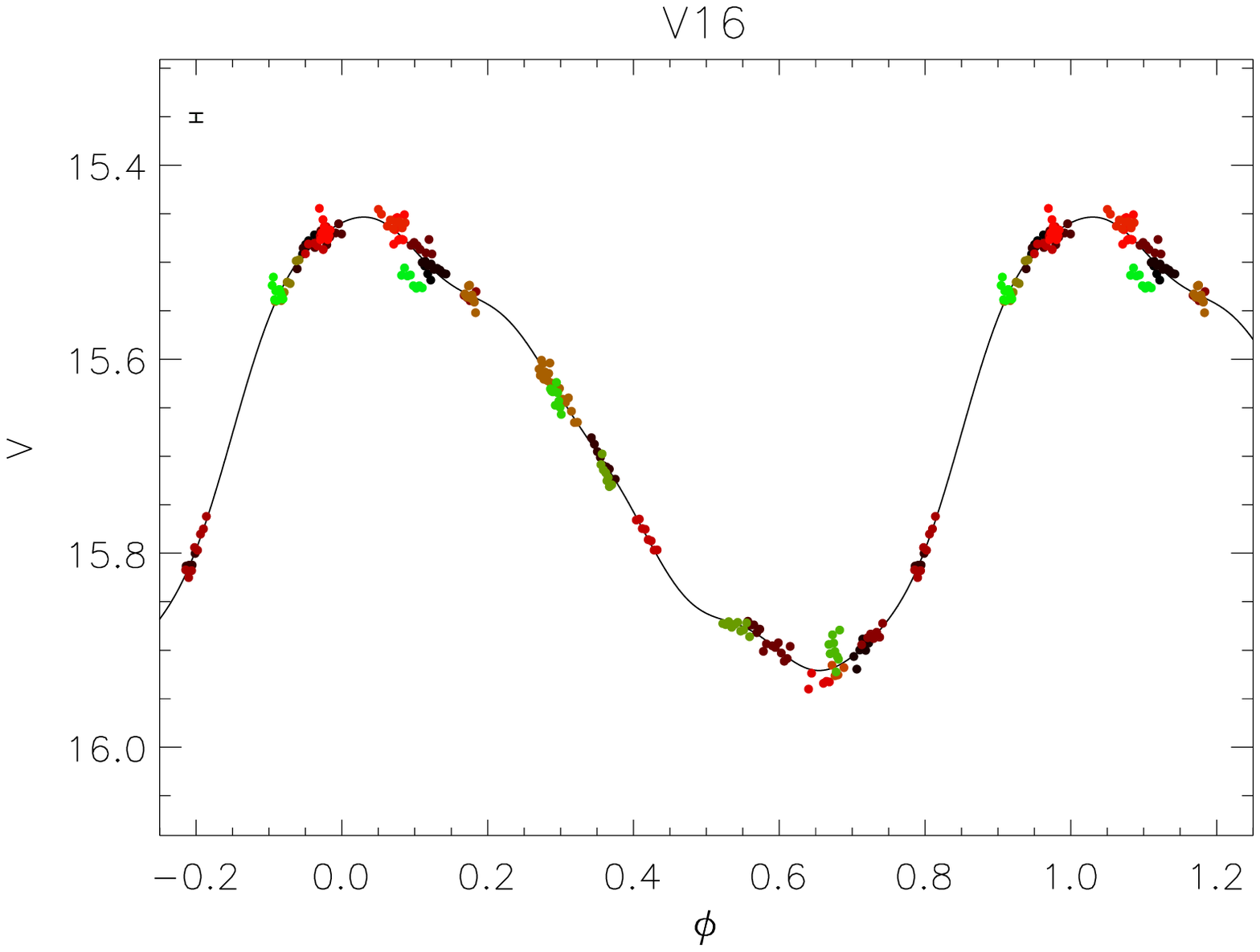}
  \includegraphics[width=4.5cm, angle=0]{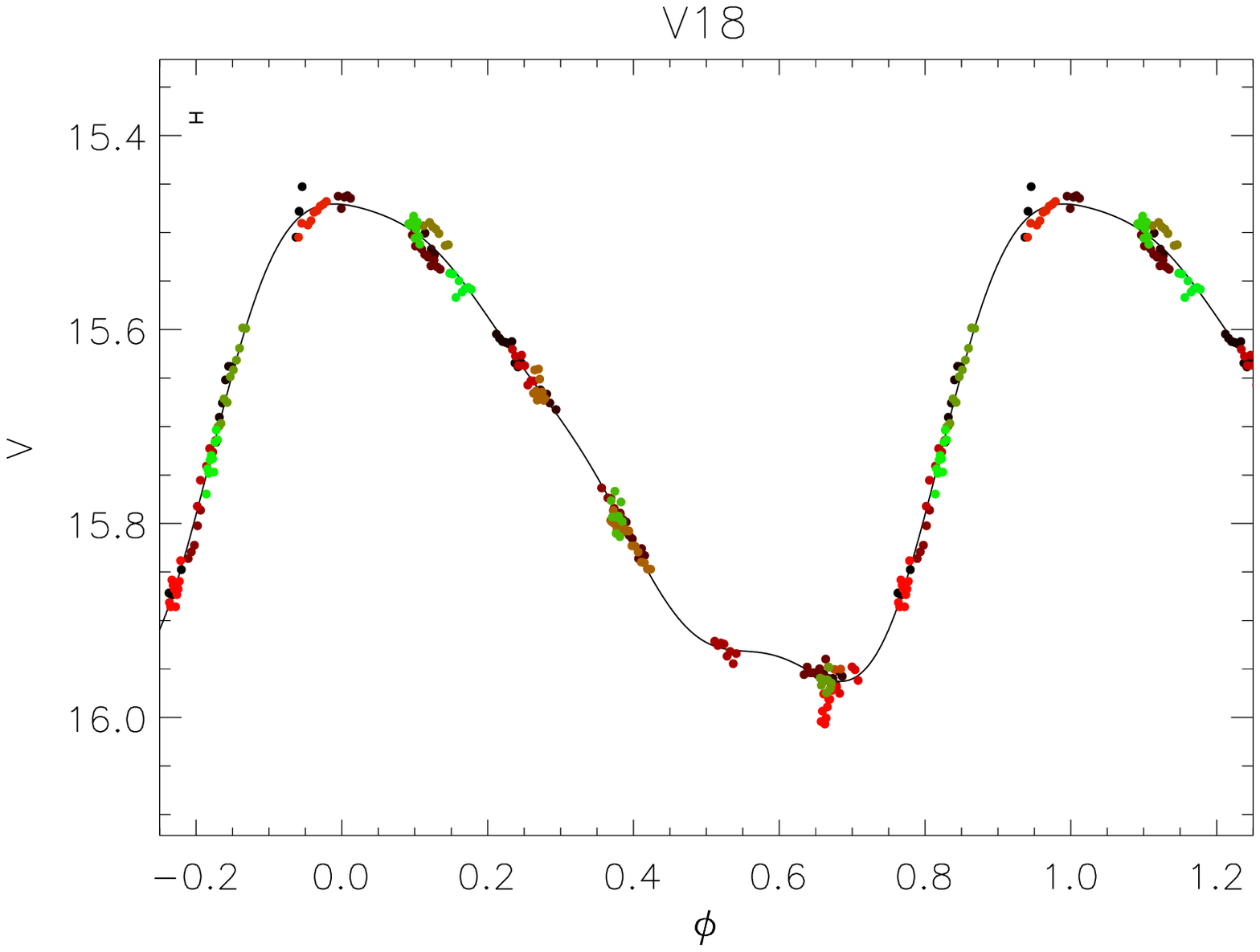}
  \includegraphics[width=4.5cm, angle=0]{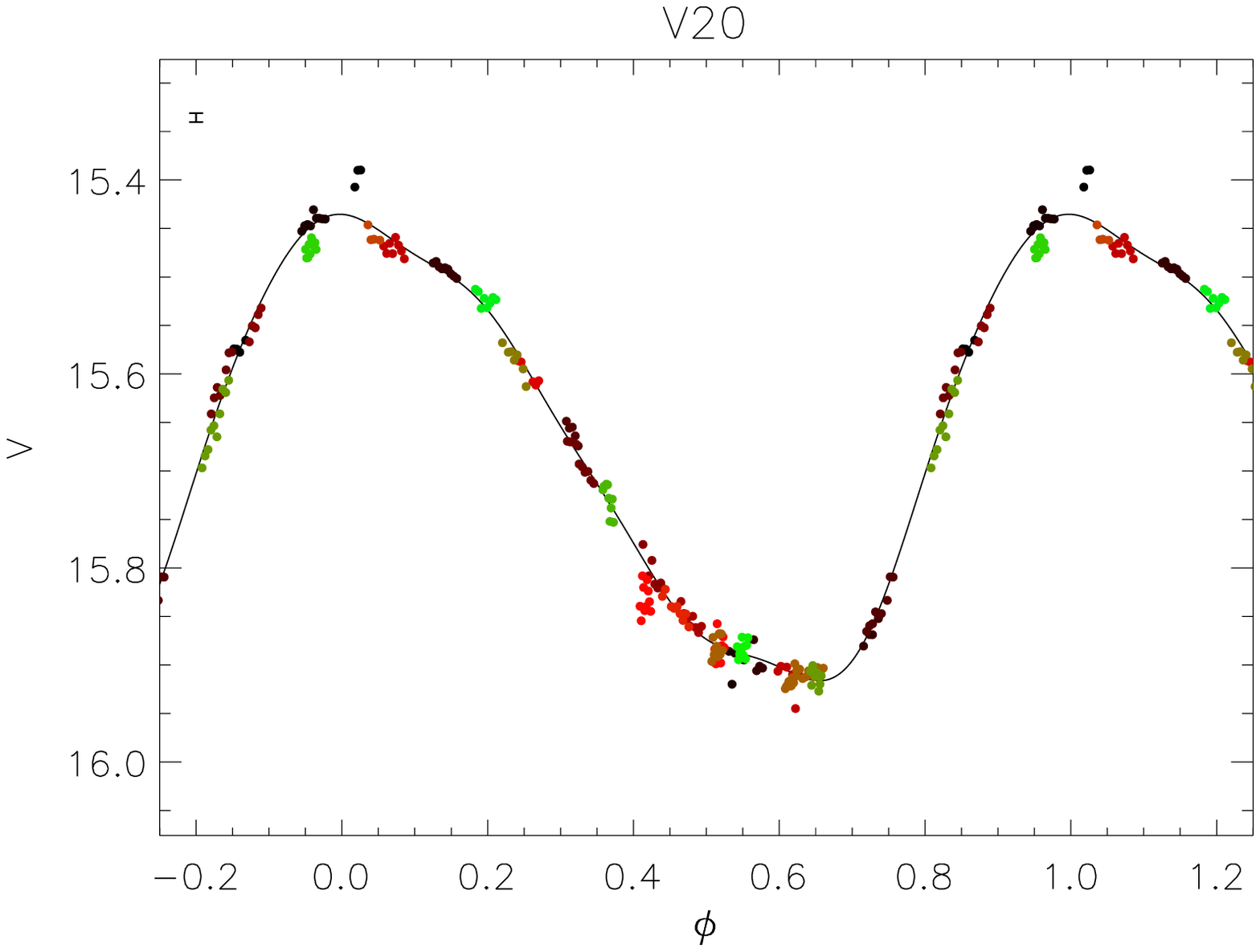}
  \includegraphics[width=4.5cm, angle=0]{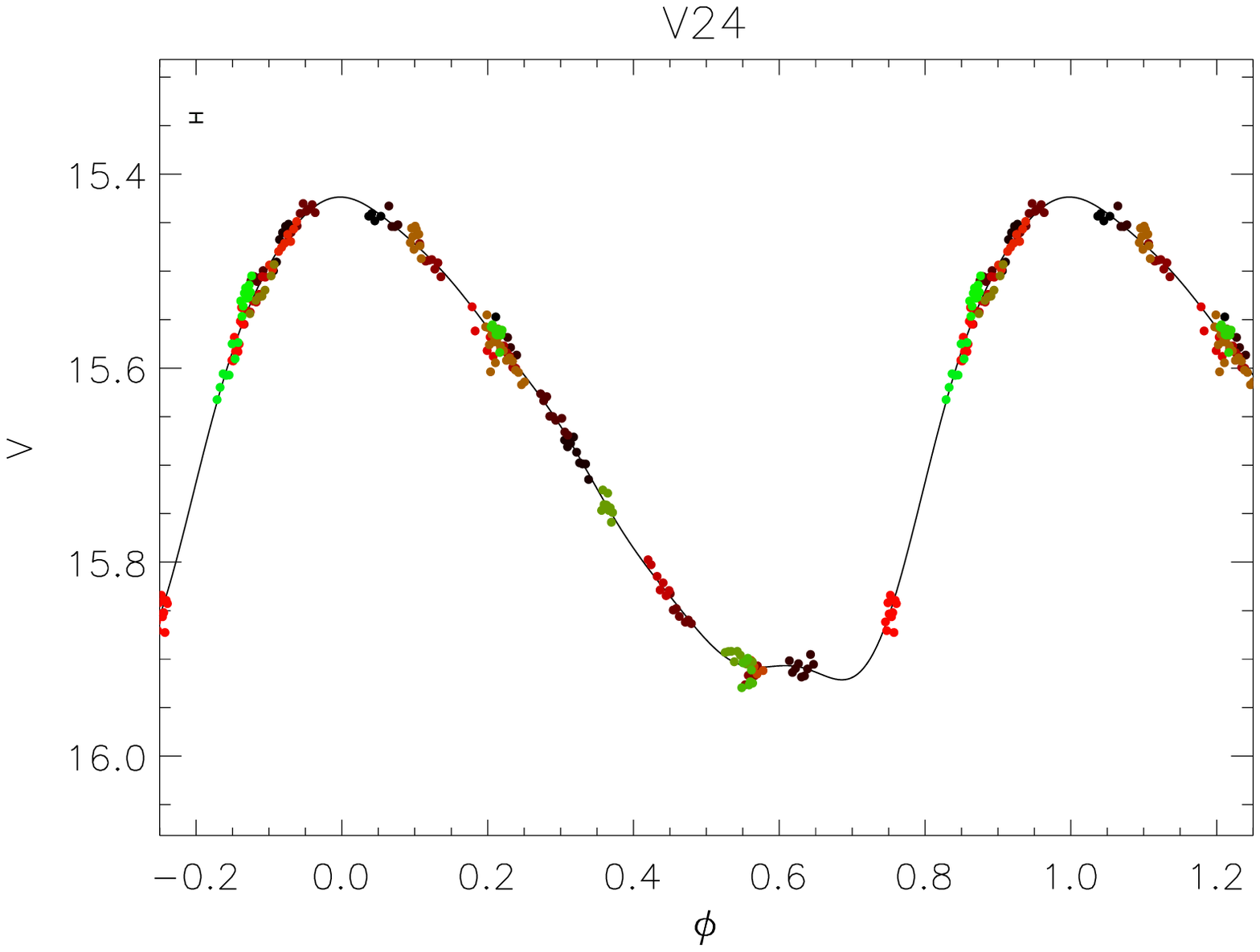}
  \includegraphics[width=4.5cm, angle=0]{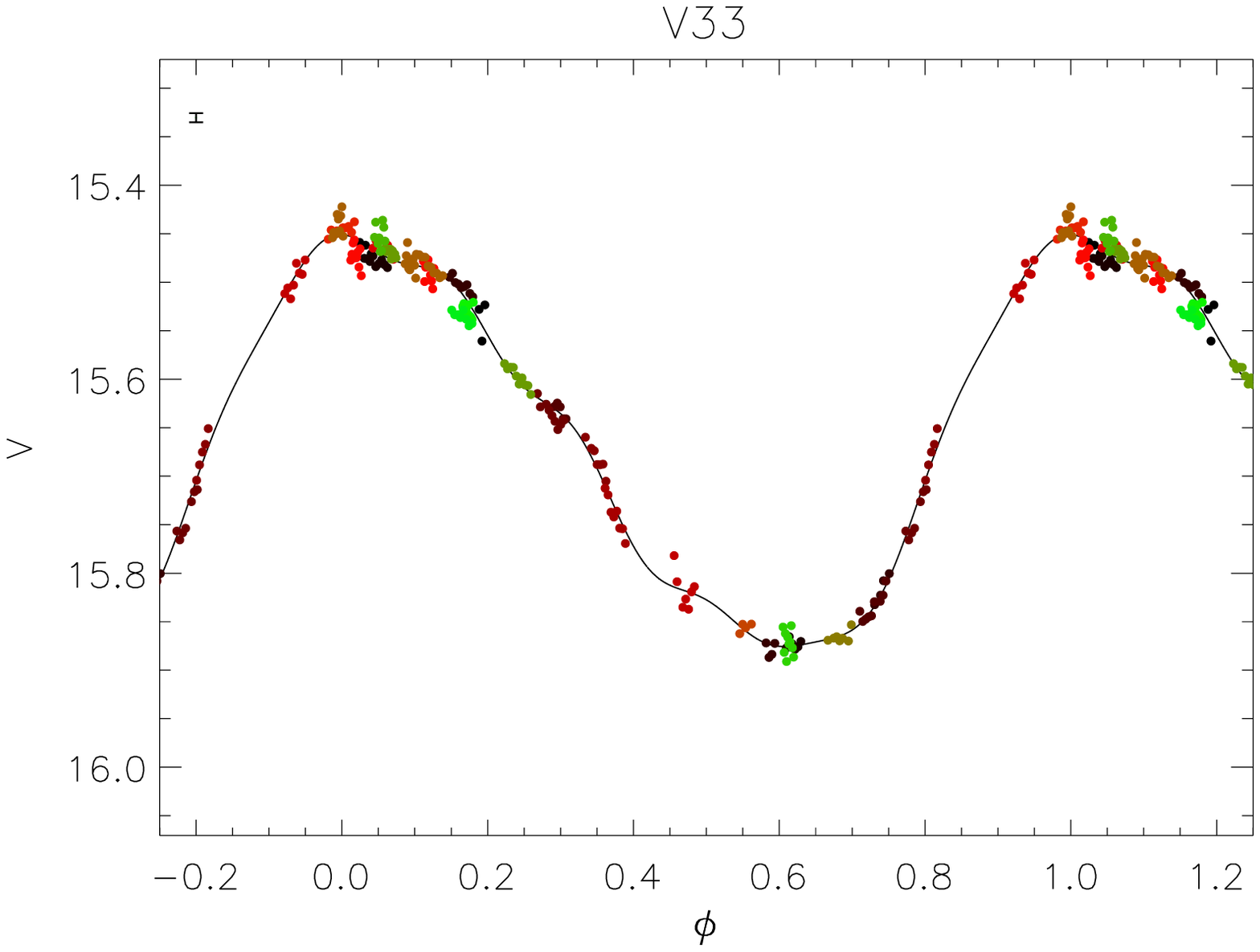}
  \includegraphics[width=4.5cm, angle=0]{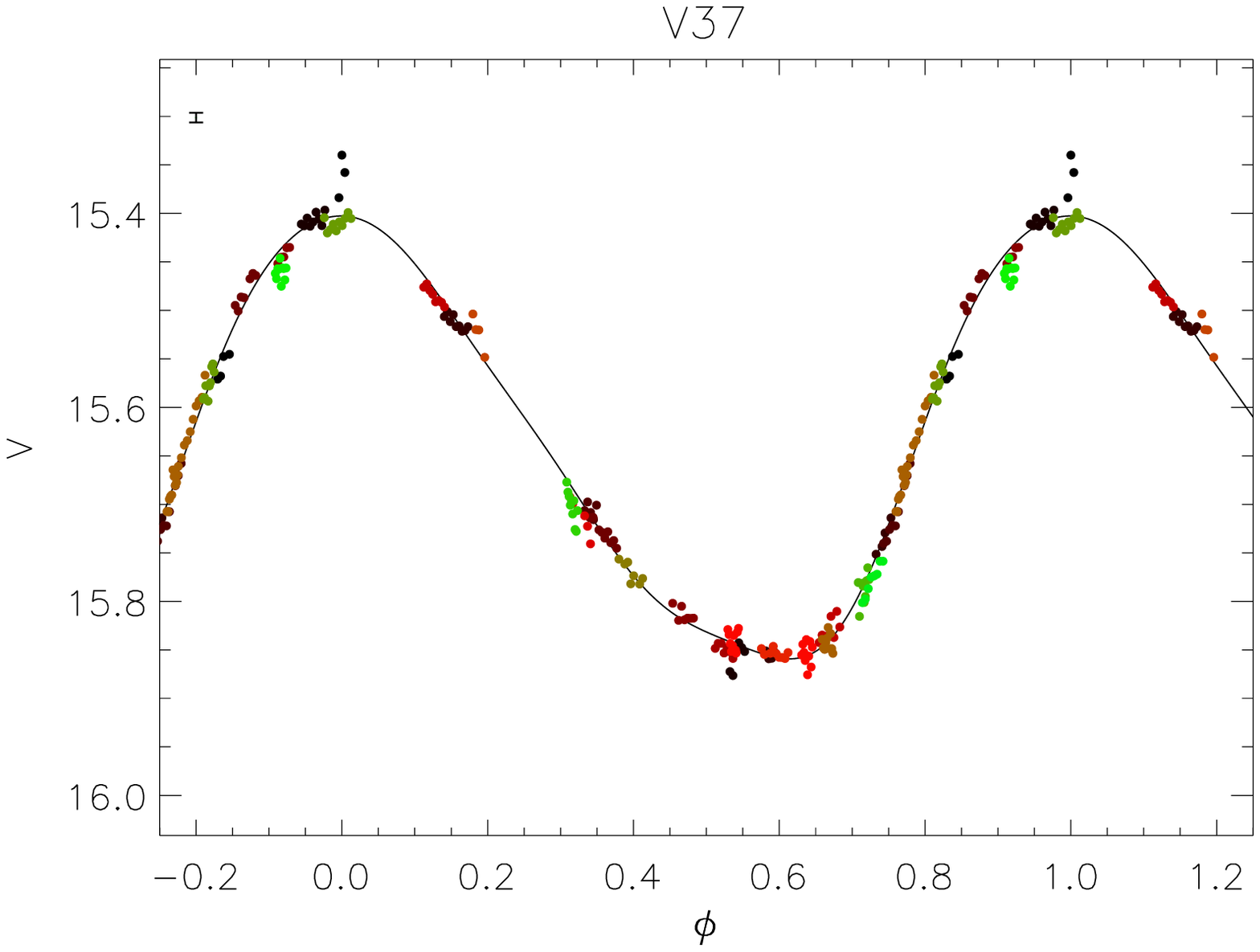}
  \includegraphics[width=4.5cm, angle=0]{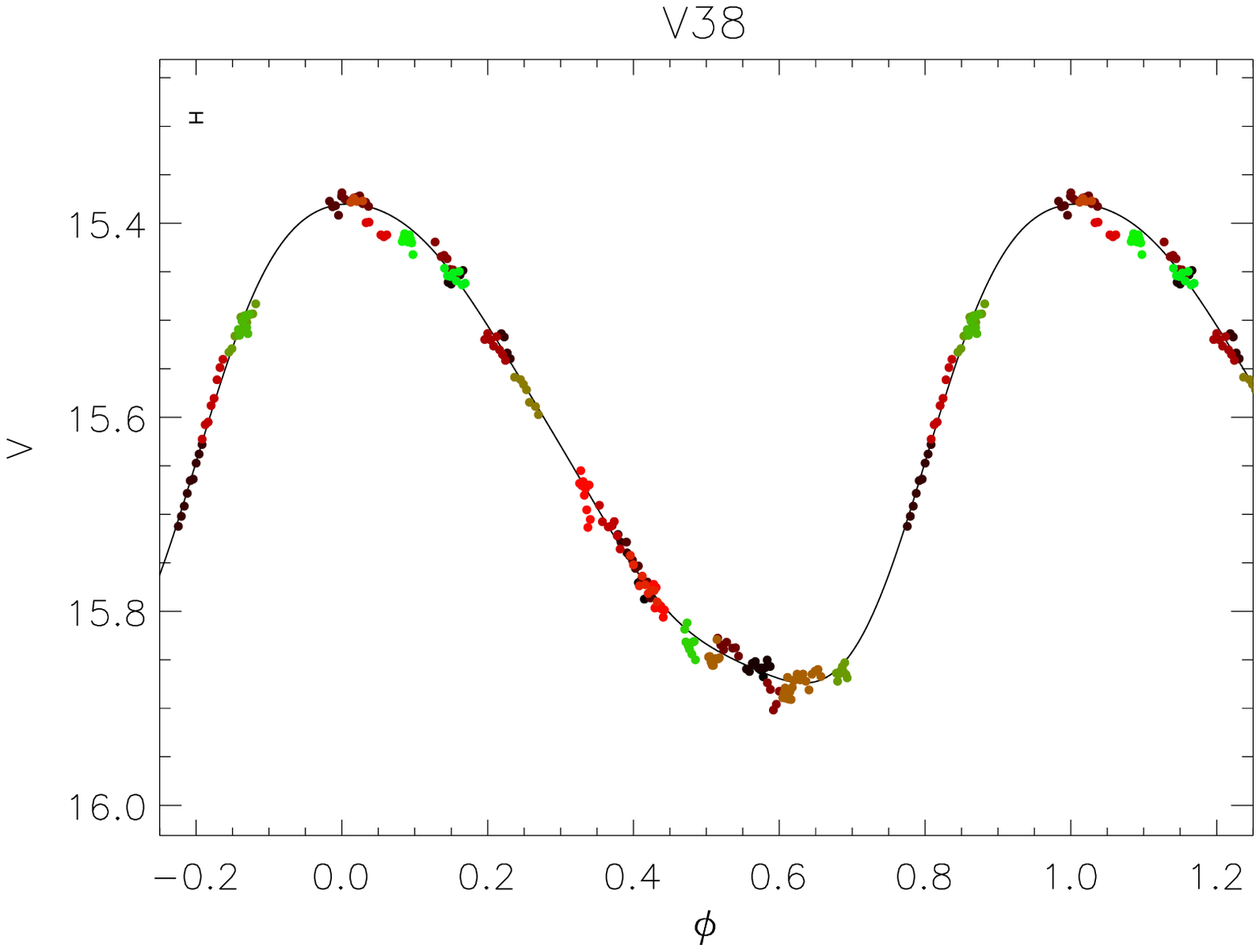}
  \includegraphics[width=4.5cm, angle=0]{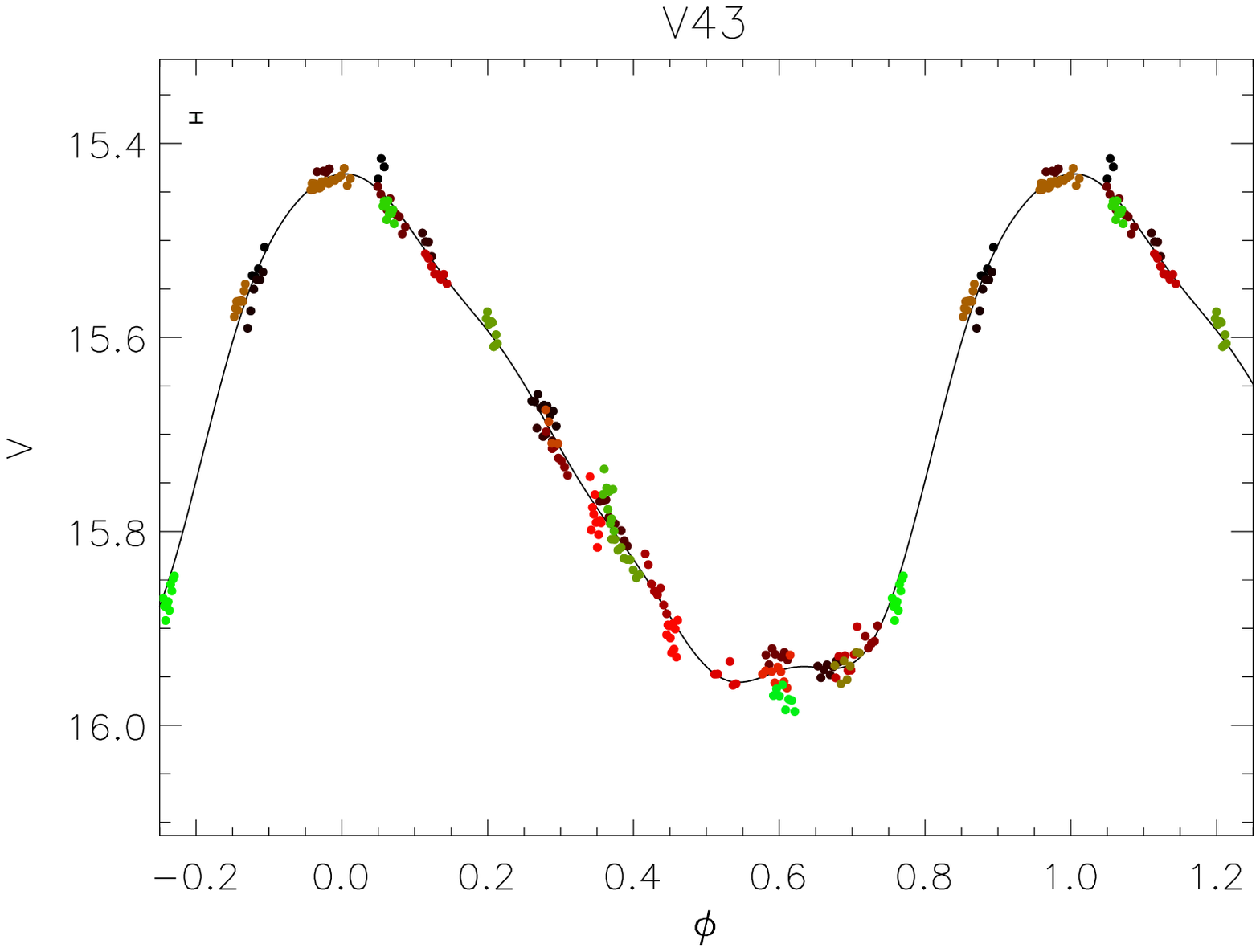}
  \includegraphics[width=4.5cm, angle=0]{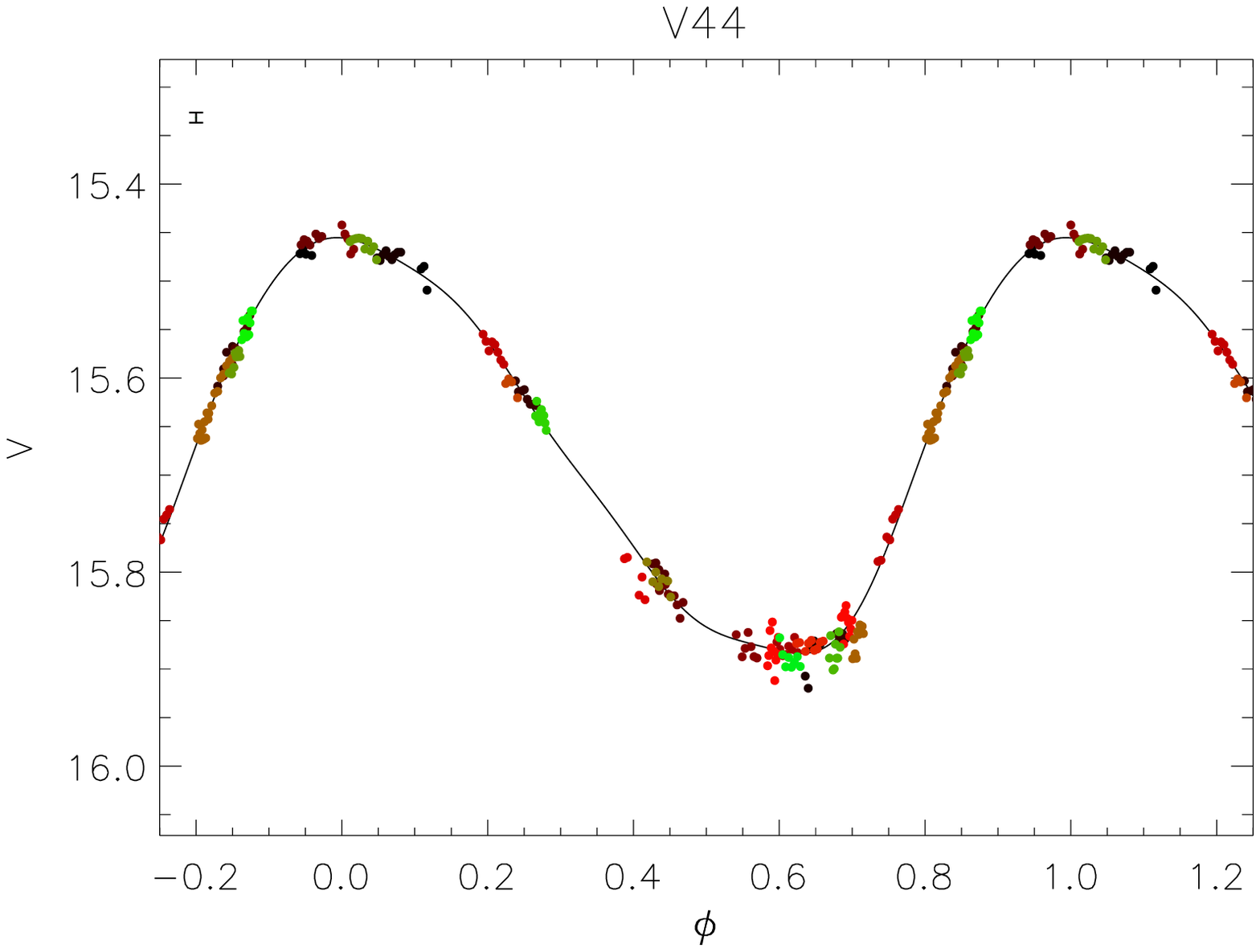}
  \includegraphics[width=4.5cm, angle=0]{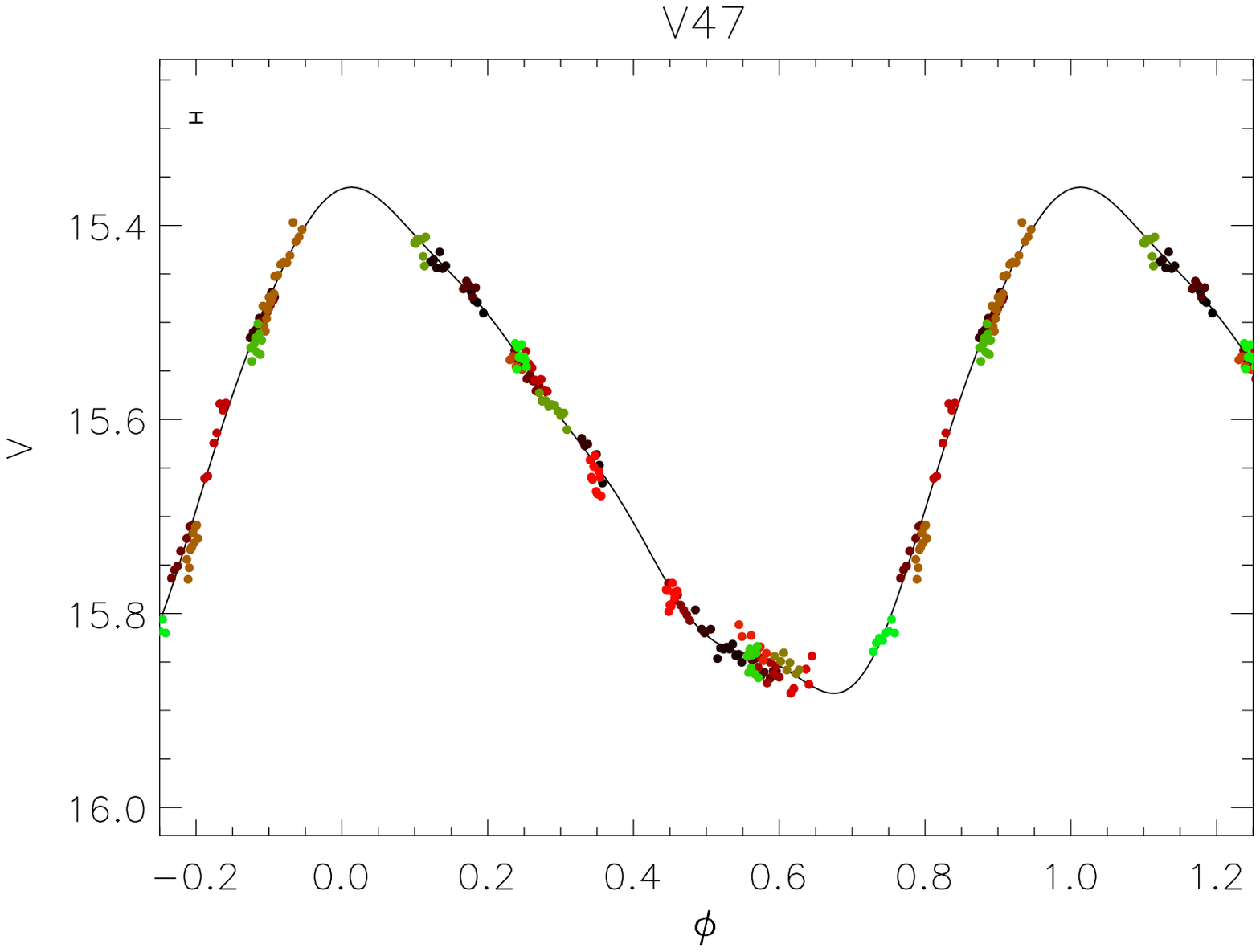}

  \caption{Same as \Fig{fig:lc_RR0V}, but for RR1 stars. \label{fig:lc_RR1V}}

\end{figure*}

\begin{figure*}
  \centering
  \includegraphics[width=14cm, angle=0]{fig/colourcode.ps}   \\
   \includegraphics[width=6cm, angle=0]{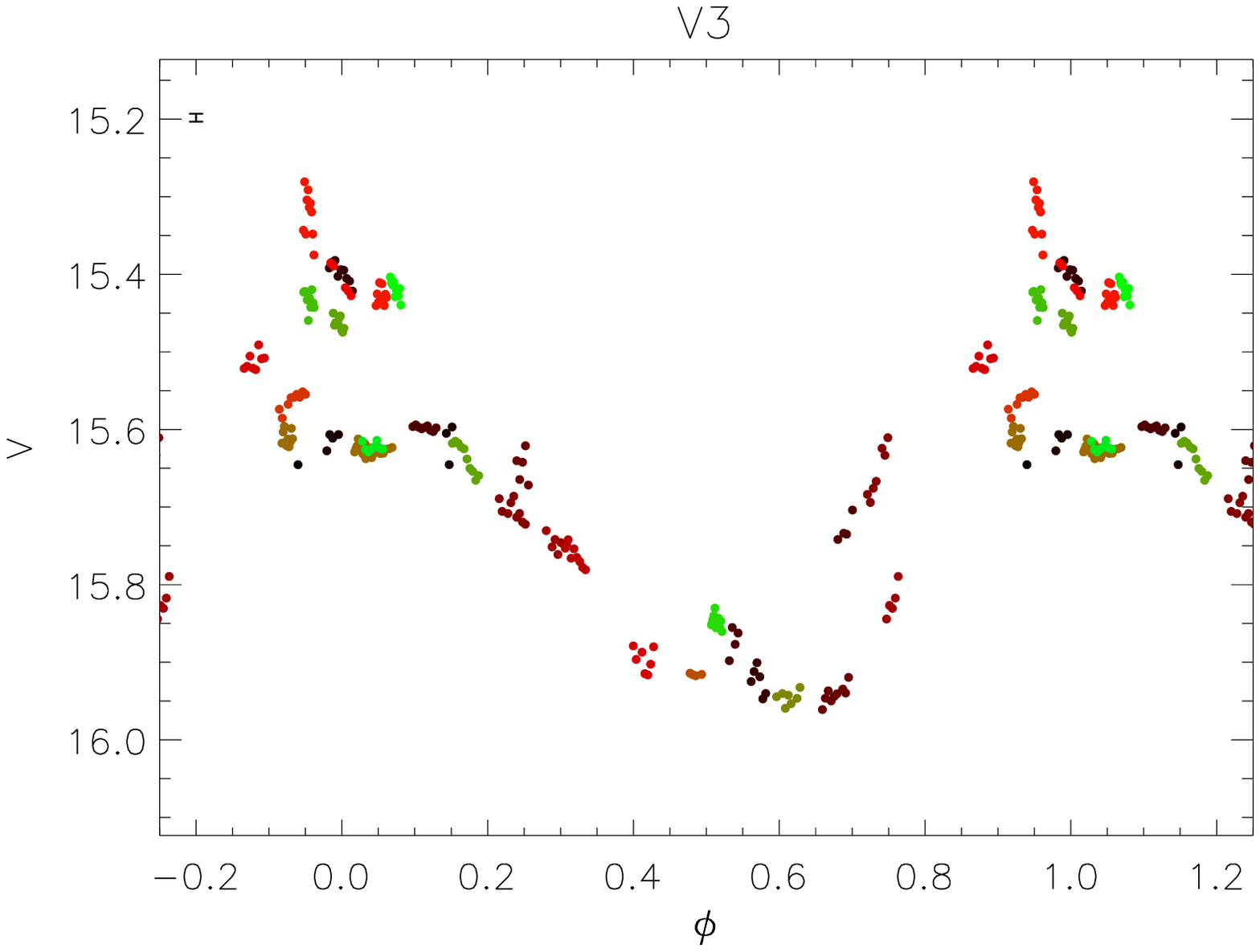}
  \includegraphics[width=6cm, angle=0]{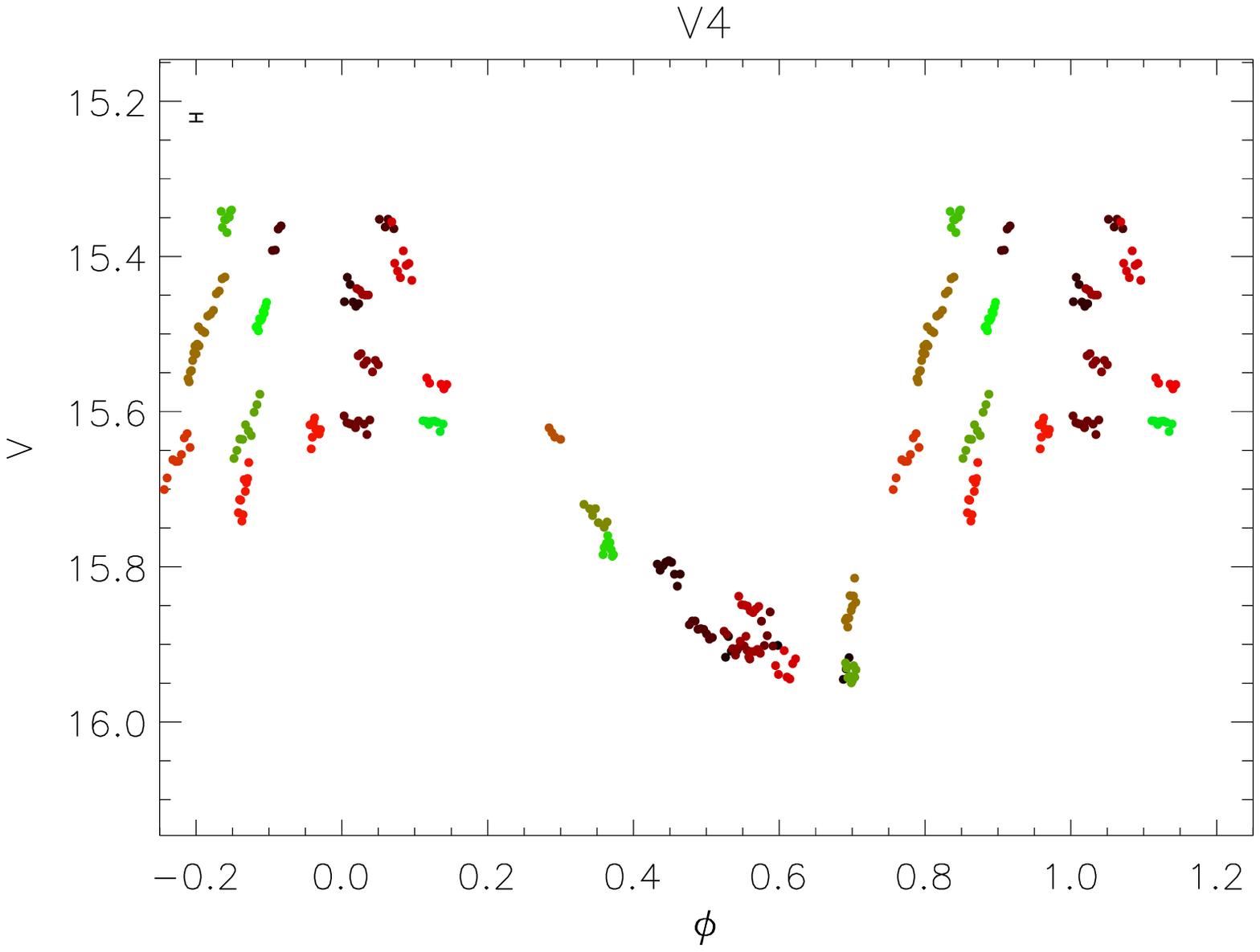}
  \includegraphics[width=6cm, angle=0]{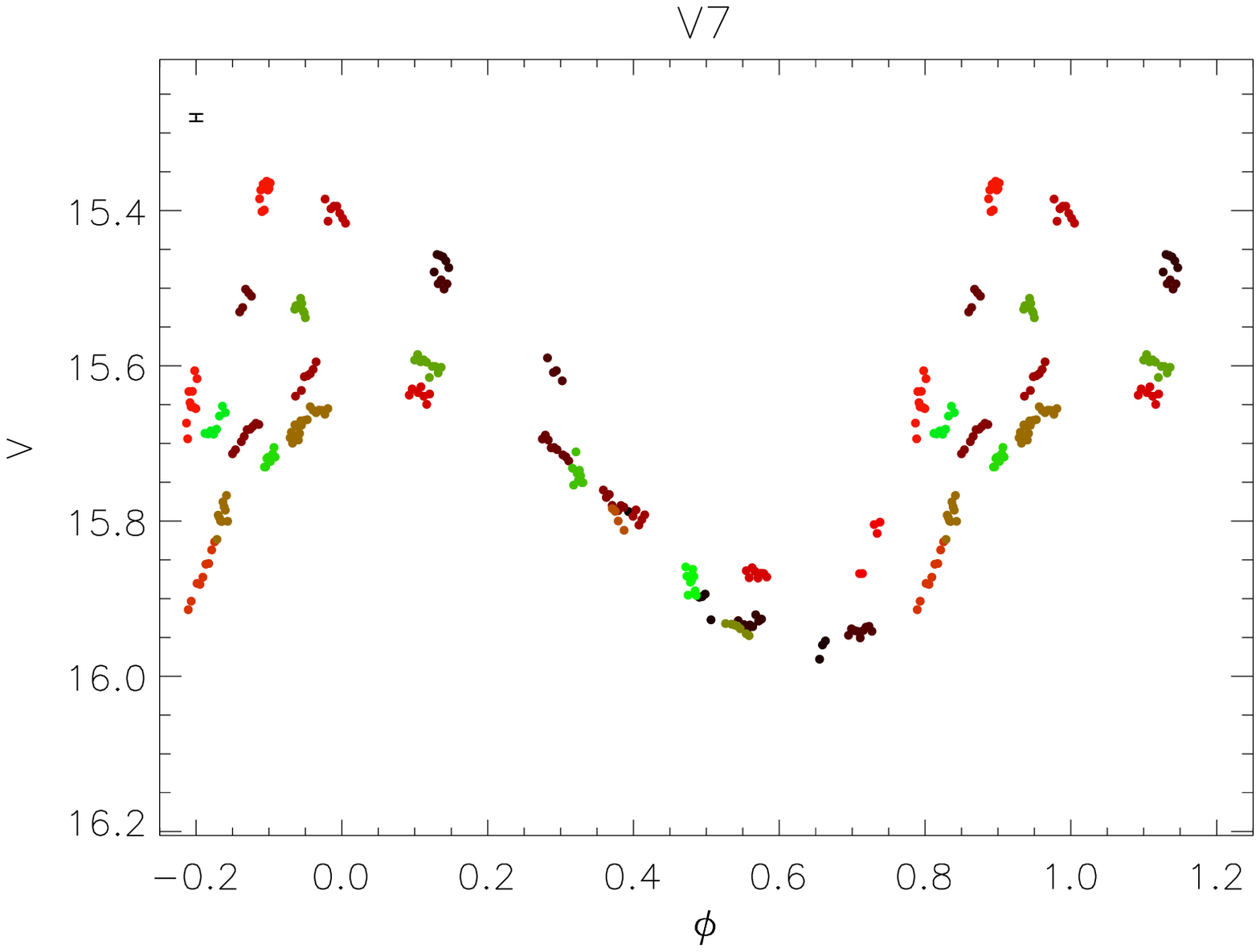}
  \includegraphics[width=6cm, angle=0]{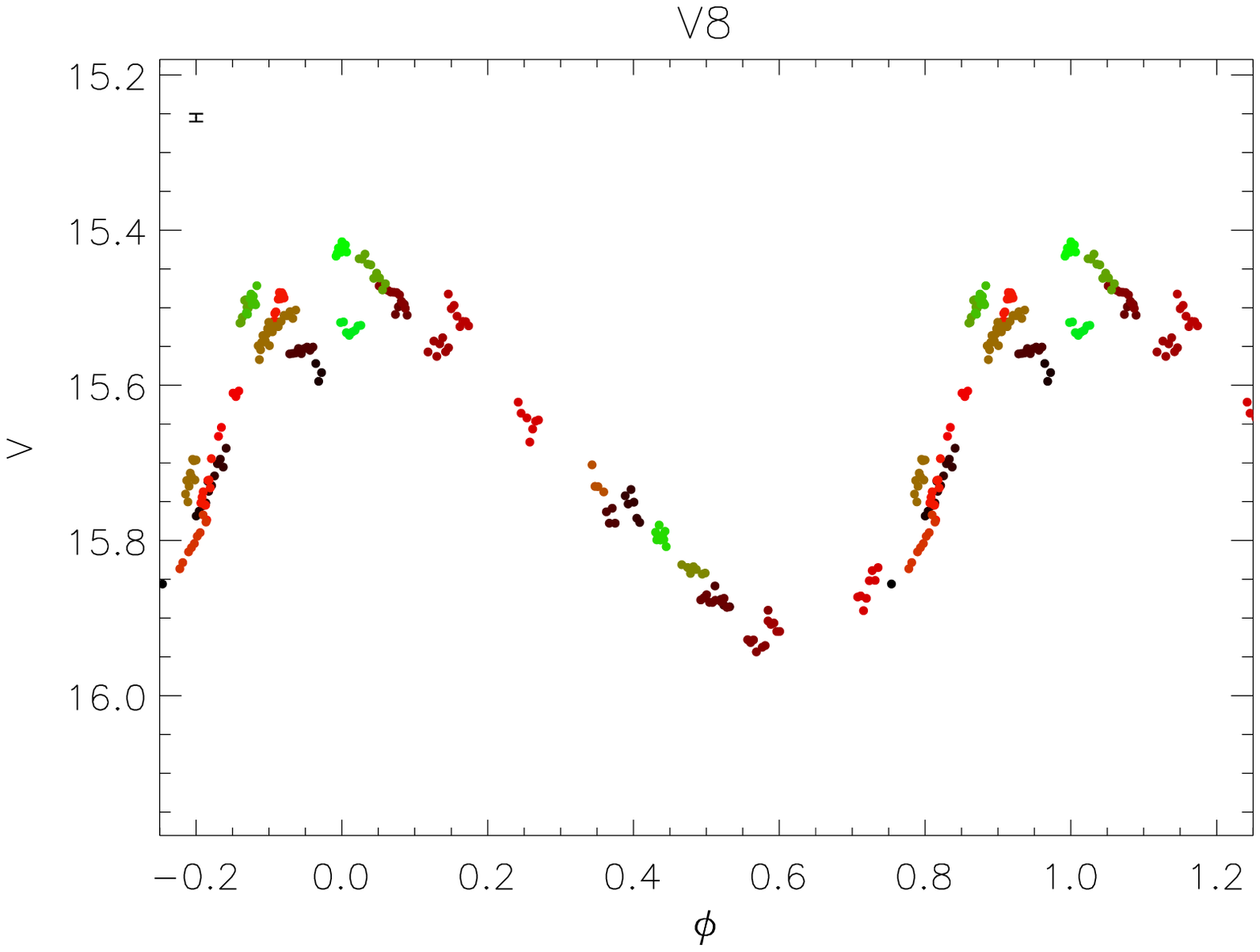}
 \includegraphics[width=6cm, angle=0]{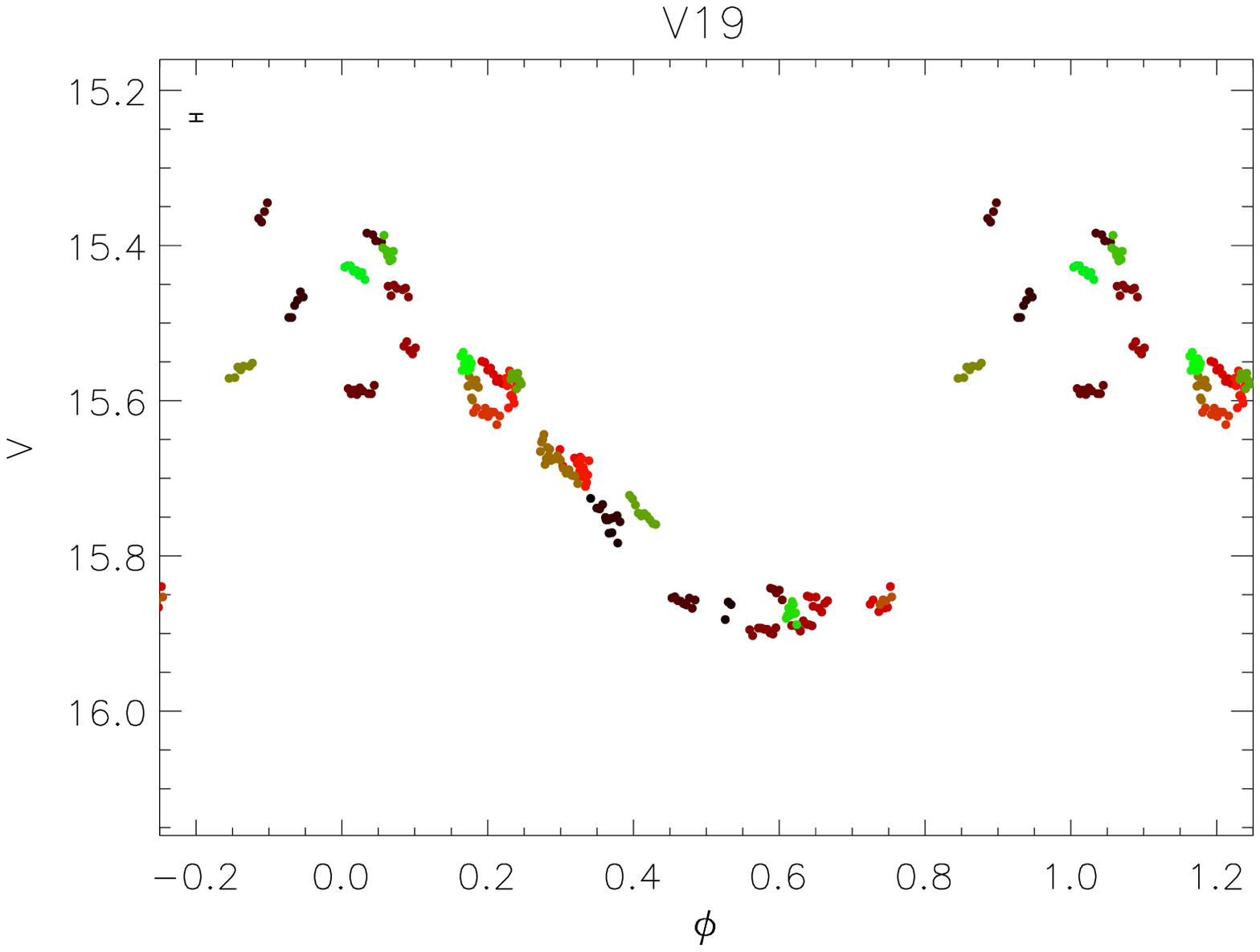}
  \includegraphics[width=6cm, angle=0]{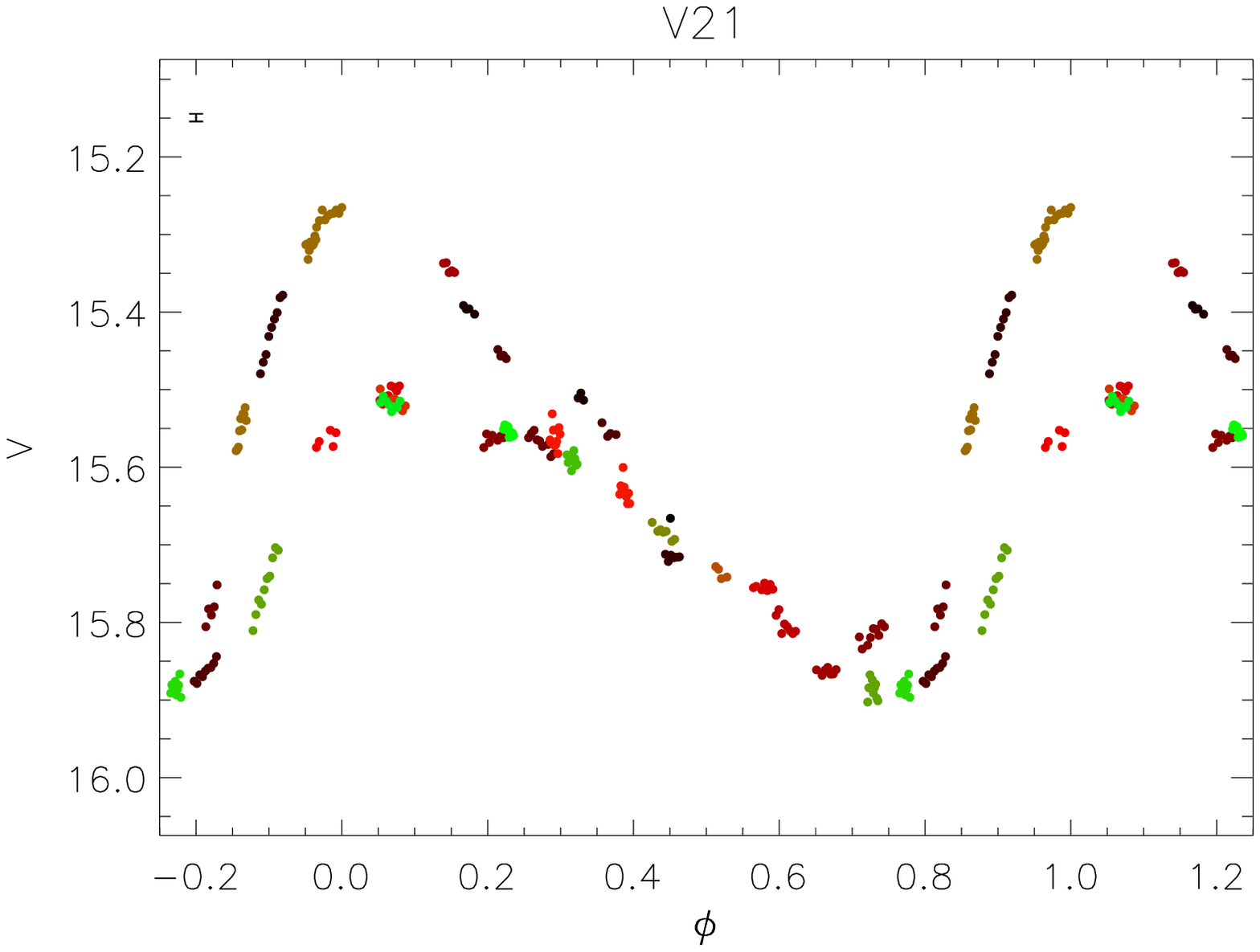}
  \includegraphics[width=6cm, angle=0]{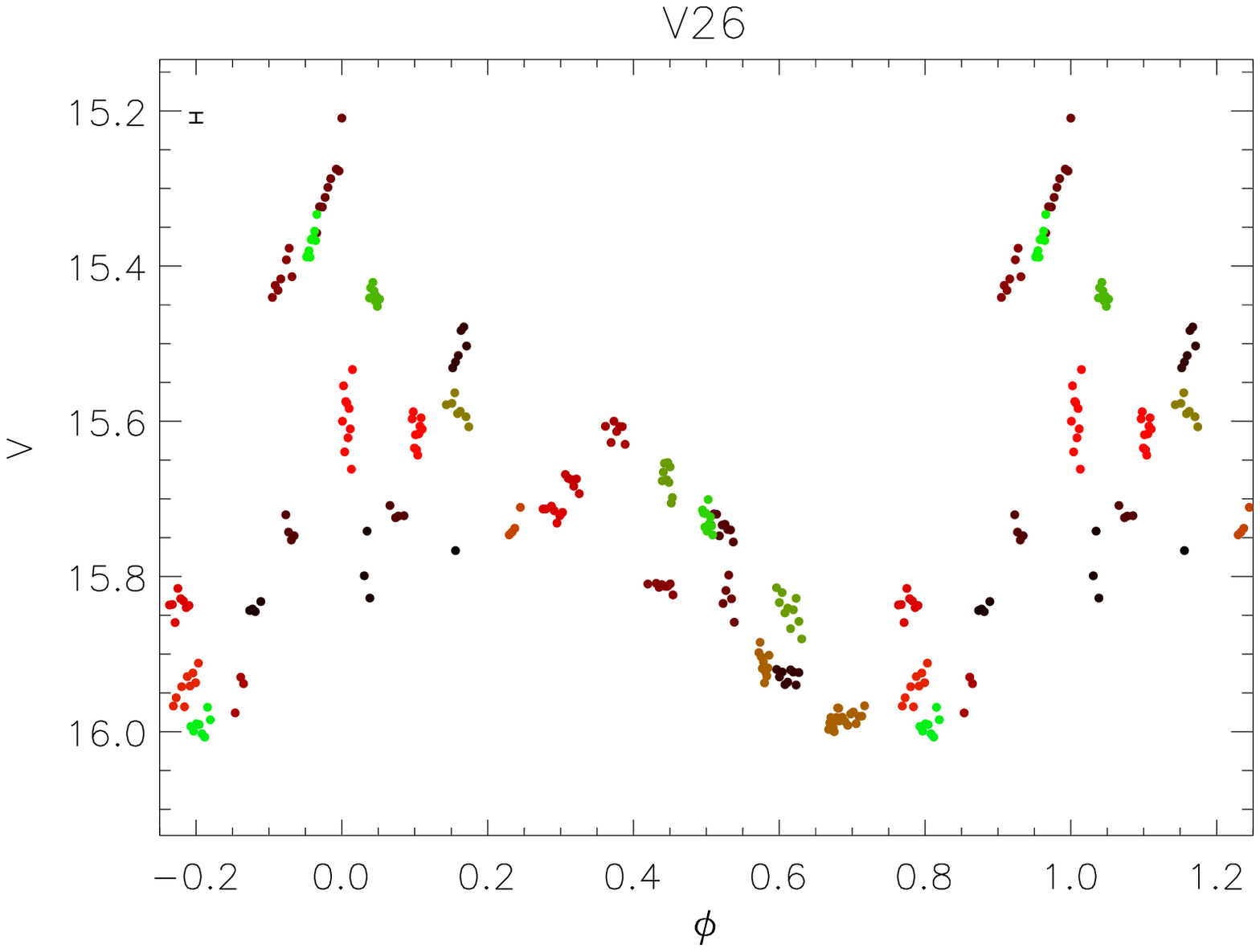}
  \includegraphics[width=6cm, angle=0]{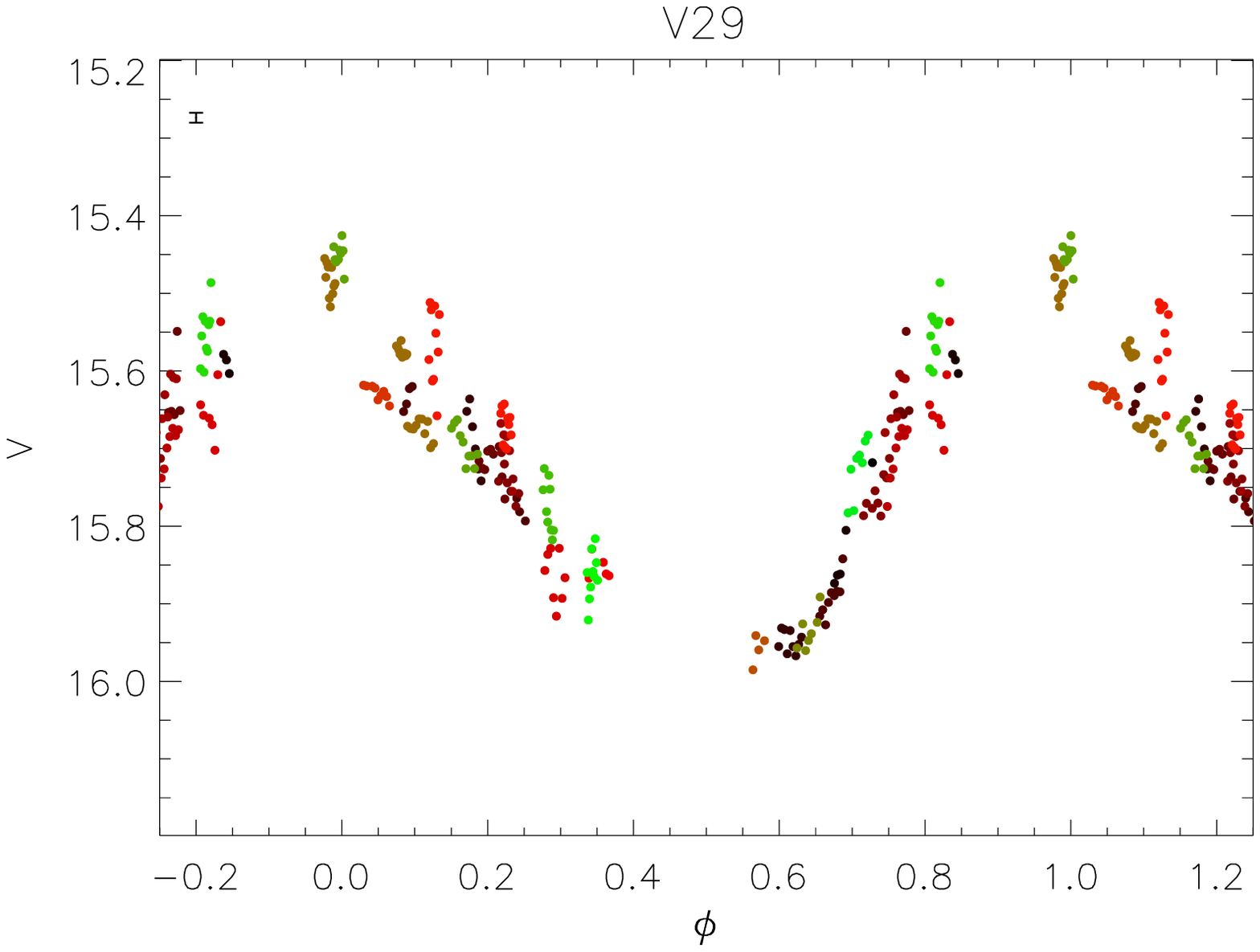}
  \includegraphics[width=6cm, angle=0]{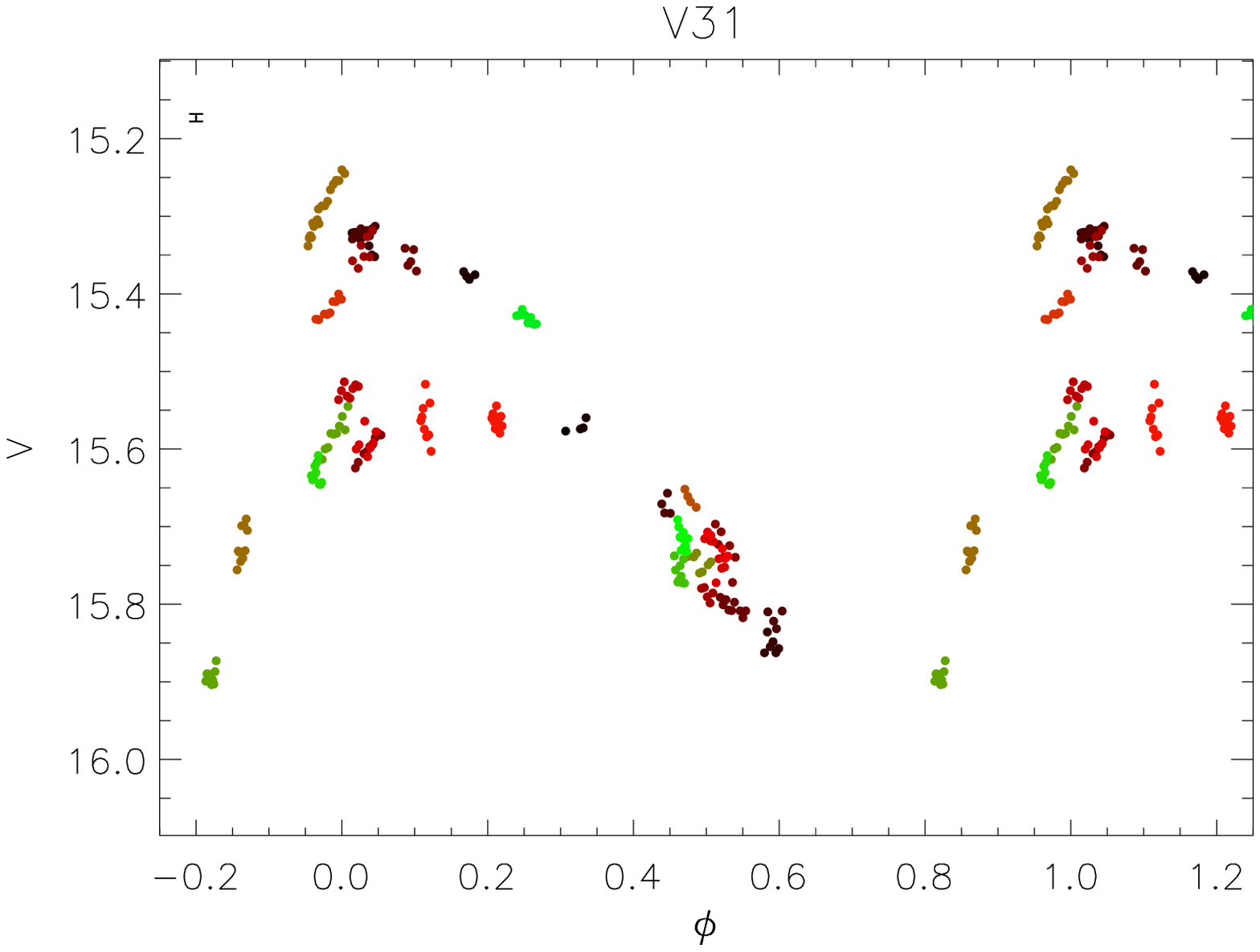}
  \includegraphics[width=6cm, angle=0]{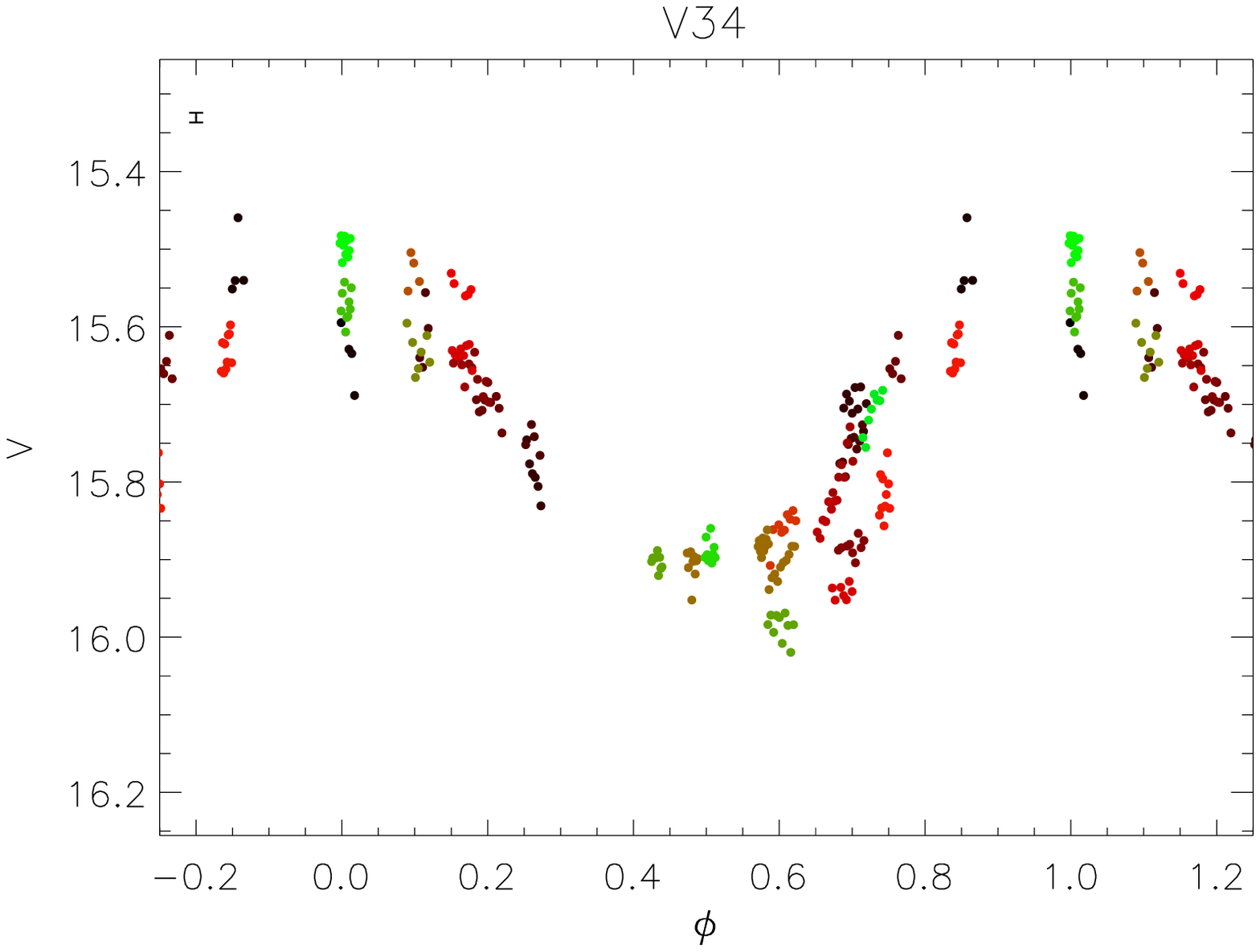}
  \includegraphics[width=6cm, angle=0]{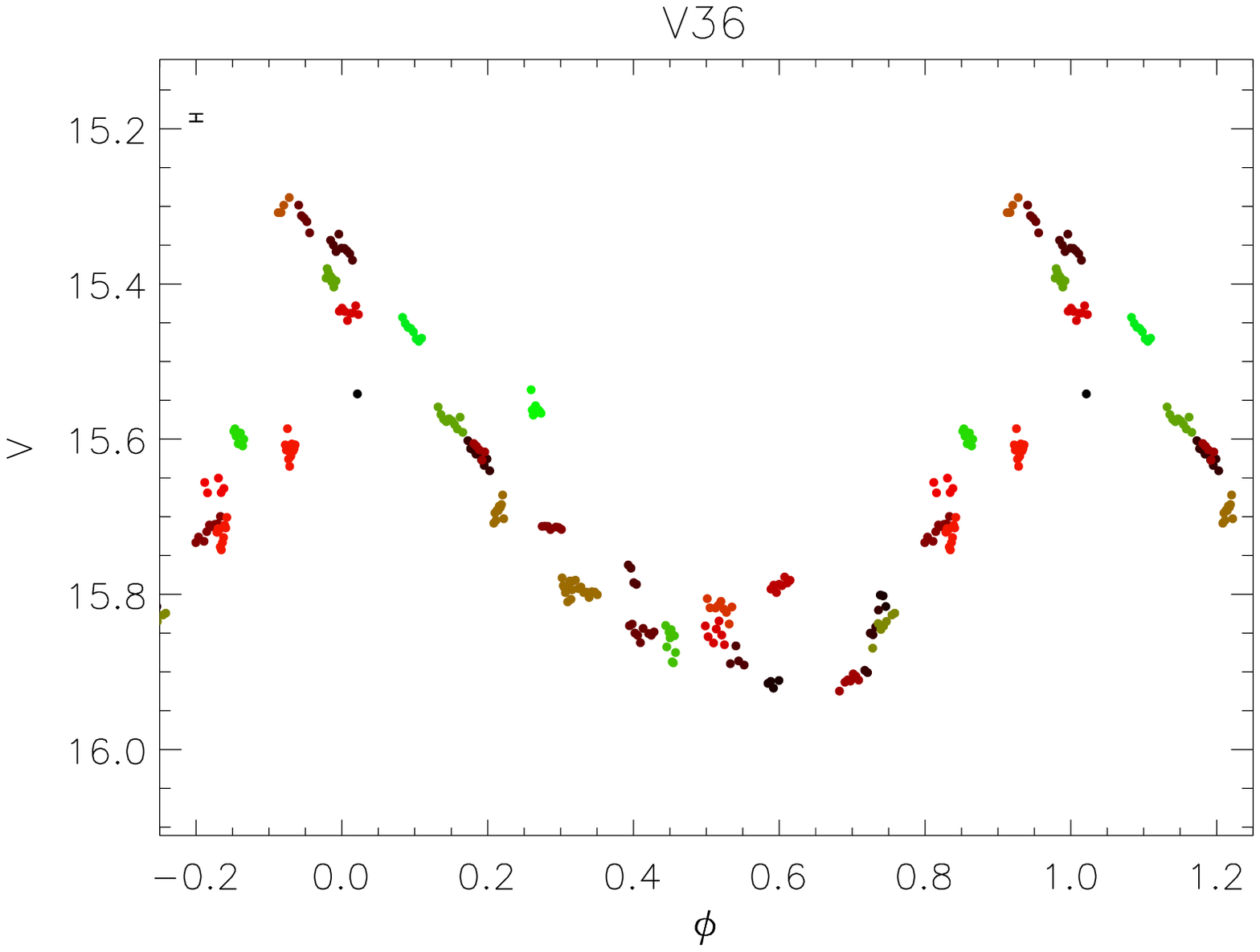}
  \includegraphics[width=6cm, angle=0]{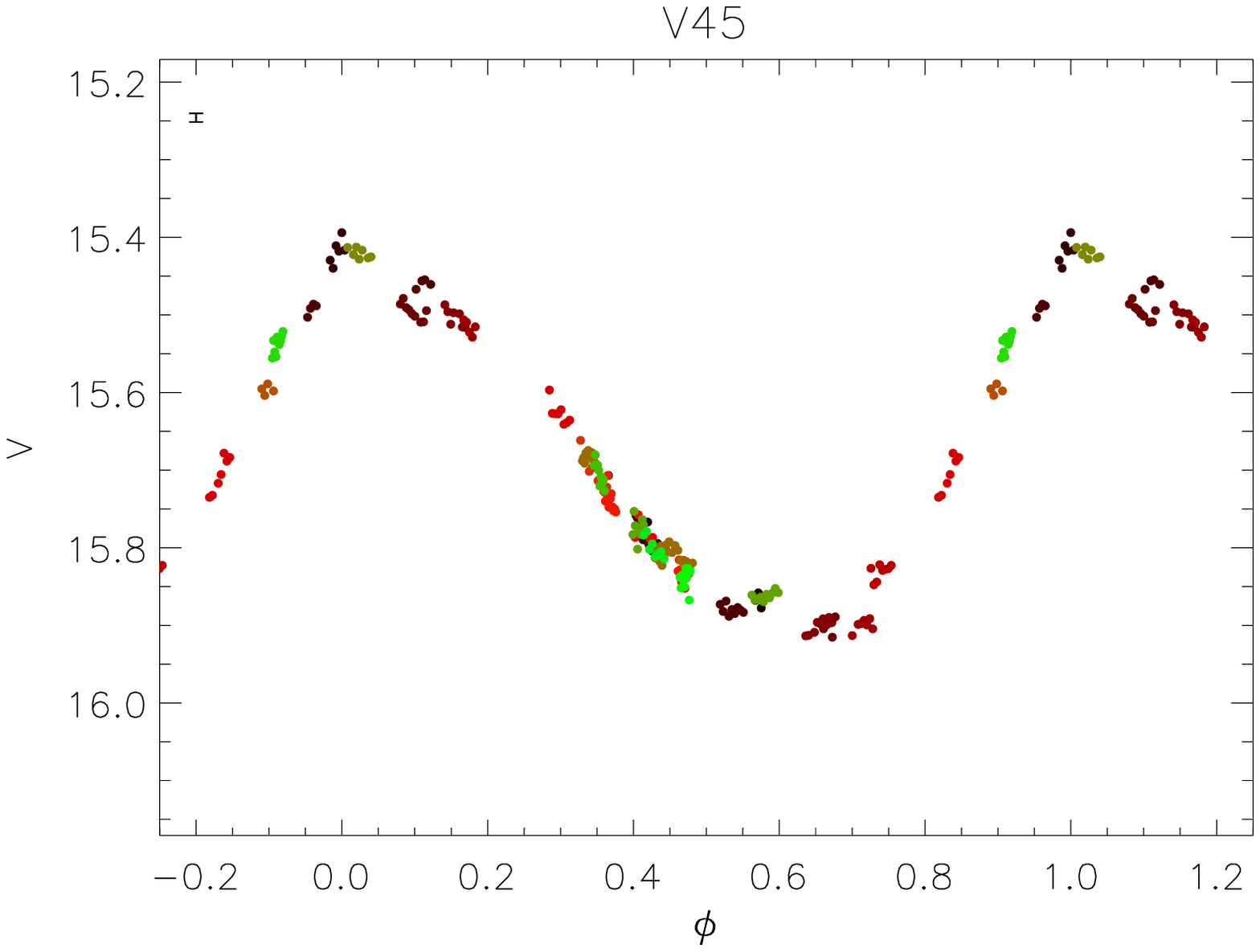}

  \caption{Phased $V-$band light curves of the RR01 stars. The light curves are phased with the first-overtone pulsation period. A typical 1-$\sigma$ error bar is plotted in the top left corner. The magnitude scale is the same on all plots in order to facilitate comparison of variation amplitude. \label{fig:lc_RR01V}}

\end{figure*}

\begin{figure*}
  \centering
  \includegraphics[width=14cm, angle=0]{fig/colourcode.ps}   \\
  \includegraphics[width=6cm, angle=0]{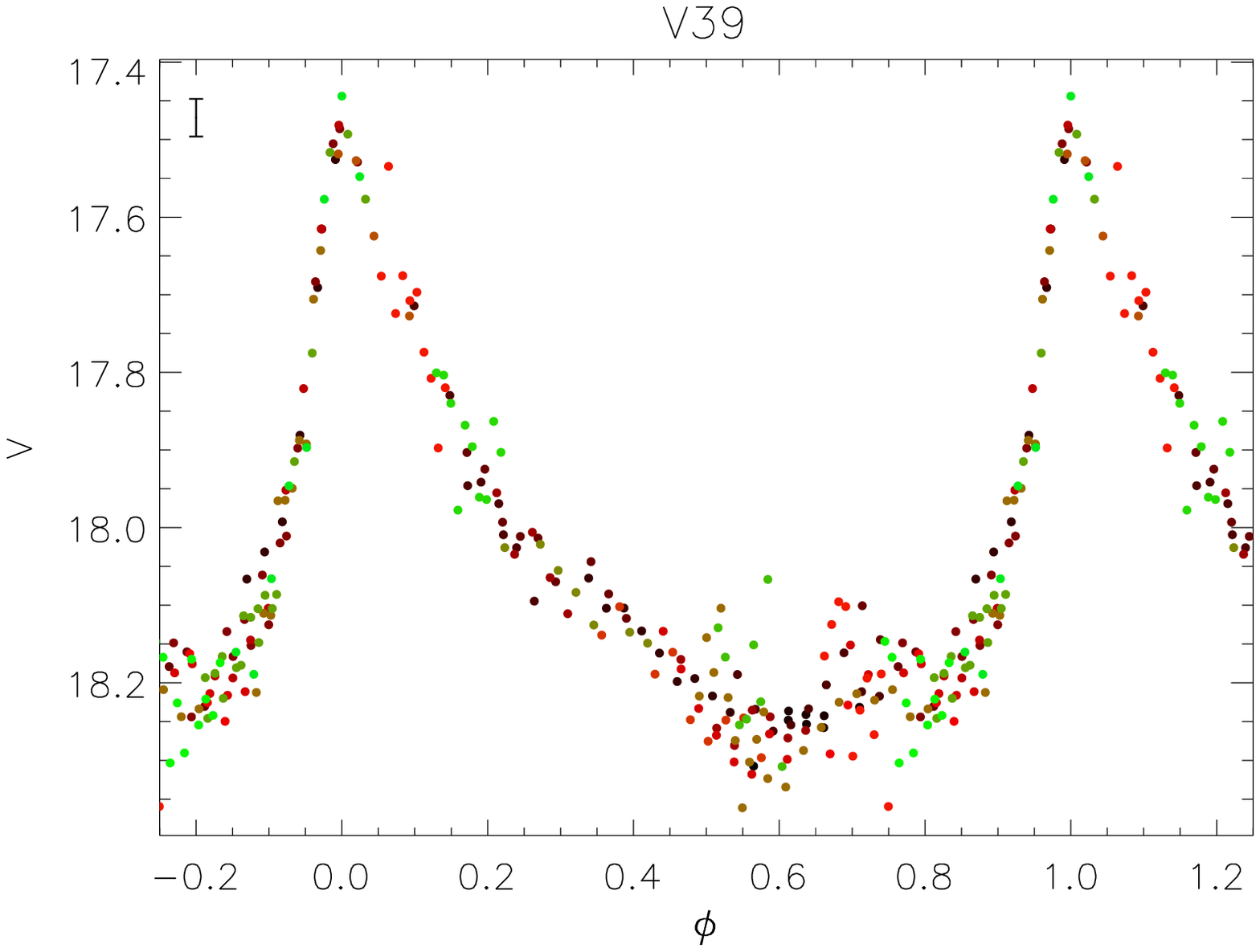}
  \includegraphics[width=6cm, angle=0]{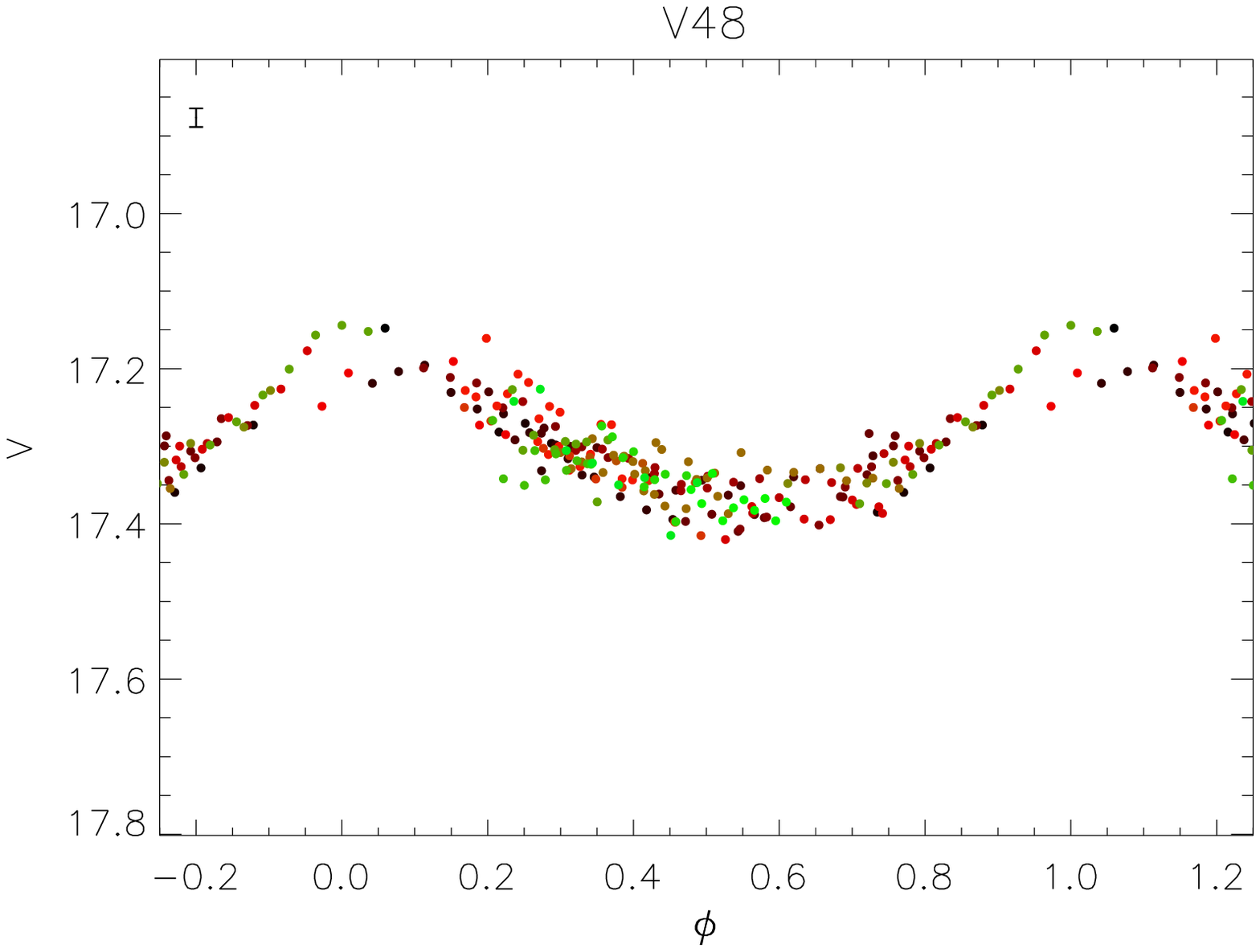}
  \includegraphics[width=6cm, angle=0]{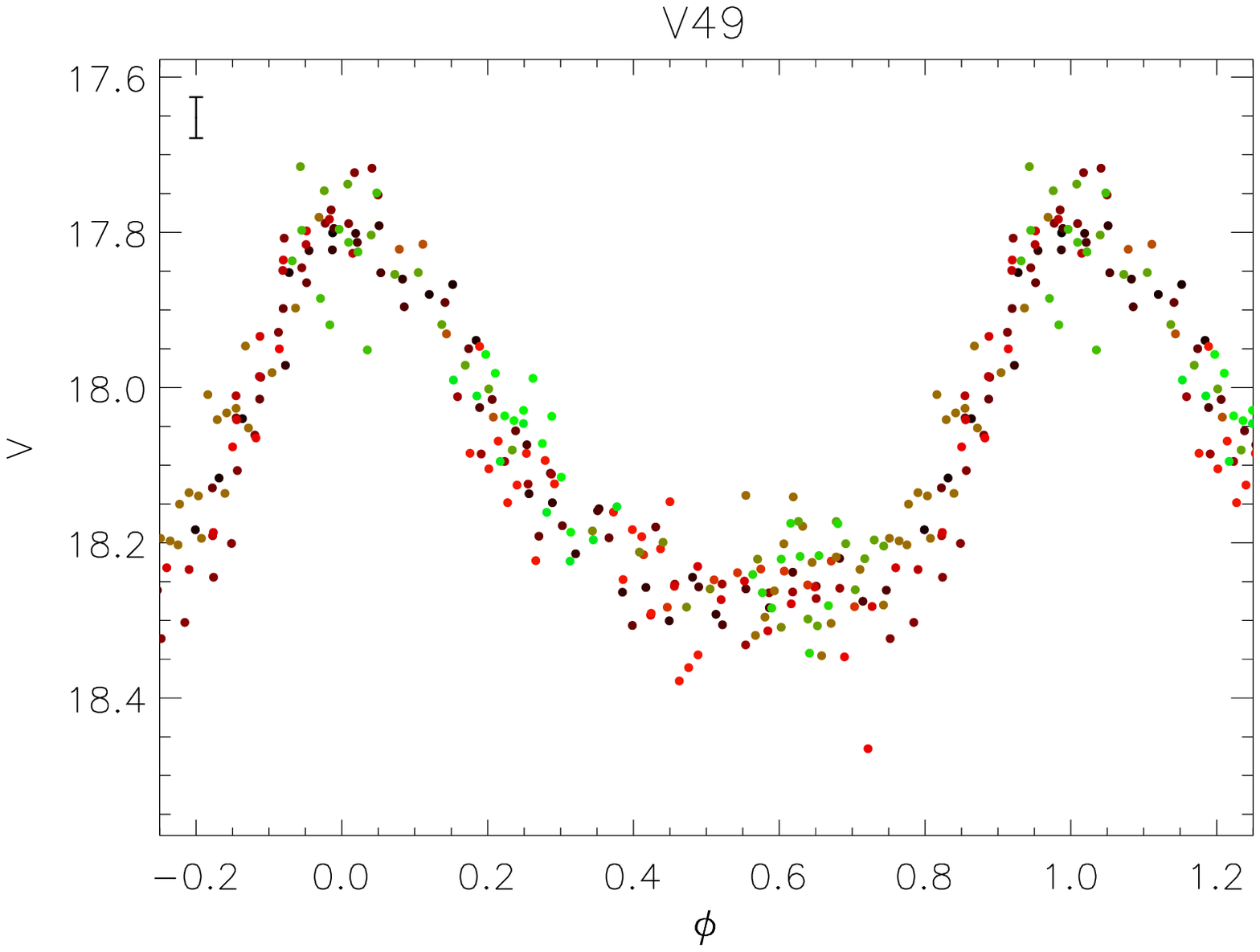}
  \includegraphics[width=6cm, angle=0]{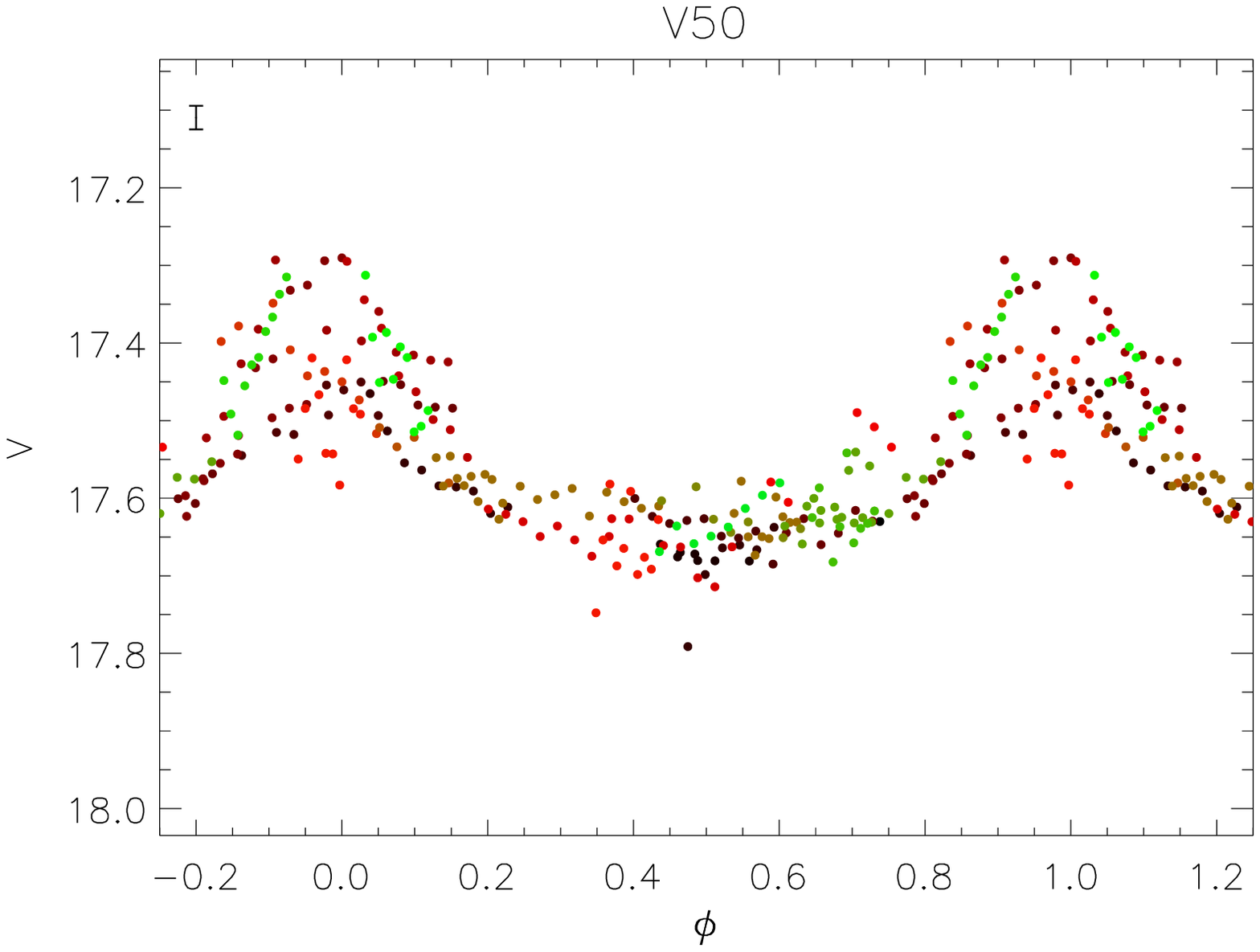}
  \includegraphics[width=6cm, angle=0]{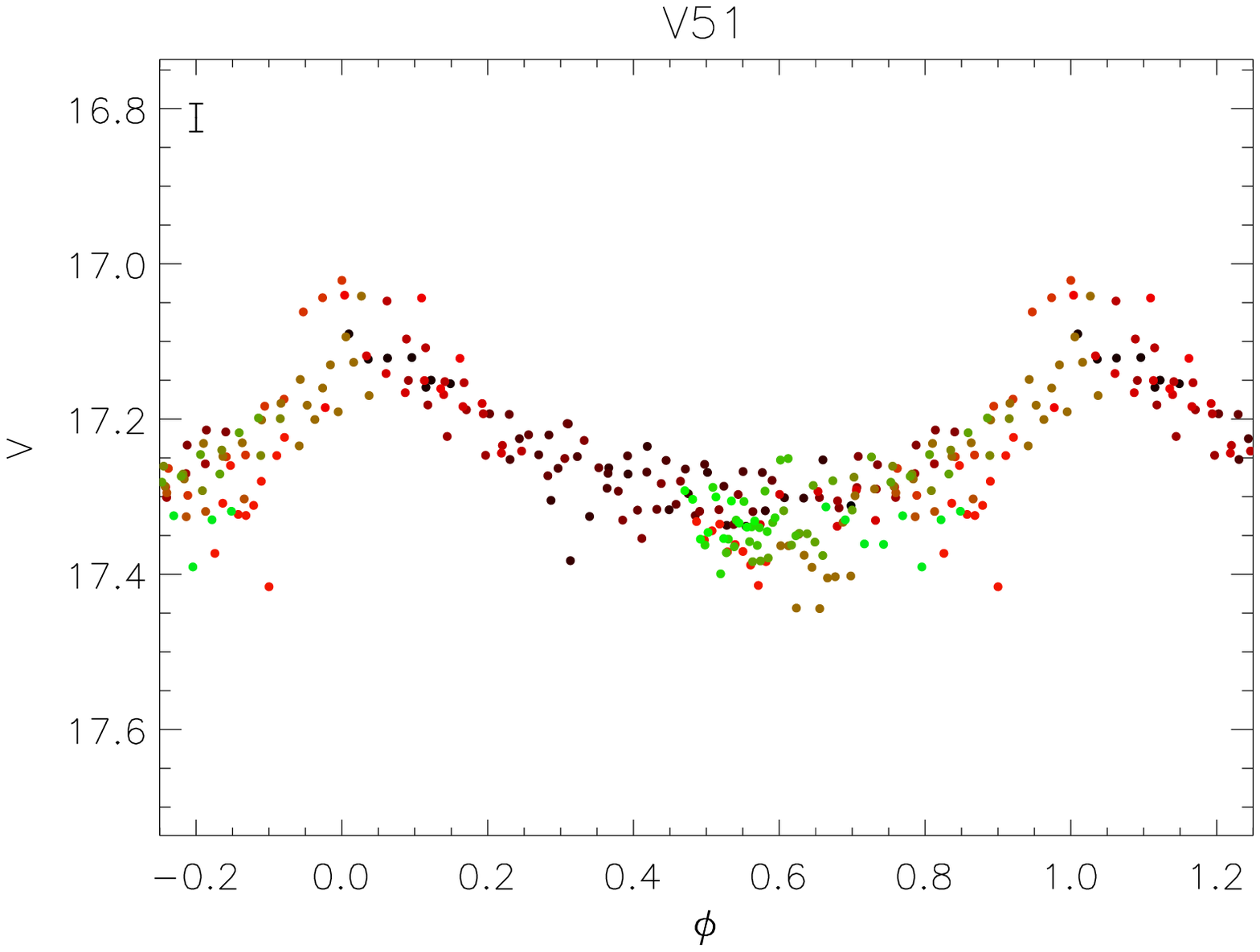}
  \includegraphics[width=6cm, angle=0]{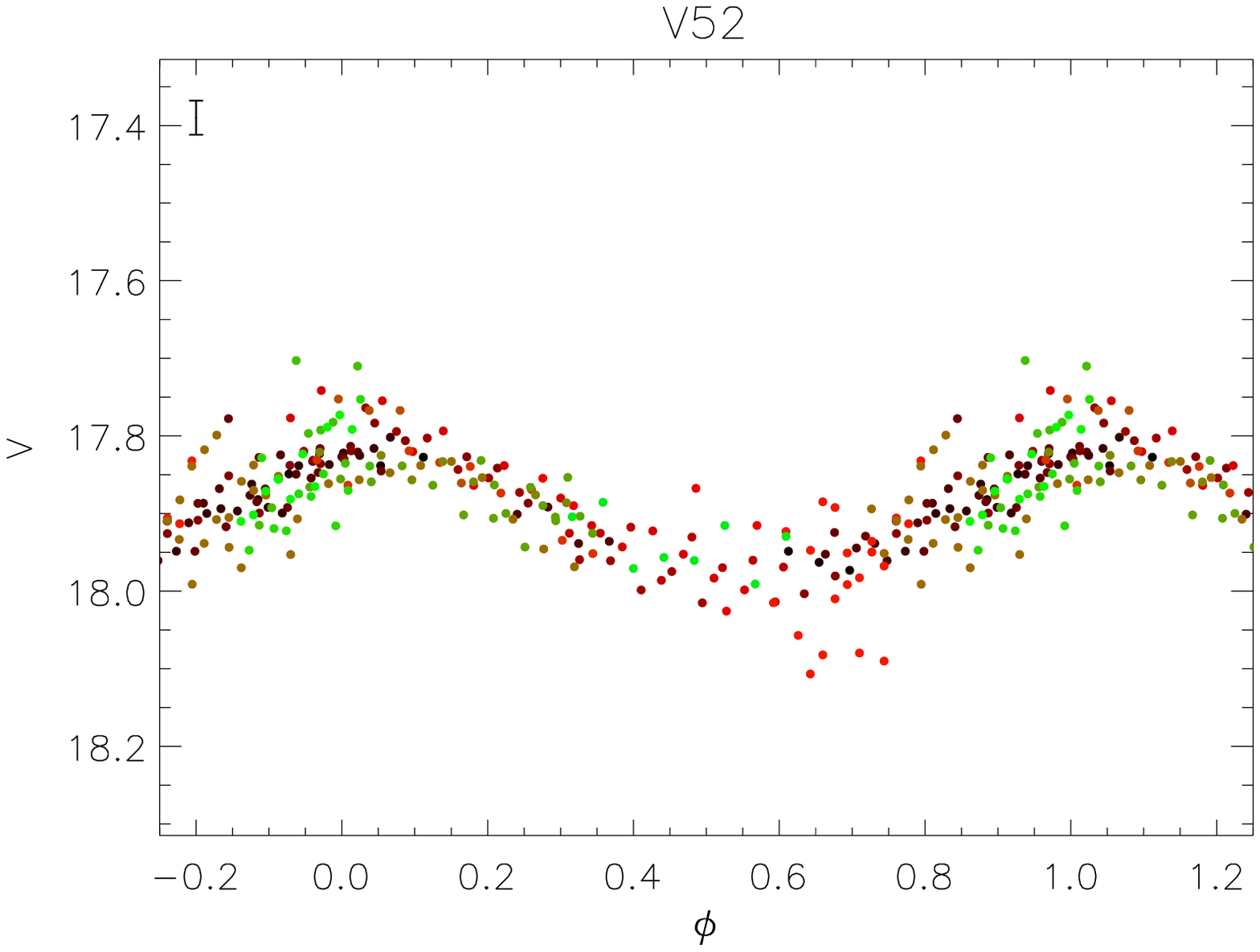}

  \caption{Phased $V-$band light curves of the confirmed SX Phe stars in M68, with a typical 1-$\sigma$ error bar plotted in the top left corner. The magnitude scale is the same on all plots in order to facilitate comparison of variation amplitude. \label{fig:lc_SXV}}

\end{figure*}

\begin{figure}
  \centering
  \includegraphics[width=8cm, angle=0]{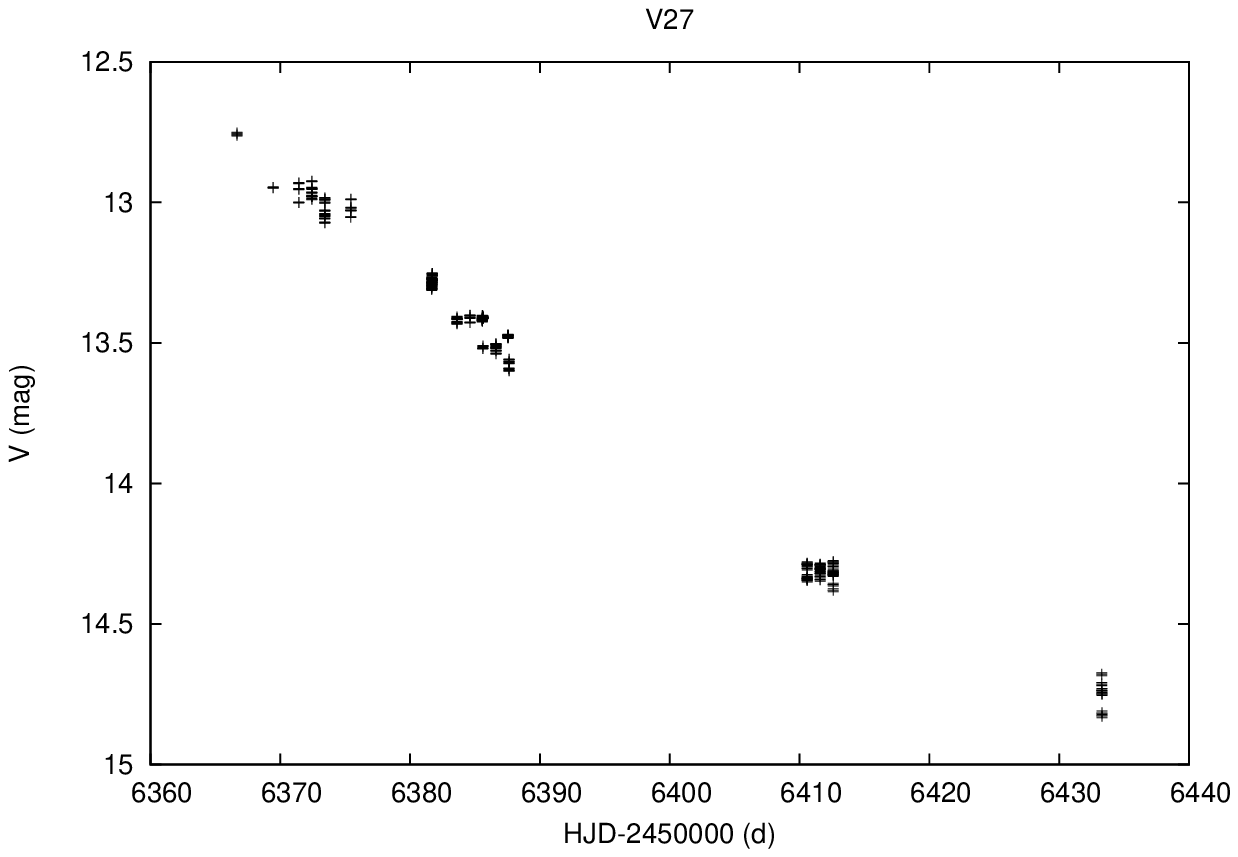}
  \includegraphics[width=8cm, angle=0]{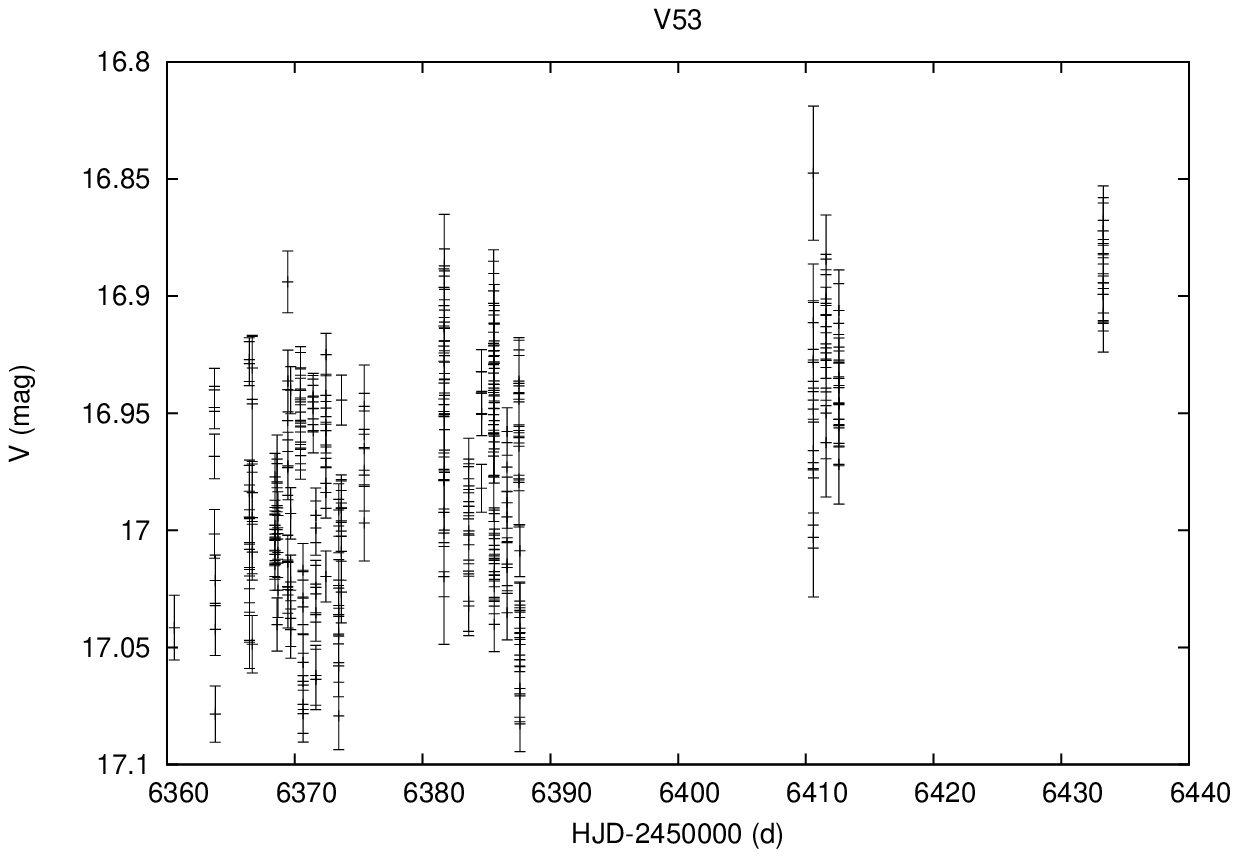}
 
  \caption{Unphased $V-$band light curves of the variables V27 (FI Hya) and V53 (unknown type), plotted with 1-$\sigma$ error bars. \label{fig:lc_others}}

\end{figure}

\begin{figure*}
  \centering
  \includegraphics[width=12cm, angle=0]{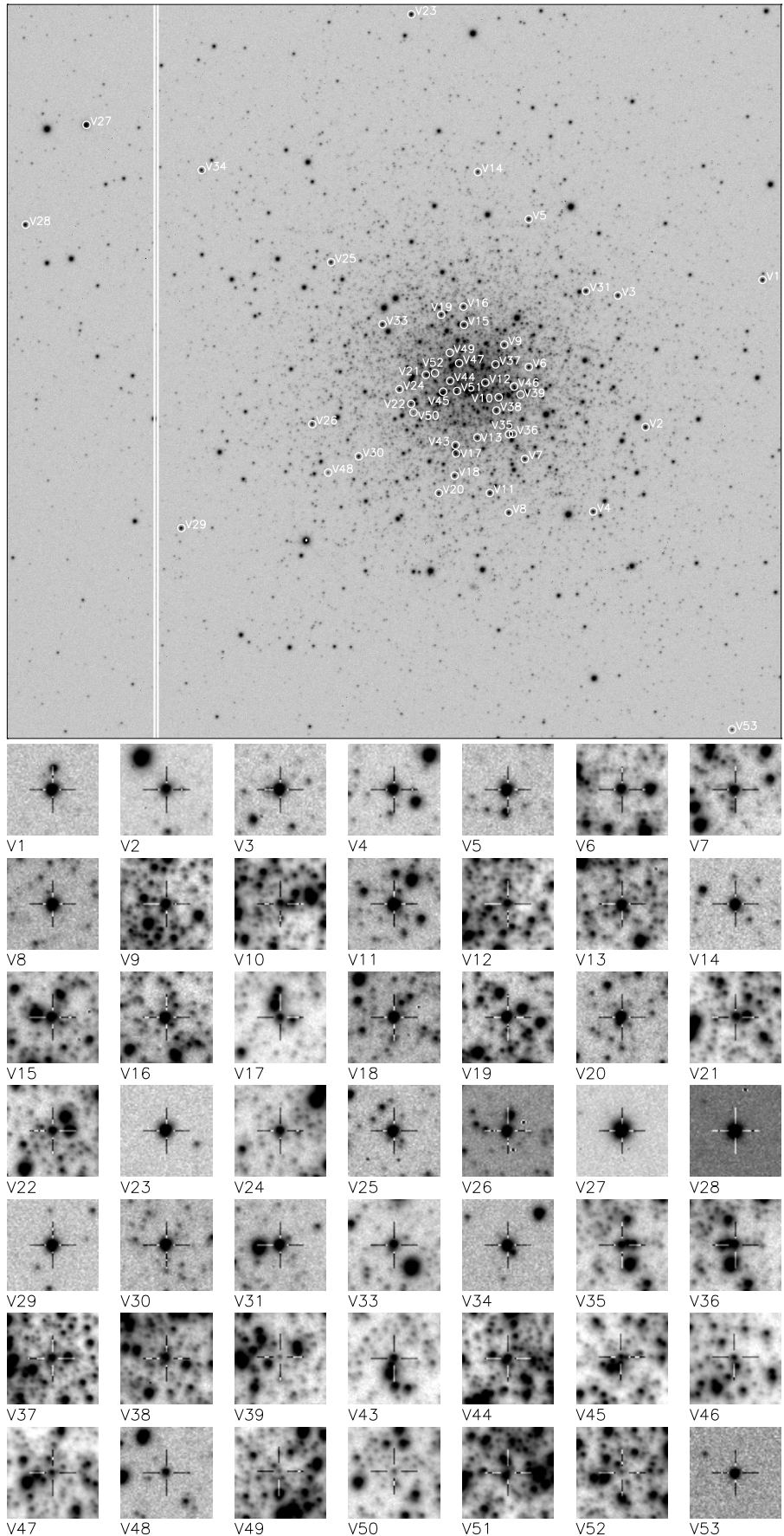}
  \caption{Finding chart for the confirmed variable objects in M68, using our $V$ reference image. North is up and East is to the left. The image size is 8.8 $\times$ 12.26 arcminutes, with each stamp 27.6$^{\prime\prime}$ $\times$ 27.6$^{\prime\prime}$. A white circle is centred on each variable and labelled with the variable number. The display scale has been modified where necessary to make the source as clear as possible on the stamps, and the location of the variable is also marked with a cross-hair. Stamps from the EMCCD reference image for V12, V44, V45 and V51 are shown in \Fig{fig:stamps_emccd}. \label{fig:fchart}}

\end{figure*}

\begin{figure*}
  \centering
  \includegraphics[width=14cm, angle=0]{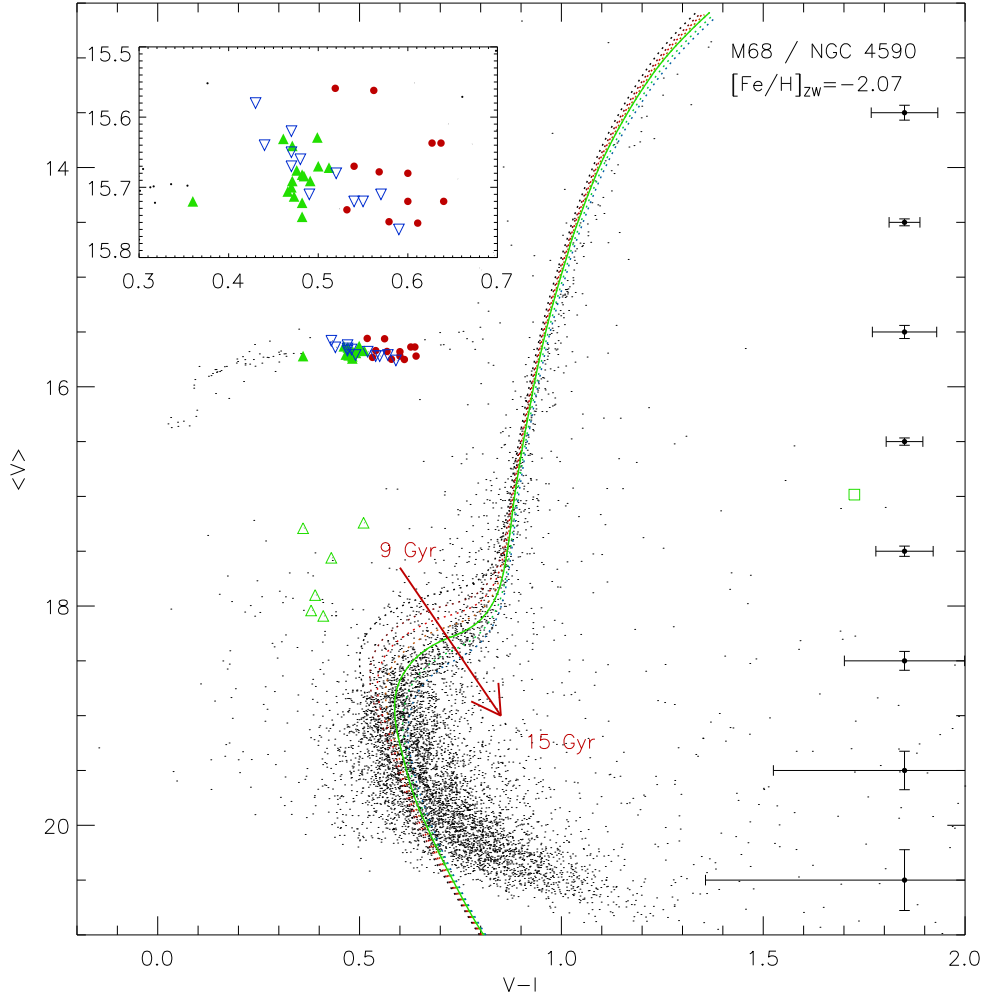}

  \caption{$(V-I), V$ colour-magnitude diagram from our photometry. The location of RR0 (red filled circles), RR1 (green filled triangles), RR01 (blue inverted triangles) and SX Phe (open green triangles) stars are shown. The variable V53 (unknown type) is shown as an open green square. Typical error bars are shown for different magnitude levels on the right-hand side of the plot. An inset in the top left corner of the plot shows a zoom of the instability strip region of the HB. For added information, isochrones for 9, 10, 11, 12, 13, 14 and 15 Gyr from \cite{dotter08} are overplotted in different colours; the best-fit isochrone (13 Gyr) is plotted as a thick green solid line. \label{fig:cmd}}

\end{figure*}

\begin{figure}
  \centering
  \includegraphics[width=8cm, angle=0]{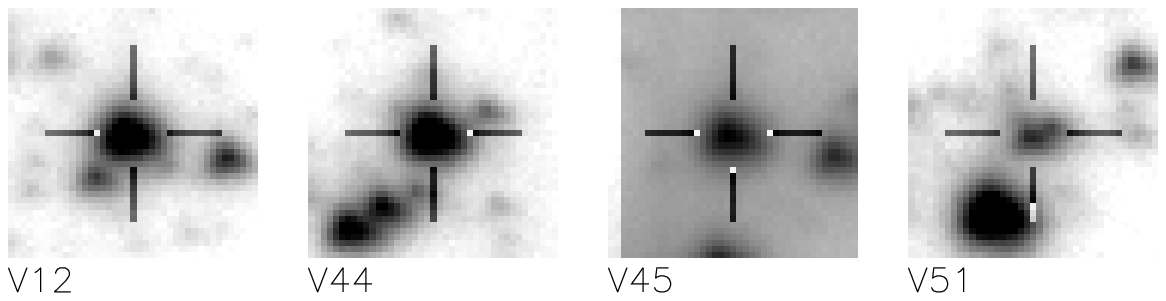}
  \caption{Stamps from the EMCCD reference image, for the stars within our EMCCD field of view, except V47, which is too close to the edge of the image. North is up and East is to the left. Each stamp is 3.6$^{\prime\prime}$ $\times$ 3.6$^{\prime\prime}$ and a cross-hair marks the location of the variable star. \label{fig:stamps_emccd}}

\end{figure}

$V-$band light curves for all of the variables objects are plotted on Figs. \ref{fig:lc_RR0V}$-$\ref{fig:lc_others}. $I-$band light curves are available for download with the electronic version of this article. A finding chart for all confirmed variables in M68 is shown in \Fig{fig:fchart} and a CMD in \Fig{fig:cmd}. The CMD confirms the classification of the confirmed variables, with RRL located on the instability strip and SX Phe stars in the blue straggler region. We also show stamps of variables detected on our EMCCD images on \Fig{fig:stamps_emccd}.

\begin{table*}
\begin{center}
  \begin{tabular}{ccccccccccc}

     \hline
    \#		 	&Epoch	&$P$ 		&$\beta$		&$\langle V \rangle$	&$\langle I \rangle$	& $A_V$	&$A_I$ &Type\\
    		 	&(HJD-2450000)&(d)  	&[$d{\rm Myr}^{-1}$]			&[mag]	&[mag]	&[mag] 	&[mag] 	&\\
  \hline  

\hline  
  RR0 \\
\hline
V2 &6411.5879  &0.5781755	&$-0.125$	&15.75 & 15.17 &0.72 &0.49 &RR0 \\
V9 &6412.5455  &0.579043	&$-$		&15.72 & 15.12 &0.69 &0.40 &RR0$b$ \\
V10 &6363.7134 &0.551920  	&$-$		&15.71 & $-$ &1.08 &$-$ &RR0$b$ \\
V12 &6369.4311 &0.615546  &$-$		&15.56 & 15.04 &0.89 &0.61 &RR0 \\
V14 &6373.4455 &0.5568499  &+1.553	&15.73 & 15.20 &1.11 &0.77 &RR0 \\
V17 &6411.5835 &0.668424  &$-$		&15.68 & 15.08 &0.82 &0.54 &RR0($b$?) \\
V22 &6373.4162 &0.5634451  &$-$		&15.67 & 15.13 &1.15 &0.75 &RR0 \\
V23 &6410.5710 &0.6588921  &$-$		&15.68 & 15.11 &1.11 &0.58 &RR0 \\
V25 &6411.5323 &0.6414842  &$-0.488$	&15.72 & 15.08 &0.79 &0.45 &RR0$b$ \\
V28 &6363.7134 &0.6067796  &$+0.102$	&15.75 & 15.14 &1.18 &0.70 &RR0 \\
V30 &6370.5680 &0.7336375  &$+0.044$	&15.64 & 15.00 &0.37 &0.25 &RR0 \\
V32 &$-$ &0.5882  &$-$	&$-$ &$-$ &$-$ &$-$ &RR0 \\
V35 &6381.6690 &0.7025348  &$-$		&15.56 & 15.00 &1.05 &0.66 &RR0 \\
V46 &6363.7743 &0.7382510  &$-$		&15.64 & 15.01 &0.54 &0.37 &RR0 \\

\hline  
  RR1 \\
\hline
V1 &6381.7111  &0.3495912  	&+0.273	&15.70 & 15.23 &$\sim$0.45 &0.40 &RR1 \\
V5 &6366.6697  &0.2821009   &$-0.497$	&15.72 & 15.36 &0.45 &0.28 &RR1$b$ \\
V6 &6385.6186  &0.3684935   &$-0.088$	&15.69 & 15.20  &0.53 &0.34 &RR1 \\
V11 &6370.6253 &0.3649338  &$+0.225$	&15.71 & 15.24 &$\sim$0.55 &0.37 &RR1 \\
V13 &6411.5829 &0.3617370   &$+0.116$	&15.74 & 15.26 &0.58 &0.36 &RR1 \\
V15 &6385.6092 &0.3722615   &$-$		&15.68 & 15.20 &0.53 &0.34 &RR1 \\
V16 &6381.6814 &0.3819671   &+0.066	&15.69 & 15.22 &0.50 &0.35 &RR1 \\
V18 &6363.7943 &0.3673459   &$-0.051$	&15.72 & 15.24 &0.55 &0.37 &RR1 \\
V20 &6363.7643 &0.3857892   &+0.234	&15.68 & 15.20 &0.56 &0.34 &RR1 \\
V24 &6370.4698 &0.3764448   &$-1.081$	&15.68 & 15.20 &0.50 &0.34 &RR1 \\
V33 &6385.5693 &0.3905647   &$-$	&15.67 & 15.17 &0.47 &0.23 &RR1 \\
V37 &6363.7727 &0.3846092   &$-$	&15.64 & 15.17 &0.47 &0.33 &RR1 \\
V38 &6370.4420 &0.3828116   &$-$	&15.63 & 15.17 &0.53 &0.33 &RR1 \\
V43 &6363.7527 &0.3706144   &$-$	&15.71 & 15.24 &0.57 &0.35 &RR1 \\
V44 &6371.4311 &0.3850912   &$-$	&15.67 & 15.16 &0.48 &0.29 &RR1 \\
V47 &6385.6436 &0.3729255   &$-$	&15.63 & 15.13 &0.49 &0.32 &RR1 \\

\hline  
  RR01 \\
\hline
V3 &6381.6890  &0.3907346  &$-$	&15.64 & 15.20 &0.68 &0.42 &RR01 \\
V4 &6410.6485  &0.3962175  &$-$	&15.67 & 15.20 &0.61 &0.40 &RR01 \\
V7 &6381.7511  &0.3879608  &$-$	&15.71 & 15.22 &0.62 &0.41 &RR01 \\
V8 &6412.5861  &0.3904076  &$-$	&15.65 & 15.18 &0.53 &0.34 &RR01 \\
V19 &6368.6564 &0.3916309   &$-$  &15.66 & 15.18 &0.56 &0.34 &RR01 \\
V21 &6385.6233 &0.4071121   &$-$  &15.62 & 15.15 &0.64 &0.42 &RR01 \\
V26 &6369.4571 &0.4070332   &$-$  &15.72 & 15.18 &0.80 &0.50 &RR01 \\
V29 &6387.5498 &0.3952413   &$-$  &15.71 & 15.14 &0.56 &0.30 &RR01 \\
V31 &6385.6217 &0.3996599   &$-$  &15.58 & 15.15 &0.66 &0.43 &RR01 \\
V34 &6363.7673 &0.4001371   &$-$  &15.76 & 15.17 &$\sim$0.60 &0.45 &RR01 \\
V36 &6384.6470 &0.415346   &$-$  &15.68 & 15.16 &0.64 &0.40 &RR01 \\
V45 &6366.6760 &0.3908187   &$-$  &15.72 & 15.17 &0.52 &0.30 &RR01 \\
\hline  
  SX Phe \\
\hline
V39 &6433.2863 &0.0640464  &$-$	&18.04 & 17.66 &0.85 &0.64 &SX \\
V48 &6387.6218 &0.043225  &$-$	&17.29 & 16.93 &0.24 &$\sim$0.11 &SX \\
V49 &6387.6120 &0.048469  &$-$	&18.09 & 17.68 &0.59 &0.40 &SX \\
V50 &6370.4561 &0.065820  &$-$	&17.56 & 17.13 &0.70 &0.30 &SXd \\
V51 &6383.6232 &0.058925  &$-$	&17.24 & 16.73 &0.35 &0.20 &SX \\
V52 &6410.5871 &0.037056  &$-$	&17.90 & 17.51 &0.25 &$\sim$0.20 &SX \\
\hline
Others\\
\hline
V27 &2697.3 &322.2342 &$-$ &9.8 & $-$ &4.03 &$-$ &Field Mira$^\dagger$\\
V53 &$-$ &$-$ &$-$ &16.98 &15.26  &$\ge 0.1$ &$-$ &?\\
\hline \hline
  \end{tabular}
  \caption{Epochs, periods, mean magnitudes and amplitudes $A$ in $V$ and $I$ for all confirmed variable stars in M68. V49, V50, V51 and V52 are newly discovered variables. A $b$ at the end of the variable type denotes stars which exhibit Blazhko modulation in their light curve. ``SX" denotes SX Phe stars, and an appended ``d" denotes double-mode pulsation. $\langle V \rangle$ and $\langle I \rangle$ are intensity-weighted mean magnitudes. Note that those are calculated from Fourier fits for all RRL stars, as the mean magnitude is stable even for unsatisfactory fits. For stars without good Fourier fits, amplitudes are calculated from the data. For V27, the data are taken from the ASAS catalogue \cite{pojmanski02}. The data for V32 are taken from C93, as that star is outside of our field of view. Values of the period-change rate parameter $\beta$ are also given, where relevant (see \Sec{sec:pchange}); in those cases, the value of $P$ given corresponds to the instantaneous period at the epoch. $^\dagger$this star is FI Hydra. \label{tab:variables}}
  \end{center}
\end{table*}

\begin{table}
\begin{center}
  \begin{tabular}{ccc}

     \hline
    \#		 &RA 	&DEC \\
  \hline  
  RR0 \\
  \hline

V2		&12:39:15.29	&-26:45:24.6	\\
V9		&12:39:25.57	&-26:44:00.8	\\
V10		&12:39:25.96	&-26:44:54.7	\\
V12		&12:39:26.918	&-26:44:39.73	\\
V14		&12:39:27.54	&-26:41:04.2	\\
V17		&12:39:29.02	&-26:45:52.4	\\
V22		&12:39:32.30	&-26:45:01.6	\\
V23		&12:39:32.41	&-26:38:23.0	\\
V25		&12:39:38.15	&-26:42:37.2	\\
V28		&12:40:00.31	&-26:41:59.8	\\
V30		&12:39:36.07	&-26:45:55.6	\\
V32		&12:39:03.29 	&-26:55:18.0    \\
V35		&12:39:25.19	&-26:45:32.5	\\
V46		&12:39:24.84	&-26:44:43.4	\\
\hline
RR1 \\
\hline
V1		&12:39:06.86	&-26:42:53.3	\\
V5		&12:39:23.83	&-26:41:52.3	\\
V6		&12:39:23.77	&-26:44:23.6	\\
V11		&12:39:26.56	&-26:46:32.7	\\
V13		&12:39:27.49	&-26:45:35.9	\\
V15		&12:39:28.50	&-26:43:40.9	\\
V16		&12:39:28.54	&-26:43:22.1	\\
V18		&12:39:29.12	&-26:46:15.2	\\
V20		&12:39:30.26	&-26:46:33.2	\\
V24		&12:39:33.15	&-26:44:46.6	\\
V33		&12:39:34.41	&-26:43:40.7	\\
V37		&12:39:26.18	&-26:44:20.9	\\
V38		&12:39:26.09	&-26:45:08.3	\\
V43		&12:39:29.06	&-26:45:43.7	\\
V44		&12:39:29.477	&-26:44:38.06	\\
V47		&12:39:28.831	&-26:44:19.92	\\

\hline  
  RR01 \\
\hline

V3		&12:39:17.33	&-26:43:09.9	\\
V4		&12:39:19.06	&-26:46:51.3	\\
V7		&12:39:24.02	&-26:45:57.7	\\
V8		&12:39:25.18	&-26:46:52.5	\\
V19		&12:39:30.15	&-26:43:30.4	\\
V21		&12:39:31.19	&-26:44:32.0	\\
V26		&12:39:39.46	&-26:45:22.9	\\
V29		&12:39:48.93	&-26:47:10.1	\\
V31		&12:39:19.65	&-26:43:05.4	\\
V34		&12:39:47.56	&-26:41:03.5	\\
V36		&12:39:24.89	&-26:45:32.1	\\
V45		&12:39:29.989	&-26:44:49.18	\\

  \hline  
  SX Phe \\
  \hline

V39		&12:39:24.34	&-26:44:51.8	\\
V48		&12:39:38.27	&-26:46:12.4	\\
V49		&12:39:29.52	&-26:44:09.3	\\
V50		&12:39:32.12	&-26:45:10.5	\\
V51		&12:39:28.986	&-26:44:48.47	\\
V52		&12:39:30.57	&-26:44:30.1	\\
\hline
Others \\
\hline
V27		&12:39:55.92	&-26:40:17.4	\\
V53		&12:39:08.91	&-26:50:33.8	\\

\hline \hline
  \end{tabular}
  \caption{Equatorial celestial coordinates of all confirmed variables in M68 at the epoch of the reference image, HJD$\sim2456385.6$ d. More precise coordinates are given for stars which are within the field of view of the EMCCD reference image, with epoch $\sim 2456451$ d. The coordinates for V32, which is outside of our field of view, are from C93. \label{tab:coordinates}}
  \end{center}
\end{table}

\begin{table}
\begin{center}
  \begin{tabular}{ccccc}

     \hline
    \#		 	&$\langle V \rangle$	&rms ($V$)	&$\langle I \rangle$ 	&rms ($I$) \\
    \#		 	&[mag]	&[mag]	&[mag] 	&[mag] \\
  \hline  

    V40		&18.32	&0.068	&17.50	&0.075\\
    V41		&18.15	&0.052	&17.36	&0.060\\
    V42		&19.05	&0.083	&18.36	&0.156\\
    SV49		&14.72	&0.008	&13.67	&0.007\\
    SV50		&15.15	&0.010	&14.14	&0.008\\
    SV51		&12.68	&0.013	&$-$		&$-$\\
    SV52		&18.17	&0.225	&$-$		&$-$\\
    SV53		&18.60	&0.077	&17.94	&0.14\\
    SV54		&18.48	&0.361	&17.00	&0.14\\
    SV55		&17.04	&0.023	&16.21	&0.03\\
    SV56		&19.81	&0.192	&18.97	&0.31\\
    SV57		&16.81	&0.016	&15.91	&0.02\\

\hline \hline
  \end{tabular}
  \caption{Light curve mean magnitudes and rms values for V40-42, for which we do not find evidence of variability, as well as for the ``new" variables published by \cite{sariya14}. Since \cite{sariya14} assigned those variables new V numbers, we add an ``S" as a prefix to avoid confusion with the confirmed variables in this paper. Note that the mean magnitudes of V41 and V42 are significantly different from those given by W94. For V41 we suggest that this might be due to blending by V16 in their data, but no explanation is offered for the difference with V42. \label{tab:rms_nonvar}}
  \end{center}
\end{table}

\subsubsection{EMCCD observations}

We repeated the method we used for CCD observations to search for variability in the EMCCD observations we obtained. Of the known variables, only V44 has a light curve, with V12, V45 and V47 also located within our FOV, but being too close to the edge to allow for photometric measurements. Furthermore, the camera was changed in May 2013, with a slightly different filter after that, meaning that measurements from images taken before and after the change need to be treated as separate light curves. 

\begin{figure}
  \centering
  \includegraphics[width=8cm, angle=0]{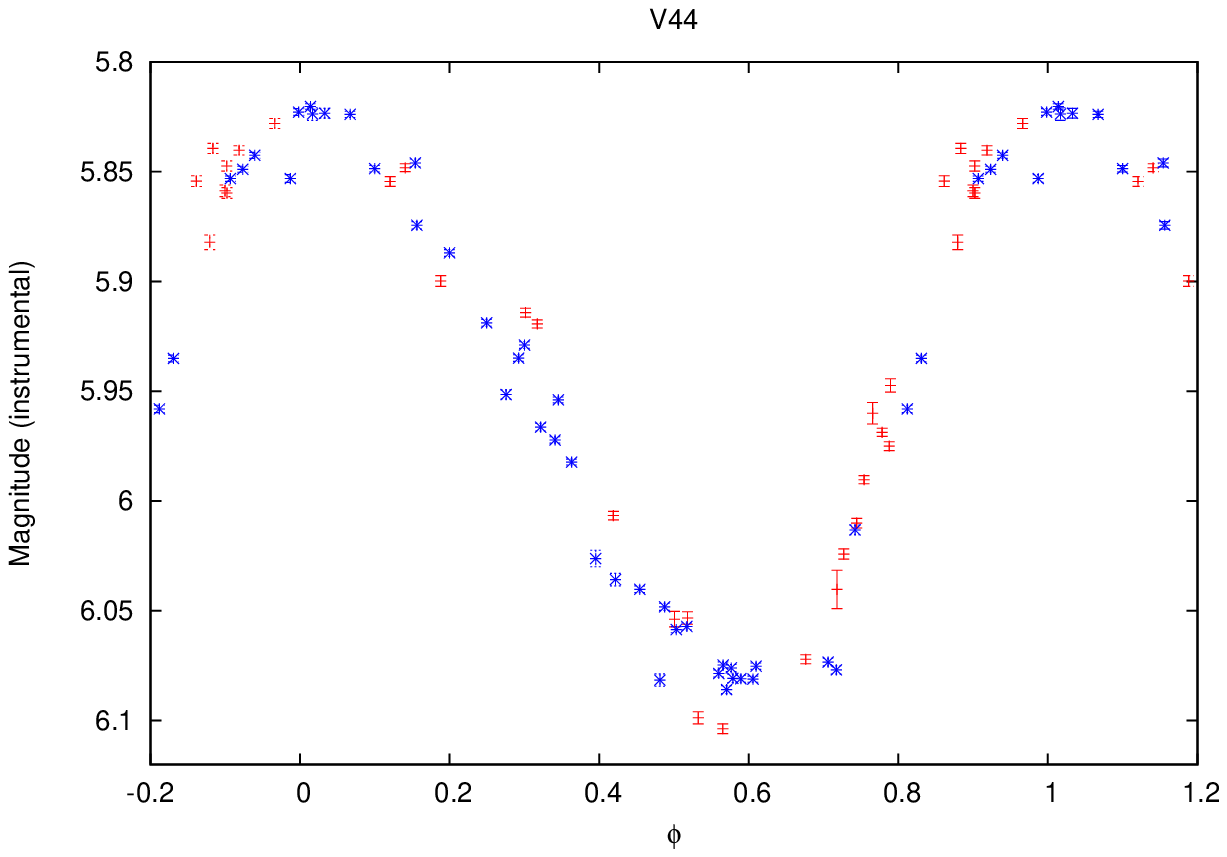}
  \includegraphics[width=8cm, angle=0]{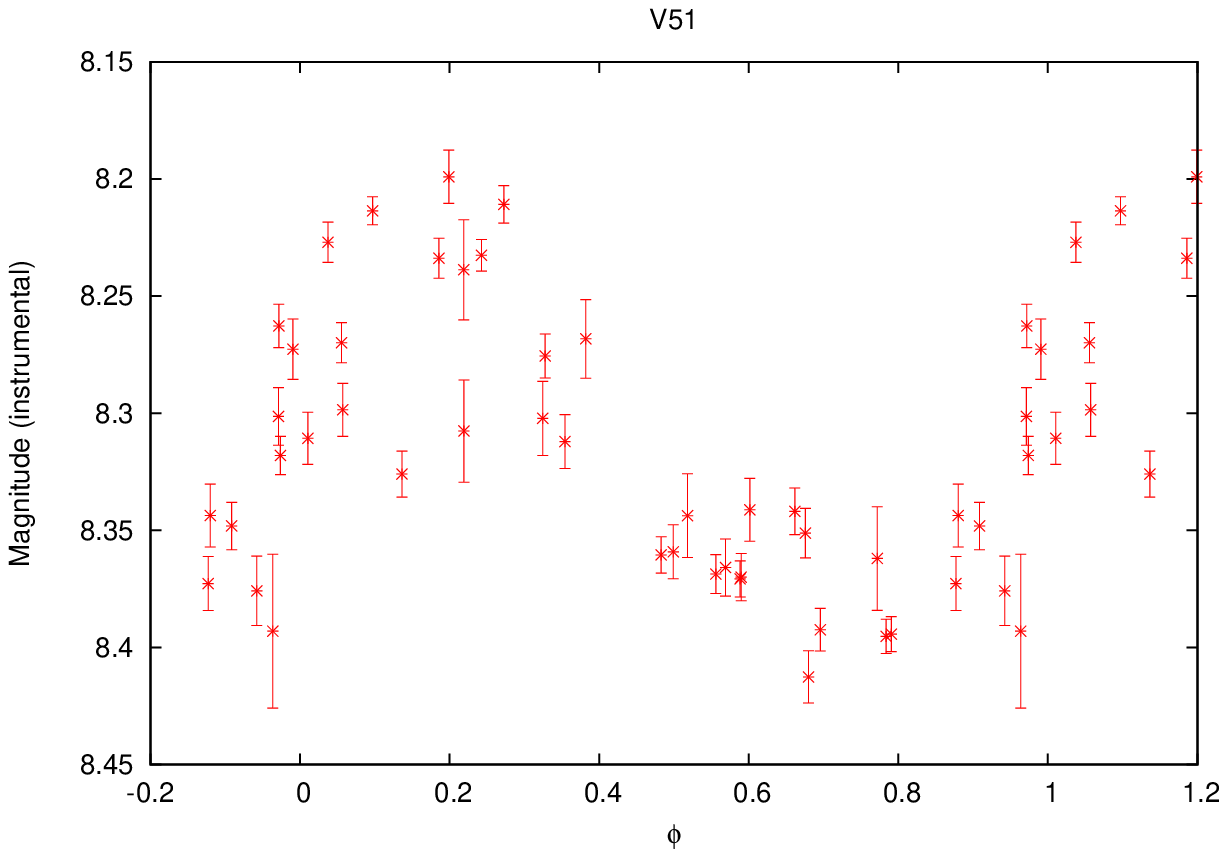}
  \caption{EMCCD light curve for variables V44 and V51. See \Tab{tab:variables} for information about these variables. Note that for V44 we show the light curves before (blue + signs) and after (red asterisks) the camera change as separate light curves, as discussed in the text. For V51, we show only the light curve after the change, as the photometry before was too poor. \label{fig:v51_LI}}
\end{figure}

We also detect the new variable V51, and confirm the period found with the CCD data for this object. The EMCCD light curves for V44 and V51 are shown in \Fig{fig:v51_LI}.

\subsection{Period changes in RRL stars}\label{sec:pchange}

Period changes have been observed in many RRL stars, both in the Galactic field and in globular clusters. Period changes are usually classed as evolutionary or non-evolutionary. Evolutionary period changes of stars on the instability strip are understood to be due to their radius increasing and contracting. These only account for slowly increasing or decreasing changes, however, and in many RRL, abrupt period changes have been observed \citep[e.g.][]{stagg80} which cannot be explained by such evolution. As yet, there is no clear explanation for such changes. 

As noted by \cite{jurcsik12}, in spite of this, period-change rates of RRL in a globular cluster can inform us about its general evolution using theoretical models. \cite{lee90} suggested that in Oosterhoff type II clusters, most of the RRL pass through the instability strip from blue to red when reaching the end of helium burning in their core. Since such a blue-to-red evolution would also lead to a period increase, \cite{lee91} argued that a mean positive value of the period change rate $\beta$ in a cluster would support this scenario. Furthermore, \cite{rathbun97} showed that the average period change should be smaller for Oosterhoff I clusters than Oosterhoff II. Few comprehensive studies of period changes in cluster RRL have been published; \cite{smith77} derived period changes for RRL in the Oosterhoff type II cluster M15, with a mean of $\beta=0.11 \pm 0.36\, \dmyr$. For the Oosterhoff type I cluster M3, \cite{jurcsik12} found a slightly positive mean value of $\beta \sim 0.01\,{\rm d\,Myr}^{-1}$, agreeing with the theoretical predictions of \cite{lee91} and the findings of \cite{rathbun97}.

The parameter $\beta$ is defined such that the period at time $t$ is given by

\begin{equation}\label{eq:beta}
P(t) = P_0 + \beta (t-E),
\end{equation}

\noindent
where $P_0$ is the period at the (arbitrary) epoch $E$, and $\beta$ as expressed in \Eq{eq:beta} is in units of $\mathrm{d\,d}^{-1}$; however $\beta$ is usually expressed in ${\rm d\,Myr}^{-1}$ as a more natural unit. The number of cycles $N_E$ elapsed at time $t$ since the epoch $E$ can then be calculated as  

\begin{equation}\label{eq:ncycles1}
N_E = \int_{E}^{t} \frac{dx}{P(x)} = \frac{1}{\beta}\ln\left[1+\frac{\beta}{P_0}(t-E) \right] \, .
\end{equation}

The phase is then
 
\begin{equation}\label{eq:phase_pchange}
\phi = N_E - \lfloor N_E \rfloor \, .
\end{equation}

\begin{figure}
  \centering
  \includegraphics[width=7cm, angle=0]{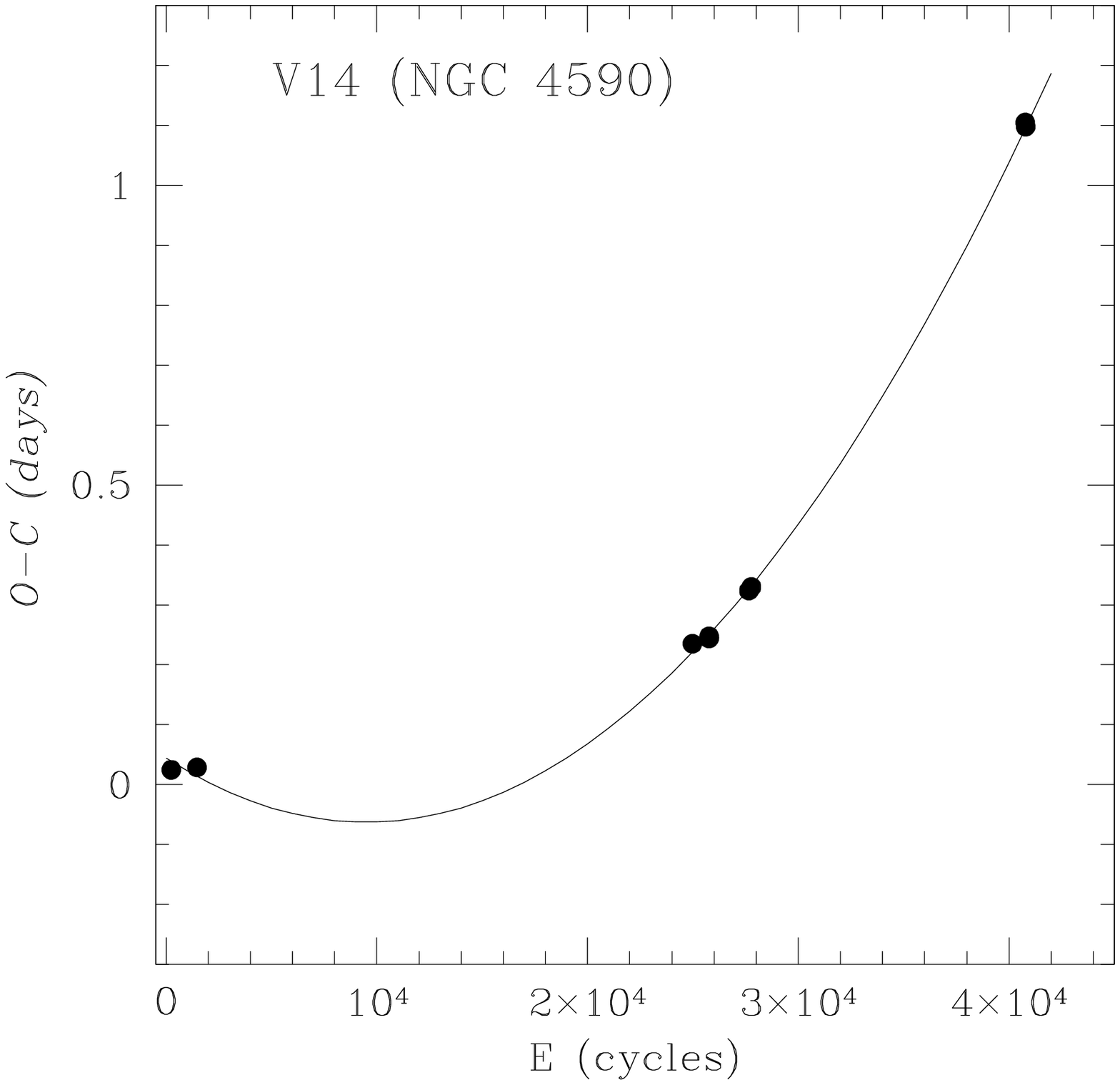}
  \includegraphics[width=7cm, angle=0]{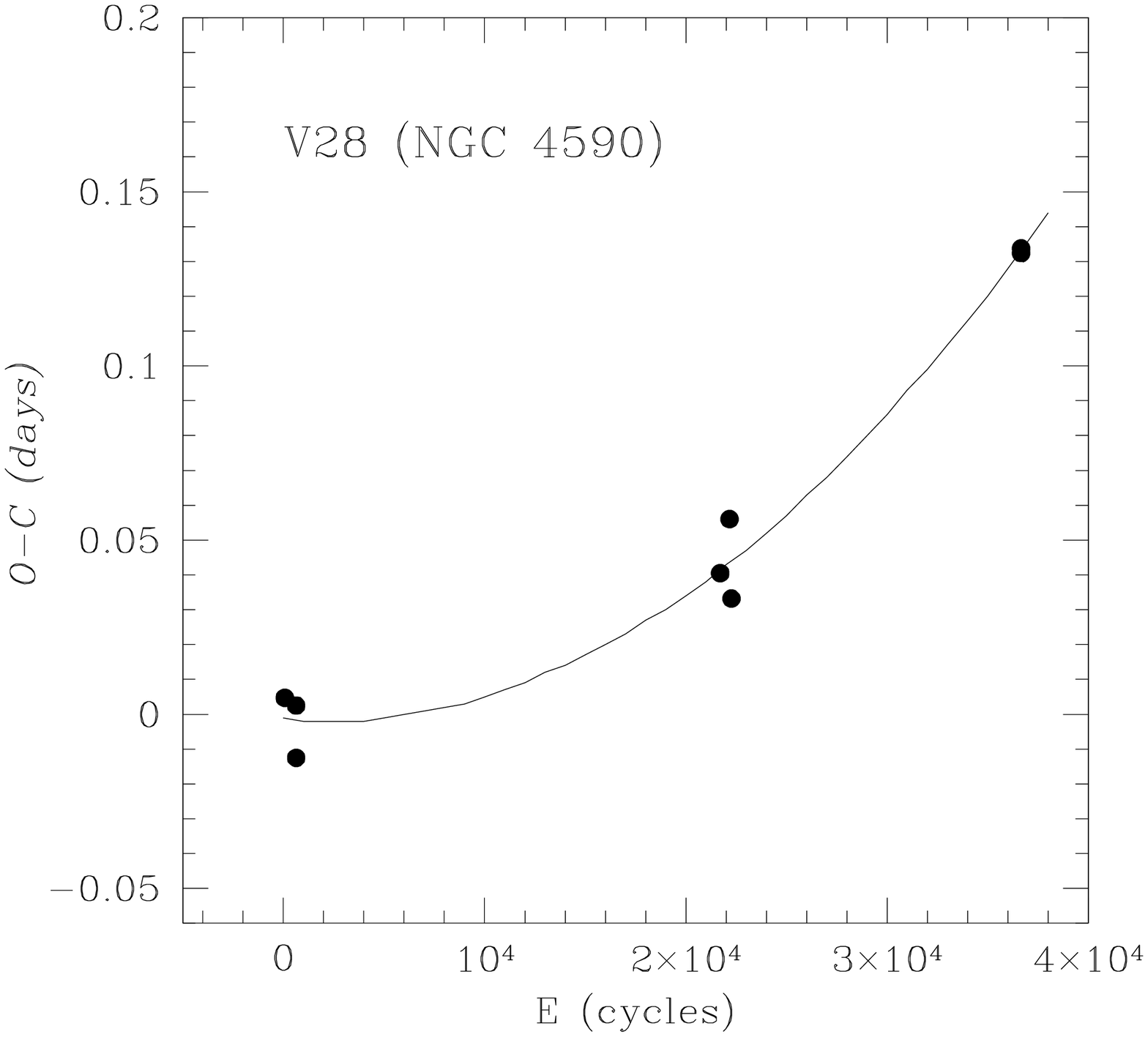}
  \caption{\textit{O-C} plots for variables V14 and V28 showing the quadratic fit to the time-dependence of the difference between observed and predicted times of light curve maxima. The values found using this quadratic fit for V14 and V28 are consistent with those derived using our grid fitting method (see text), $\beta$=+1.553 d Myr$^{-1}$ (V14)  and $\beta=$+0.102 d Myr$^{-1}$ (V28). \label{fig:oc}}

\end{figure}

Here we use data stretching back to 1951 (see Sec. \ref{sec:variables}) to derive period changes for RR0 and RR1 stars for which data sets are not well phased and/or aligned with a single constant period. To do this we performed a grid search in the ($P_0, \beta$) plane by minimising the string length of our data combined with those of W94 (both taken in $V-$band); we then searched manually around the best-fit solution incorporating other data sets taken in different passbands. As a consistency check, we also calculated period-change parameters using the \textit{O-C} method. This consists of using an ephemeris to predict the times of maxima in the light curves, and then plotting the difference between observed and predicted times of maxima against time. By fitting a quadratic function to this, a value of $\beta$ can be derived; two such fits are shown in \Fig{fig:oc} for variables V14 and V28, which produced values of $\beta$ consistent with the values derived using the grid method. The interested reader is referred to the papers of, e.g., \cite{belserene64} and \cite{nemec85b} for further details. We found here that the values we found for $\beta$ using the \textit{O-C} method did not always produce well-phased light curves for all data sets. This is most likely due to the small number of observed maxima available for the variables in this cluster. For the cases where the two methods were not in agreement, we used the value found with the grid search. C93 derived period-change rates by computing periods for their light curves, and for archival light curves, and subtracting one from another. The method used here, and the availability of more data, mean that our period-change calculations should be more robust. The signs of our values of $\beta$ agree with those of C93 except for V2 and V18, which, however, had very large associated error bars in that study.

\begin{figure}
  \centering
  \includegraphics[width=6cm, angle=270]{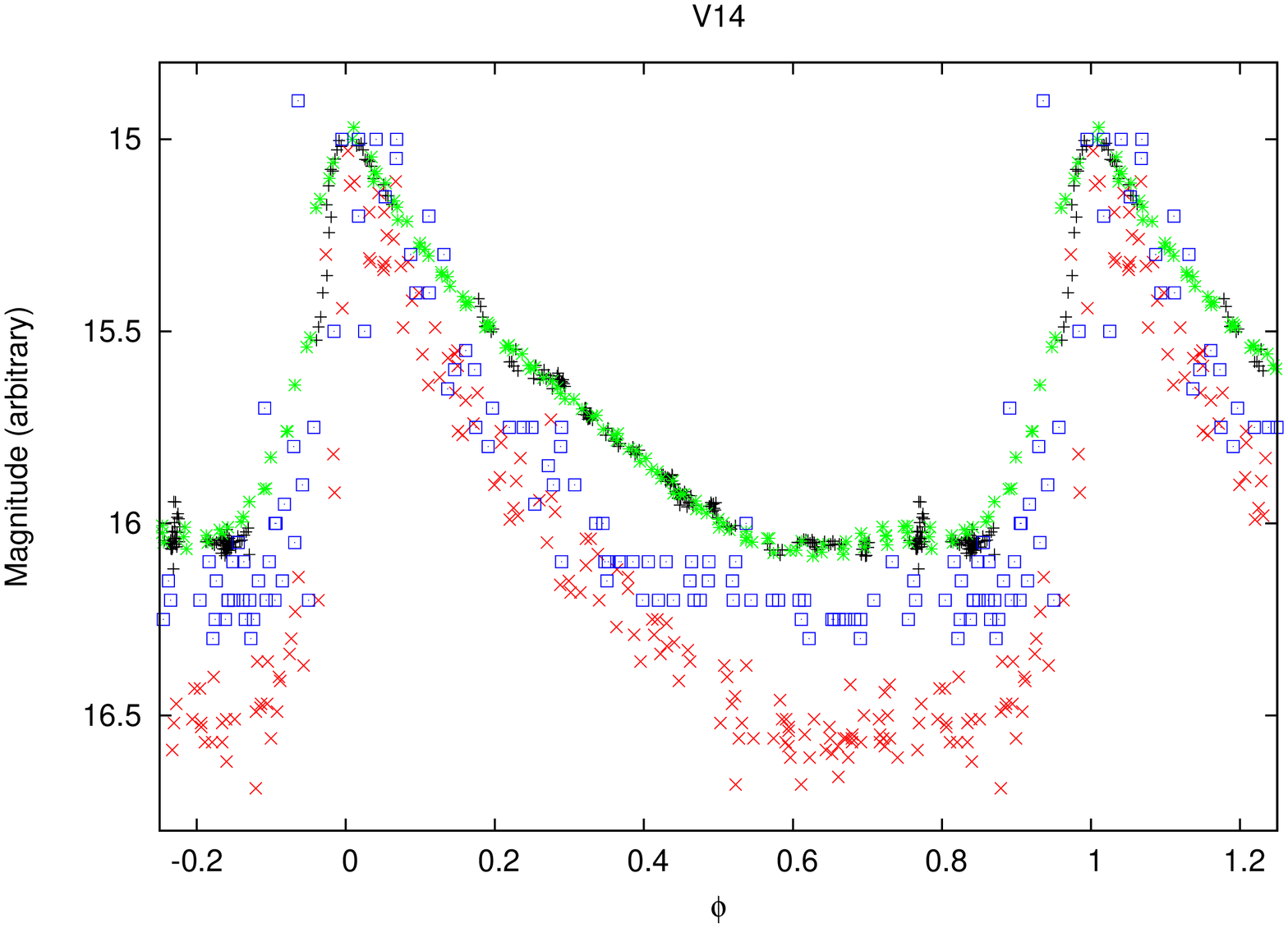}
  \includegraphics[width=6cm, angle=270]{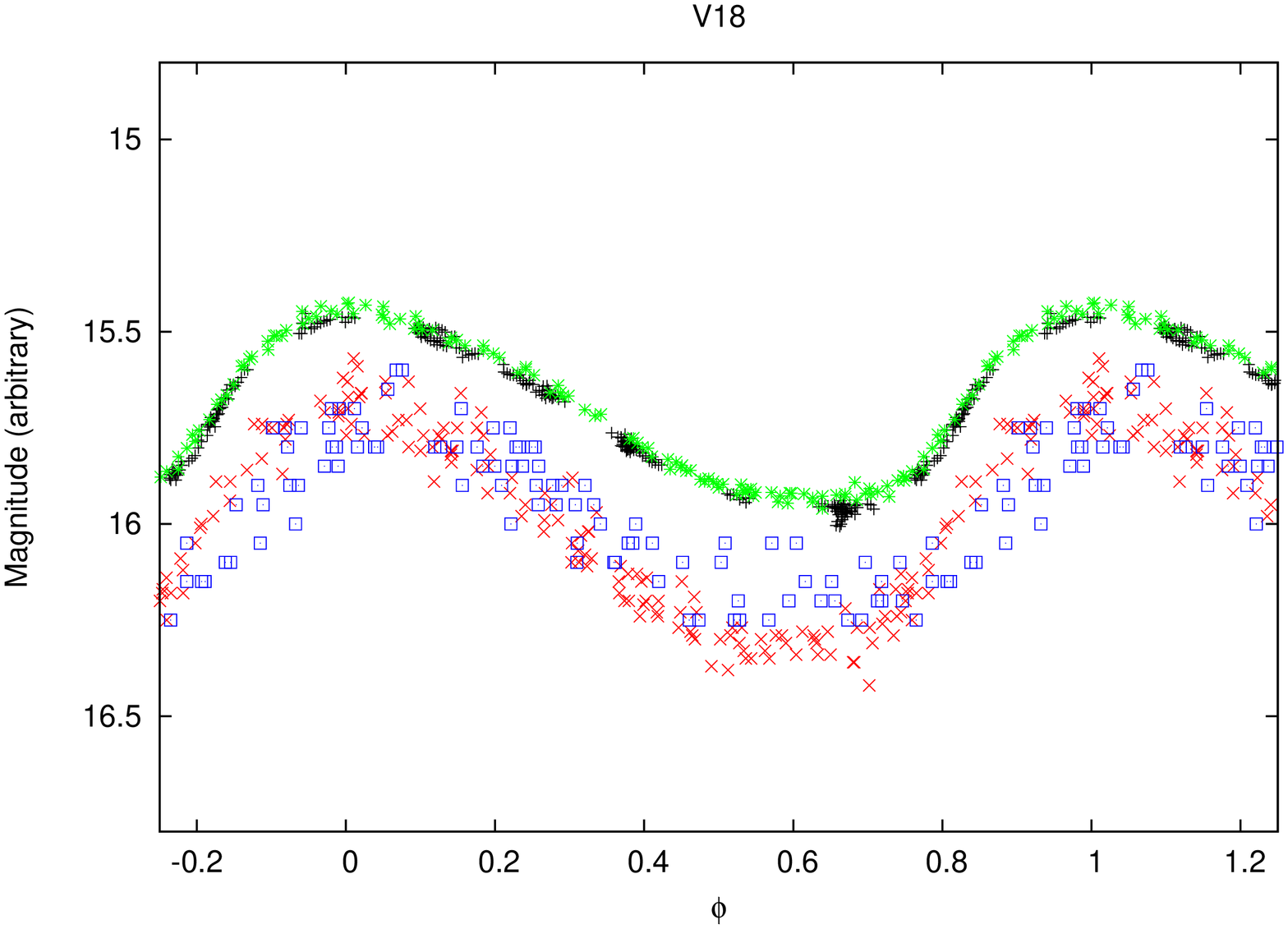}
  \caption{Phased light curves for V14 (top) and V18 (bottom), showing the data sets of this paper (black, + symbols), W94 (green asterisks), C93 (red crosses) and Rosino \& Pietra (1954, blue squares). The light curves were phased using a period change of +1.553 d Myr$^{-1}$ (V14) and $-0.051$ d Myr$^{-1}$ (V18). \label{fig:v14_pchange}}

\end{figure}

The values of $\beta$ from our analysis are listed in \Tab{tab:variables}. Examples of phased light curves are shown in \Fig{fig:v14_pchange} for V14 and V18. The large spread in values found for $\beta$ means that we are unable to draw any firm conclusion about the general evolution of M68. We find a mean value of $\langle \beta \rangle=0.02 \pm 0.57$ d Myr$^{-1}$.

\subsection{Discussion of individual RRL variables}\label{sec:individual}

Note that RR0 variables are plotted in \Fig{fig:lc_RR0V}, RR1 in \Fig{fig:lc_RR1V}, and RR01 in \Fig{fig:lc_RR01V}. Details of period-change calculations are given in \Sec{sec:pchange}.

\begin{figure}
  \centering
  \includegraphics[width=8cm, angle=0]{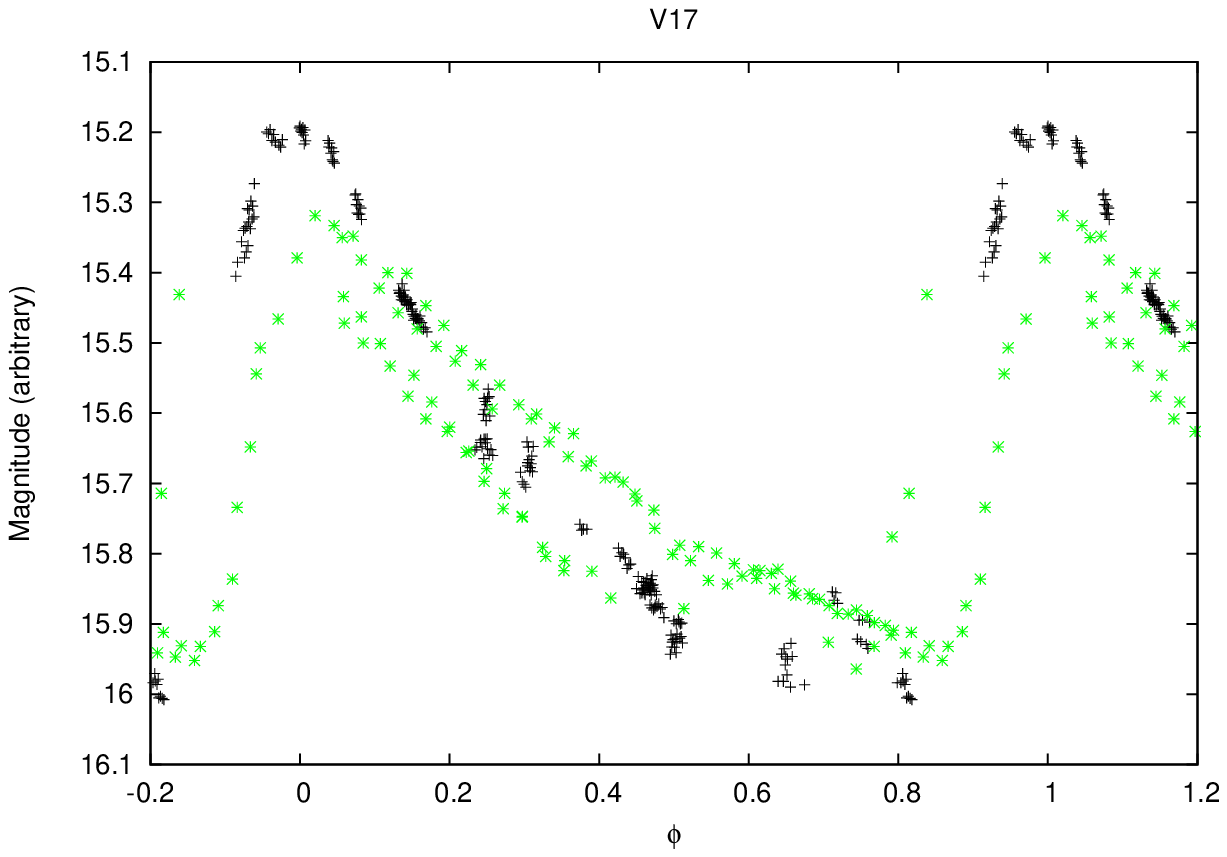}
  \caption{Phased light curves for V17, showing the $V-$band data sets of this paper (black, + symbols) and W94 (green asterisks). The amplitude is larger in our data by $\sim 0.15$ mag compared to the observations of W94,; we suggest that this might be due to slow amplitude modulation. \label{fig:v17}}

\end{figure}

\begin{itemize}
\item {\bf V1}: We could only phase the different data sets by including a period-change parameter $\beta=0.273$ d Myr$^{-1}$. 
\item {\bf V2}: W94 noted that this star features Blazhko modulation, but our light curves do not enable us to confirm this. We found that a negative period change parameter $\beta=-0.125$ d Myr$^{-1}$ was needed to phase-fold the various data sets.
\item {\bf V5}: This star requires a linear period change $\beta=-0.497$ d Myr$^{-1}$ to phase the different data sets; however, the data from C93 are not phased well with our best-fit period and period change parameter $\beta$; according to W94, this star also has slow amplitude variations with a period of several days; the scatter in our data is consistent with that assessment.
\item {\bf V6}: The data from \cite{rosino54} are not aligned with the other data sets with a constant period; we find a period-change parameter of $\beta=-0.088$ d Myr$^{-1}$. The light curve may suggest some Blazhko modulation, but our data do not allow us to make a strong claim about this.
\item {\bf V9}: Our light curve shows Blazhko modulation, as proposed by W94.
\item {\bf V10}: Our light curve for this object shows clear Blazhko amplitude modulation; interestingly, W94 found a constant amplitude but a variable period.
\item {\bf V11}: We could only phase the data sets simultaneously by including a period-change parameter $\beta=0.225$ d Myr$^{-1}$. 
\item {\bf V12}: We find no evidence of Blazhko modulation for this star, contrary to W94, who found strong cycle-to-cycle variations; this may be due do the baseline of W94 being longer by $\sim50$ days.
\item {\bf V13}: We find that the inclusion a period-change parameter is necessary to phase all the light curves; we find $\beta=0.116$ d Myr$^{-1}$
\item {\bf V14}: W94 noted a variation in shape for the bump at minimum brightness which we cannot confirm in our light curve. We also found that a rather large period change parameter $\beta=1.553$ d Myr$^{-1}$ was necessary to phase-fold all the data sets (see \Fig{fig:v14_pchange}).
\item {\bf V15}: Our observations show some slight residual scatter, which might suggest Blazhko modulation; some evidence of similar scatter is visible in the data of W94.
\item {\bf V16}: A period-change parameter $\beta=0.066$ d Myr$^{-1}$ was included to improve the phase-folding of the various data sets.
\item {\bf V17}: We do not find evidence for the Blazhko effect in this star as suggested by W94 on the timescales covered by our baseline. However, comparison of our light curve with that of W94 (\Fig{fig:v17}) shows clearly that the amplitude of the light curve is larger in our data set, which might indicate that modulation is present but slow.
\item {\bf V18, V20, V24}: These stars all required the inclusion of a period-change parameter for satisfactory phase-folding of available data sets.
\item {\bf V25}: We find a period-change parameter to phase all the data sets of $\beta=-0.488$ d Myr$^{-1}$. The light curve also shows Blazhko modulation, as was already noted by W94.
\item {\bf V28}: We tentatively suggest that this star might be affected by the Blazhko effect. A period change parameter of $\beta=0.102$ d Myr$^{-1}$ was found to improve the phase-folding of all data sets.
\item {\bf V30}: A small positive period-change parameter $\beta=0.044$ d Myr$^{-1}$ was found to be necessary to phase all data sets.
\item {\bf V33}: This star is now RR1, but used to be RR01, as first noted by C93. Like C93, we fail to detect signs of secondary pulsations which were visible in the data of \cite{vanagt59}.
\item {\bf V47}: We do not confirm the suggestion of W94 of this star being affected by Blazhko modulation; this may be due to the longer baseline of the W94 data. 

\end{itemize}

\section{Fourier decomposition of RR Lyrae star light curves}\label{sec:fourdec}

We performed a Fourier decomposition of the $V-$band light curves of RRL variables in order to derive several of their properties with well-established empirical relations. We can then use individual stars' properties to estimate the parameters of the host cluster. Fourier decomposition equates to fitting light curves with the Fourier series

\begin{equation}\label{eq:fourcos}
m(t) = A_0 + \sum_{k=1}^N A_{k} \cos \left[ \frac{2\pi k}{P}(t - E) + \phi_k \right] \, ,
\end{equation}

\noindent
where $m(t)$ is the magnitude at time $t$, $N$ is the number of harmonics used in the fit, $P$ is the period of the variable, $E$ is the epoch, and $A_k$ and $\phi_k$ are the amplitude and phase of the $k^{th}$ harmonic respectively. The epoch-independent Fourier parameters are then defined as

\begin{eqnarray}
R_{ij} &=& A_i / A_j \\
\phi_{ij} &=& j\phi_i - i\phi_j\, .
\end{eqnarray}

In each case we used the lowest number of harmonics that provided a good fit, to avoid over-fitting variations in the light curves due to noise. We also checked the sensitivity of the parameters we derived to the number of harmonics $N$, and we used only light curves with stable parameters, i.e. ones that showed little variation with $N$, to estimate cluster parameters. Double-mode pulsators and objects exhibiting signs of Blazhko modulation are also excluded from the following analysis.

The $A_k$ coefficients for the first four harmonics and the Fourier parameters $\phi_{21}, \phi_{31}$ and $\phi_{41}$ are given in \Tab{tab:fourier} for the light curves for which we could obtain a Fourier decomposition. We used the deviation parameter $D_m$, defined by \cite{jurcsik96}, as an estimate of the reliability of derived parameters, and use $D_m < 5$ \citep[e.g.][]{cacciari05} as a selection criterion. The value of $D_m$ for each of the successful Fourier decompositions is given in \Tab{tab:fourier}. 

We did not fit RR0 stars V9, V10, and V25, and RR1 star V5, because they all exhibit Blazhko-type modulation. Furthermore, we did not use the fits of V14, V28, V30 and V46 to derive star properties because those fits have a value of $D_m > 5$. V17 was not fitted either because the parameters varied significantly with the number of harmonics used in the fit, as well as showing signs of slow amplitude modulation (see \Sec{sec:individual}). This leaves us with 5 RR0 and 15 RR1 stars for which we derive individual properties in the next section. We also note that for V1, V2, V14, V23, and V30, we combined our data with that of W94 in order to derive fits, due to insufficient phase coverage of our data alone.

\begin{table*}
\begin{center}
  \begin{tabular}{ccccccccccc}

     \hline
    \#		&$A_0$	&$A_1$	&$A_2$ 	&$A_3$	&$A_4$	&$\phi_{21}$	&$\phi_{31}$	&$\phi_{41}$	&$N$	&$D_m$	 \\
  \hline
  RR0\\
  \hline  
V2& 15.749(2)&  0.281(2)&  0.086(2)&  0.057(2)&  0.042(2)&  4.031(17)&  8.039(23)&  6.073(32) & 9 & 4.34 \\ 
V12& 15.559(2)&  0.325(2)&  0.145(2)&  0.113(2)&  0.075(2)&  3.870(10)&  7.940(12)&  5.774(19) &11 & 3.11 \\
V14& 15.732(2)&  0.421(2)&  0.154(3)&  0.116(3)&  0.068(3)&  3.926(13)&  7.926(18)&  5.765(27) &10 & 5.04 \\
V22& 15.670(2)&  0.424(2)&  0.179(2)&  0.136(2)&  0.100(2)&  3.836(9)&  7.780(13)&  5.717(18) &11 & 2.36 \\
V23& 15.678(3)&  0.338(3)&  0.159(3)&  0.118(3)&  0.078(3)&  3.899(20)&  8.157(28)&  6.163(40) & 8 & 1.83 \\
V28& 15.751(2)&  0.371(2)&  0.178(2)&  0.146(2)&  0.076(2)&  3.842(8)&  8.363(11)&  6.064(17) & 8 & 6.44 \\
V30& 15.637(2)&  0.165(2)&  0.059(2)&  0.029(2)&  0.011(2)&  4.121(24)&  8.510(43)&  7.114(104) & 7 &17.59 \\
V35& 15.562(2)&  0.338(2)&  0.163(2)&  0.113(2)&  0.081(2)&  4.041(8)&  8.411(11)&  6.478(14) & 9 & 2.87 \\
V46& 15.637(2)&  0.209(2)&  0.091(2)&  0.055(2)&  0.026(3)&  4.090(25)&  8.684(35)&  7.187(47) &10 & 7.94 \\

\hline
RR1\\
\hline
V1& 15.700(2)&  0.296(3)&  0.059(3)&  0.023(3)&  0.016(3)&  4.558(39)&  2.581(93)&  1.461(125) & 4 &$-$ \\
V6& 15.691(1)&  0.251(1)&  0.040(2)&  0.021(1)&  0.009(1)&  4.619(15)&  2.962(31)&  2.102(73) & 4 &$-$ \\
V11& 15.706(1)&  0.265(2)&  0.052(2)&  0.025(2)&  0.007(2)&  4.554(13)&  2.521(25)&  1.837(78) & 4 \\
V13& 15.742(1)&  0.274(2)&  0.049(2)&  0.028(2)&  0.013(2)&  4.536(15)&  2.722(23)&  1.066(47) & 5 &$-$\\
V15& 15.684(1)&  0.241(1)&  0.041(2)&  0.019(2)&  0.007(2)&  4.636(15)&  3.074(29)&  1.883(73) & 5 &$-$\\
V16& 15.691(1)&  0.229(2)&  0.039(2)&  0.023(2)&  0.001(2)&  4.706(16)&  2.865(31)&  1.936(410) & 6 &$-$\\
V18& 15.722(1)&  0.249(2)&  0.046(2)&  0.027(2)&  0.015(2)&  4.566(15)&  2.873(24)&  1.930(48) & 4 &$-$\\
V20& 15.676(1)&  0.239(1)&  0.036(1)&  0.019(2)&  0.004(1)&  4.806(15)&  3.129(27)&  2.355(126) & 5 &$-$\\
V24& 15.682(2)&  0.247(2)&  0.048(2)&  0.022(2)&  0.011(2)&  4.637(17)&  2.894(35)&  1.978(68) & 6 &$-$\\
V33& 15.670(1)&  0.211(2)&  0.030(2)&  0.013(2)&  0.006(2)&  4.627(23)&  2.983(51)&  2.787(115) & 6 &$-$\\
V37& 15.641(1)&  0.227(2)&  0.036(2)&  0.011(2)&  0.006(2)&  4.392(21)&  3.013(59)&  2.440(98) & 4 &$-$\\
V38& 15.631(1)&  0.250(2)&  0.038(2)&  0.016(2)&  0.006(2)&  4.537(16)&  3.180(37)&  2.052(93) & 4 &$-$\\
V43& 15.713(1)&  0.261(2)&  0.048(2)&  0.021(2)&  0.011(2)&  4.439(17)&  2.408(35)&  1.378(62) & 5 &$-$\\
V44& 15.672(2)&  0.216(2)&  0.032(2)&  0.012(2)&  0.004(2)&  4.706(31)&  3.205(62)&  1.854(147) & 5 &$-$\\
V47& 15.629(2)&  0.248(2)&  0.048(2)&  0.016(2)&  0.007(2)&  4.645(21)&  2.799(62)&  2.676(123) & 5 &$-$\\

\hline \hline
  \end{tabular}
  \caption{Parameters from the Fourier decomposition. Numbers in parentheses are the 1-$\sigma$ uncertainties on the last decimal place. \label{tab:fourier}}
  \end{center}
\end{table*}

\begin{table*}
\begin{center}
  \begin{tabular}{cccccc}

     \hline
    \#		&$\fehzw$ &$\fehuves$ &$M_V$	&$\log(L/L_{\bigodot})$ 	&$\log\teff$	\\
 \hline
RR0 \\
 \hline
V2&  -1.85(2) &  -2.17(6) &  0.578(30) &  1.679(12) &3.804(2) \\
V12&  -2.09(1) &  -2.81(3) &  0.522(24) &  1.704(10) &3.800(2) \\
V22&  -2.04(1) &  -2.72(4) &  0.498(27) &  1.709(11) &3.807(2) \\
V23&  -2.04(3) &  -2.58(8) &  0.455(27) &  1.733(11) &3.797(2) \\
V35&  -1.97(1) &  -2.21(3) &  0.399(30) &  1.758(12) &3.795(2) \\

\hline
RR1 \\
\hline
V1&  -2.06(3) & $-$ &  0.521(23) &  1.658(10) &3.855(2) \\
V6&  -2.06(2) & $-$ &  0.533(21) &  1.655(9) &3.854(1) \\
V11&  -2.12(1) & $-$ &  0.548(21) &  1.650(9) &3.852(1) \\
V13&  -2.08(1) & $-$ &  0.524(21) &  1.658(9) &3.854(1) \\
V15&  -2.05(2) & $-$ &  0.538(21) &  1.653(9) &3.854(1) \\
V16&  -2.11(1) & $-$ &  0.550(21) &  1.650(9) &3.851(1) \\
V18&  -2.07(1) & $-$ &  0.511(21) &  1.664(9) &3.854(1) \\
V20&  -2.08(1) & $-$ &  0.529(21) &  1.658(9) &3.852(1) \\
V24&  -2.10(2) & $-$ &  0.517(21) &  1.663(9) &3.852(1) \\
V33&  -2.11(2) & $-$ &  0.526(21) &  1.661(9) &3.851(1) \\
V37&  -2.10(2) & $-$ &  0.540(21) &  1.654(9) &3.852(1) \\
V38&  -2.06(2) & $-$ &  0.536(21) &  1.655(9) &3.853(1) \\
V43&  -2.14(1) & $-$ &  0.531(21) &  1.658(9) &3.851(1) \\
V44&  -2.07(2) & $-$ &  0.533(21) &  1.656(9) &3.853(1) \\
V47&  -2.10(2) & $-$ &  0.536(21) &  1.655(9) &3.852(1) \\

\hline \hline
\end{tabular}

  \caption{Physical parameters for the RRL variables calculated using the Fourier decomposition parameters and the relations given in the text. Numbers in parentheses are the 1-$\sigma$ uncertainties on the last decimal place. The values of $\fehuves$ listed in this table for RR0 stars are derived using the relation of \cite{nemec13}. In addition, we note that \cite{smith83} found $\fehzw \approx -2.15$ for V2 and $\fehzw \approx -2.07$ for V25, using the $\Delta$S method (see \Sec{sec:rrmet}). Note that errors quoted for $\log\teff$ are statistical only and not systematic; we could not calculate systematic errors for $\log\teff$ as no errors are given on the empirical coefficients in Eqs. (\ref{eq:teffrr0}) and (\ref{eq:teffrr1}). Errors on $M_V$ and $\log(L/L_{\bigodot})$ for RR1 stars are dominated by the error on the zero-point $K_0$ in \Eq{eq:mvrr1}. \label{tab:starpar}}
  \end{center}
\end{table*}

\subsection{Metallicity}\label{sec:rrmet}

In this section we derive metallicities for each variable for which a good Fourier decomposition, according to our selection criteria discussed above, could be obtained. To do this, we use empirical relations from the literature. For RR0 stars, we used the relation of \cite{jurcsik96}, which expresses [Fe/H] as a function of the period and of the Fourier parameter $\phi^s_{31}$; the $s$ superscript denotes that \cite{jurcsik96} derived their relations using \textit{sine} series, whereas we fit cosine Fourier series (Eq. \ref{eq:fourcos}). The Fourier parameters can be easily converted using the equation

\begin{equation}\label{eq:phis}
\phi^s_{ij} = \phi_{ij} - (i-j)\, \frac{\pi}{2}\, .
\end{equation}

[Fe/H] can then be expressed as

\begin{equation}\label{eq:metrr0}
\mathrm{[Fe/H]_J} = -5.038 - 5.394\, P + 1.345\, \phi^s_{31}\, ,
\end{equation}

\noindent
where the subscript J denotes a non-calibrated metallicity, and the period $P$ is in days. This value can be transformed to the metallicity scale of \cite{zinn84} (hereafter ZW) via the simple relation from \cite{jurcsik95},

\begin{equation}\label{eq:zw}
\fehzw = \frac{\rm [Fe/H]_J - 0.88}{1.431}\, .
\end{equation}

\cite{kovacs02} investigated the validity of this relation, and found that \Eq{eq:metrr0} yields metallicity values that are too large by $\sim 0.2$ dex for metal-poor clusters. This was supported by the findings of \cite{gratton04} and \cite{difabrizio05}, who compared metallicity values for RRL stars in the Large Magellanic Cloud (LMC) using both spectroscopy and Fourier decomposition. Therefore we include a shift of -0.20 dex (on the $\mathrm{[Fe/H]_J}$ scale) or -0.14 dex (on the ZW scale) to the metallicity values derived for RR0 stars using \Eq{eq:metrr0}. 

More recently, \cite{nemec13} derived a relation for the metallicity of RR0 stars, using observations of RRL taken with the \textit{Kepler} space telescope,

\begin{equation}\label{eq:fehnemec}
\fehuves = b_{0} + b_{1}P + b_{2}\phi^s_{31} + b_{3}\phi^s_{31}P + b_{4}\left(\phi^s_{31}\right)^2\, ,
\end{equation}

\noindent
where $\fehuves$ is the metallicity on the widely-used scale of \cite{carretta09}, and the constant coefficients were determined by \cite{nemec13} as $b_0=-8.65 \pm 4.64$, $b_1=-40.12 \pm 5.18$, $b_3=6.27 \pm 0.96$ and $b_4=-0.72 \pm -0.72 \pm 0.17$. The scale of \cite{carretta09} was derived using spectra of red giant branch (RGB) stars obtained using GIRAFFE and UVES. ZW metallicity values can be transformed to that scale (hereafter referred to as the UVES scale) using

\begin{equation}\label{eq:fehuves}
\fehuves = -0.413 + 0.130\,\fehzw - 0.356\,\fehzw^2\, .
\end{equation}

In the discussion that follows, we considered RR0 metallicity values obtained both with the scale of \cite{jurcsik96} and that of \cite{nemec13}\footnote{However, we calculated errors on metallicity values derived from the relation of \cite{nemec13} ignoring the errors on the coefficients $b_i$, as those would lead to very large errors.}. 

For the RR1 variables, we used the relation of \cite{morgan07} to derive metallicity values; this relates [Fe/H], $P$ and $\phi_{31}$ as

\begin{eqnarray}\label{eq:metrr1}
\fehzw &=& 2.424 - 30.075\, P +  52.466\, P^2 \\ \nonumber
&&+ 0.982\, \phi_{31} + 0.131 \phi_{31}^2 - 4.198\, \phi_{31}P \, .
\end{eqnarray}

Metallicity values calculated using Eqs. (\ref{eq:metrr0}) \& (\ref{eq:zw}) and (\ref{eq:metrr1}) are listed in \Tab{tab:starpar}.  

We note that the metallicities of only two RRL stars in M68 have previously been measured in the literature. \cite{smith83} measured the spectroscopic metallicity index $\Delta S$ for the RR0-type variables V2 and V25 as 11.0 and 10.5, respectively, which yield metallicities $\fehzw \approx -2.15$ and $-2.07$, respectively, when converted to the ZW scale via the following relation from \cite{suntzeff91}:

\begin{equation}\label{eq:deltas}
\fehzw = -0.158 \Delta {\rm S} - 0.408\, .
\end{equation}

We include these metallicity measurements as a footnote to \Tab{tab:starpar} for the sake of completeness.

\subsection{Effective Temperature}

\cite{jurcsik98} derived empirical relations to calculate the effective temperature of fundamental-mode RRL stars, relating the $(V-K)_0$ colour to $P$ and to several of the Fourier coefficients and parameters:

\begin{eqnarray}
(V-K)_0 &=&  1.585 + 1.257\, P - 0.273\, A_1 - 0.234\, \phi^s_{31} \\ \nonumber
&&+ 0.062\, \phi^s_{41} \\\label{eq:teffrr0}
\log\,\teff &=& 3.9291 - 0.1112\,(V-K)_0 \\ \nonumber
&& - 0.0032 \, \mathrm{[Fe/H]_J} \, .
\end{eqnarray}

\cite{simon93} used theoretical model to derive a corresponding relation for RR1 stars, 

\begin{equation}\label{eq:teffrr1}
\log\,\teff = 3.7746 - 0.1452 \log\, P + 0.0056\, \phi_{31} \, .
\end{equation}

We use those relations to derive $\log\teff$ for all of our RR0 and RR1 stars, and give the resulting values in \Tab{tab:starpar}. As we mentioned in previous analyses \citep[e.g.][]{arellano08b}, there are some important caveats to consider when estimating temperatures with Eqs. (\ref{eq:teffrr0}) and (\ref{eq:teffrr1}). The values of $\log\teff$ for RR0 and RR1 stars are on different absolute scales \citep[e.g.][]{cacciari05}, and temperatures derived using these relations show systematic deviations from those predicted by \cite{castelli99} in their evolutionary models, or by the temperature scales of \cite{sekiguchi00}. However, we use these in order to have a comparable approach to that taken in our previous studies of clusters.

\subsection{Absolute Magnitude}\label{sec:rrmag}

We use the empirical relations of \cite{kovacs01} to derive $V$-band absolute magnitudes for the RR0 variables,

\begin{equation}\label{eq:mvrr0}
M_V = -1.876 \log P - 1.158\, A_1 + 0.821 \, A_3 + K_0 \, ,
\end{equation}

\noindent
where $K_0$ is the zero-point of the relation. Using the absolute magnitude of the star RR Lyrae of $M_V=0.61 \pm 0.10$ mag derived by \cite{benedict02}, and a Fourier decomposition of its light curve, \cite{kinman02} determined a zero-point for \Eq{eq:mvrr0} of $K_0=0.43$ mag. Here, however we adopt a slightly different value of $K_0=0.41 \pm 0.02$ mag, as in several of our previous studies \citep[e.g.][]{arellano10}, in order to maintain consistency with a true distance modulus of $\mu_0=18.5$ mag for the LMC \citep{freedman01}. A nominal error of 0.02 mag on $K_0$ was adopted in the absence of uncertainties on $K_0$ in the literature.

For RR1 variables, we use the relation of \cite{kovacs98},

\begin{equation}\label{eq:mvrr1}
M_V = -0.961\, P - 0.044\, \phi^s_{21} - 4.447 \, A_4 + K_1 \, ,
\end{equation}

\noindent
where we adopted the zero-point value of $K_1=1.061 \pm 0.020$ mag \citep{cacciari05}, with the same justification as for our choice of $K_0$.

We also converted the magnitudes we obtained to luminosities using

\begin{equation}\label{eq:logl}
\log\left( L/L_{\bigodot} \right) = -0.4\, \left[M_V + B_C(\teff) - M_{\rm bol, \bigodot }\right] \, ,
\end{equation}

\noindent
where $M_{\rm bol, \bigodot}$ is the bolometric magnitude of the Sun, $M_{\rm bol, \bigodot }=4.75$ mag, and $B_C(\teff)$ is a bolometric correction which we estimate by interpolating from the values of \cite{montegriffo98} for metal-poor stars, and using the value of $\log\teff$ we derived in the previous section. Values of $M_V$ and $\log(L/L_{\bigodot})$ for the RR0 and RR1 variables are listed in \Tab{tab:starpar}. Using our average values of $M_V$, in conjunction with the average values of $\fehzw$ (\Sec{sec:rrmet}), we find a good agreement with the $M_V - \fehzw$ relation derived in the literature \cite[e.g.][see Fig.~9 of that paper]{kains12b}.

\subsection{Masses}\label{sec:rrmasses}

Empirical relations also exist to derive masses of RRL stars from the Fourier parameters, although as noted by \cite{simon93}, such relations are better suited to derive average values for RRL stars in clusters. We therefore use mean parameters to derive mean masses for our RRL stars. For RR0 stars, we use the relation of \cite{vanalbada71},

\begin{equation}\label{eq:massrr0}
\begin{split}
\log(\mathcal{M}/\masssun)=16.907 \, -\, 1.47\log(P)\, +\, 1.24 \log(L/\lsun)\,\\
-\, 5.12 \log\teff \, ,
\end{split}
\end{equation}

\noindent
where we use the symbold $\mathcal{M}$ to denote masses in order to avoid confusion with absolute magnitudes elsewhere in the text. Using this, we find an average mass $\langle {\rm \mathcal{M}}_{\rm RR0}/\masssun\rangle = 0.74 \pm 0.14$. 

For RR1 stars, we used the relation derived by \cite{simon93} from hydrodynamic pulsation models,

\begin{equation}\label{eq:massrr1}
\log(\mathcal{M}/\masssun)= 0.52 \log P - 0.11 \phi_{31} + 0.39\, ,
\end{equation}

\noindent
which yield a mean mass $\langle \mathcal{M}_{\rm RR1}/\masssun\rangle = 0.70 \pm 0.05$. This is lower than the value of $\langle\mathcal{M}/\masssun\rangle=0.79$ found by \cite{simon93} for M68, but confirms the value of $\langle\mathcal{M}/\masssun\rangle=0.70 \pm 0.01$ found by W94 using the same method.

\section{Double-mode pulsations}\label{sec:doublemode}

\subsection{RR Lyrae}\label{sec:rrd}

M68 contains 12 identified double-mode RRL (RR01) stars. RR01 stars are of particular interest, because, given metallicity measurements, the double-mode pulsation provides us with a unique opportunity to measure the mass of these objects without assuming a stellar evolution model, and to study the mass-metallicity distribution of field and cluster RRL stars \citep[e.g.][]{bragaglia01}. ``Canonical" RR01 have first-overtone-to-fundamental pulsation period ratios ranging from $\sim 0.74$ to 0.75 \citep[e.g.][]{soszynski09} for radial pulsations. Other period ratios can be interpreted as being evidence for non-radial pulsation, or for higher-overtone secondary pulsation, with studies reporting a period ratio of $\sim$ 0.58-0.59 between the second overtone and the fundamental radial pulsation \citep[e.g.][]{poretti10}.

We analysed the light curves of RR01 stars in our images using the time-series analysis software {\tt Period04} \citep{lenz05}. In \Tab{tab:rr2}, we list the periods we found for each variable, as well as the period ratio. We found ratios within the range of canonical value for all stars except for V45, for which we did not detect a second pulsation period, unlike W94.

\begin{table*}
\begin{center}
  \begin{tabular}{ccccccc}

  \hline
    \#		 &$P_1 [d]$	&$A_1$ 	&$P_0 [d]$	&$A_0$	  &$P_1/P_0$ 	&$\mathcal{M}/\masssun$\\
  \hline  

V3		&0.3907346	&0.206	&0.523746	&0.101	&0.7460	&$0.789 \pm 0.119$	\\
V4		&0.3962175	&0.205	&0.530818	&0.111	&0.7464	&$0.787 \pm 0.118$	\\
V7		&0.3879608	&0.201	&0.520186	&0.107	&0.7458	&$0.789 \pm 0.119$	\\
V8		&0.3904076	&0.219	&0.522230	&0.046	&0.7476	&$0.784 \pm 0.118$	\\
V19		&0.3916309	&0.202	&0.525259	&0.061	&0.7456	&$0.790 \pm 0.119$	\\
V21		&0.4071121	&0.221	&0.545757	&0.138	&0.7460	&$0.789 \pm 0.119$	\\
V26		&0.4070332	&0.200	&0.546782	&0.147	&0.7444	&$0.793 \pm 0.120$	\\
V29		&0.3952413	&0.236	&0.530525	&0.023	&0.7450 	&$0.792 \pm 0.119$	\\
V31		&0.3996599	&0.200	&0.535115	&0.121	&0.7469	&$0.786 \pm 0.118$	\\
V34		&0.4001371 	&0.187	&0.537097	&0.050	&0.7450	&$0.792 \pm 0.119$	\\
V36		&0.415346   	&0.203	&0.557511	&0.106	&0.7450	&$0.792 \pm 0.119$	\\
V45		&0.3908187	&0.224	&$-$	&$-$	&$-$	&$-$\\
   
\hline \hline
  \end{tabular}
  \caption{Fundamental and first-overtone pulsation periods and the period ratio for the double-mode RRL detected in M68; also listed are semi-amplitudes for each pulsation mode. \label{tab:rr2}}
  \end{center}
\end{table*}

In the past, theoretical model tracks and the ``Petersen" diagram ($P_1/P_0$ vs. $P_0$, \citealt{petersen73}) have been used to estimate the masses of RR01 pulsators. Using this approach, W94 estimated average RR01 masses of $\sim0.77\,\masssun$ and $\sim0.79\,\masssun$, using the models of \cite{cox91} and \cite{kovacs91} respectively. Here we use the new relation of Marconi et al. (2015, submitted), who used new non-linear convective hydrodynamic models to derive a relation linking the stellar mass with the period ratio and the metallicity, through

\begin{equation}\label{eq:mrrd}
\begin{split}
\log\,(\mathcal{M}/\masssun) = (-0.85 \pm 0.05) - (2.8 \pm 0.3)\log\,(P_1/P_0)\\
 - (0.097 \pm 0.003)\log\,Z  \, .
\end{split}
\end{equation}

In order to convert metallicities from $\fehzw$ to $Z$, we used the relation of \cite{salaris93}, 

\begin{equation}\label{eq:zfeh}
\log\,Z = \fehzw - 1.7 + \log(0.638\, f + 0.362) \, ,
\end{equation}

\noindent
where $f$ is the $\alpha$-element abundance, for which we adopt a value of 0.3 for M68 \citep[e.g.][]{carney96}. We also adopted a value of $\fehzw=\clustermet$ corresponding to the mean metallicity of the RRL stars listed in \Tab{tab:starpar}. Using Eqs. (\ref{eq:mrrd}) and (\ref{eq:zfeh}), we find the masses given in \Tab{tab:rr2}; we find a mean mass for RR01 stars of $\langle\mathcal{M}/\masssun\rangle = 0.789 \pm 0.003 \mathrm{~(statistical)} \pm 0.119 \mathrm{~(systematic)}$, in excellent agreement with the findings of W94.

\subsection{SX Phoenicis}\label{sec:sxp_double}

We also carried out a search for double-mode pulsation in the light curves of the SX Phe stars in our sample. For V50, we find a second pulsation period $P_1=0.051745$ d in addition to the fundamental period $P_0=0.065820$ d, giving a ratio $P_1/P_0 \sim 0.786$, within the range of expected values for a metal-poor double-radial-mode SX Phe star \citep{santolamazza01}. This is confirmed by the location of that secondary period on the $\log P - V_0$ diagram, which is further discussed in \Sec{sec:sxp_dist}.

\section{Cluster properties from the variable stars}\label{sec:clusterprop}

\subsection{Oosterhoff Type}

In order to determine the Oosterhoff type of this cluster, we calculate the mean periods of the fundamental-mode RRL, as well as the ratio of RR1 to RR0 stars. We find $\langle P_{\rm RR0} \rangle=0.63 \pm 0.07$ d and $\langle P_{\rm RR1} \rangle= 0.37 \pm 0.02$ d. RR1 stars make up 54\% of the single-mode RRL stars. Although $\langle P_{\rm RR0} \rangle$ is somewhat lower than the usual canonical value of $\sim 0.68$, these values, as well as the very low metallicity, point to M68 being an Oosterhoff II cluster, in agreement with previous classifications \citep[e.g.][]{lee99}. As mentioned earlier, the fact that M68 is a clear Oosterhoff II type is in disagreement with studies concluding that M68 could be of extragalactic origin \citep[e.g.][]{yoon02}, since globular clusters in satellite dSph galaxies of the Milky Way fall mostly within the gap between Oosterhoff types I and II. 

\begin{figure}
  \centering
  \includegraphics[width=8cm, angle=0]{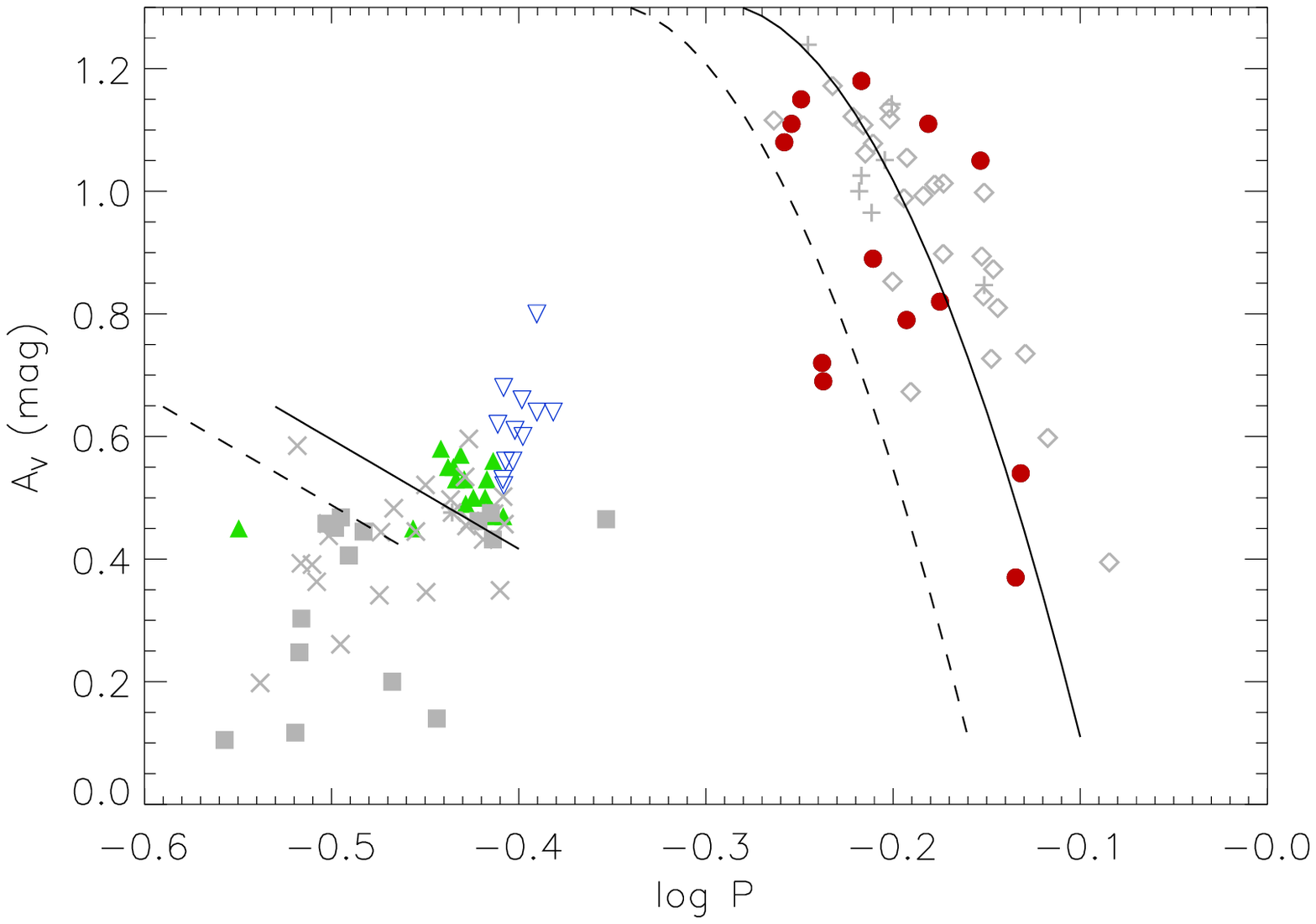}
  \includegraphics[width=8cm, angle=0]{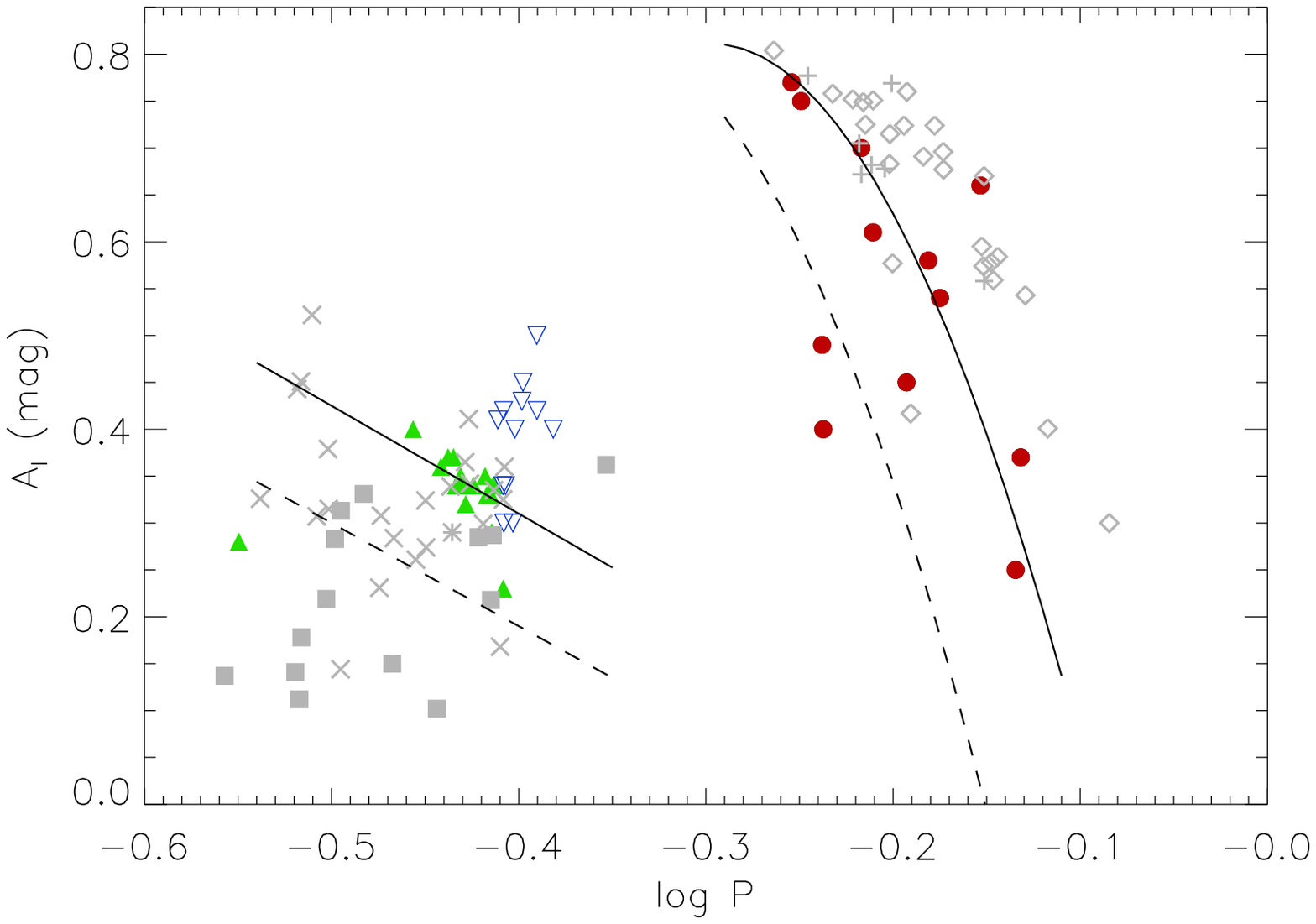}

  \caption{Bailey diagram of the light curve amplitude $A$ against $\log P$ for RRL stars in M68, in the $V$ (top) and $I$ (bottom) bands. The locations of confirmed RRL stars are marked in red with filled circles (RR0), green filled triangles (RR1) and blue open inverted triangles (RR01). For the RR01 stars, the amplitude is that of the dominant mode, which is the first overtone in all cases here. For the $V-$band diagram, we also plot the tracks derived by \cite{zorotovic10} for evolved stars (solid lines), found by shifting the relations obtained by fitting data for the Oosterhoff type I cluster M3 (dashed lines) by 0.06 in $\log P$. For the $I-$band diagram, we plot the rescaling of those relations by \cite{kunder13} for Oosterhoff type I (dashed lines) and type II (solid lines) clusters. On both diagrams we also plot the RRL stars in Oosterhoff type II clusters NGC~5024 \citep{arellano11} and M9 \citep{arellano13} in light grey. For those, RR0 and RR1 stars are plotted as open diamonds and filled squares respectively, with RRL stars exhibiting Blazhko modulation plotted as + signs (RR0) and crosses (RR1). One double-mode star is also plotted as an asterisk. \label{fig:bailey}}

\end{figure}

Plotting the RRL stars on a ``Bailey" diagram ($\log P - A$, where $A$ denotes the amplitude of the RRL light curve; \Fig{fig:bailey}) also allows us to confirm the Oostheroff II classification, by comparing the location of our RRL stars to the analytical tracks derived by \cite{zorotovic10} by fitting the loci of normal and evolved stars of \cite{cacciari05}. \Fig{fig:bailey} shows that the locations of our RRL stars are in good agreement with these tracks (for the $V$ band) and those of \cite{kunder13} (who rescaled those tracks for the $I$ band). However, it is obvious from the $I-$band plot (bottom panel of Fig. \ref{fig:bailey}) that the RRL loci for M68 are different from those of Oosterhoff type II clusters NGC~5024 \citep{arellano11} and M9 \citep{arellano13}, but are in agreement with the loci for NGC~2808 \citep{kunder13}.

\subsection{Cluster metallicity}\label{sec:clustermet}

Although \cite{clement01}, found a correlation between the metallicity of a globular cluster and the mean period of its RR0 stars, $\langle P_{\rm RR0} \rangle$, they noted that this relation does not hold when $\langle P_{\rm RR0} \rangle$ is larger than 0.6 d, i.e. for Oosterhoff type II clusters. Instead, we estimate the metallicity of M68 by calculating the average of the metallicities we derived for RRL stars in this cluster. As in our previous papers \citep[e.g.][]{kains13b}, we assume that there is no systematic offset between the metallicities derived for RR0 and RR1 stars. Using this method, we find a mean metallicity of $\fehzw=\clustermet$ (using \Eq{eq:zw} to derive RR0 metallicities) and $\fehzw=-2.13 \pm 0.11$ (using \Eq{eq:fehnemec}), corresponding to $\fehuves=\clustermetuves$ and $\fehuves=-2.30 \pm 0.17$ respectively. Both values are in excellent agreement with the value of \cite{carretta09b}, who found $\fehuves=-2.27\pm0.01\pm0.07$ (statistical and systematic errors respectively), and with most values in the literature, listed in \Tab{tab:lit_met}. In the rest of this paper, we will use $\fehzw=\clustermet$ as the cluster metallicity, due to the larger scatter in RR0 metallicities derived using \Eq{eq:fehnemec}, and the problem with uncertainties in the relation of \cite{nemec13} already mentioned in \Sec{sec:rrmet}.


\begin{table*}
\begin{center}
  \begin{tabular}{cccc}

     \hline
    Reference		&$\fehzw$	&$\fehuves$	&Method	\\
 \hline
This work			&$\metrrzw$		&$\metrruves$	& Fourier decomposition of RRL light curves\\
This work			&$\metmvfeh$		&$\metmvfehuves$	& $M_V-$[Fe/H] relation\\
\cite{carretta09b}	&-2.10 $\pm$ 0.04	&$-2.27 \pm 0.07$	& UVES spectroscopy of red giants\\
\cite{carretta09a}	&-2.08 $\pm$ 0.04	&$-2.23 \pm 0.07$	& FLAMES/GIRAFFE spectra of red giants\\
\cite{rutledge97}	&-2.11 $\pm$ 0.03	&$-2.27 \pm 0.05$	&CaII triplet \\
\cite{harris96}		&-2.23 	&-2.47	&Globular cluster catalogue \\
\cite{suntzeff91}	&-2.09 	&-2.24	&Spectroscopy of RR Lyrae/ $\Delta$S index \\
\cite{gratton89}		&-1.92 $\pm$ 0.06	&$-1.97 \pm 0.09$	&High-resolution spectroscopy \\
\cite{minniti93}		&-2.17 $\pm$ 0.05	&$-2.37 \pm 0.08$	&FeI and FeII spectral lines\\
\cite{zinn84}		&-2.09 $\pm$ 0.11	&$-2.24 \pm 0.18$	&$Q_{39}$ index \\
\cite{zinn80}		&-2.19 $\pm$ 0.06	&$-2.41 \pm 0.10$	&$Q_{39}$ index \\

\hline \hline
  \end{tabular}

  \caption{Different metallicity estimates for M68 in the literature. Values were converted using \Eq{eq:fehuves} where necessary. \label{tab:lit_met}}
  \end{center}
\end{table*}

\subsection{Reddening}\label{sec:rrl_red}

Individual RR0 light curves can be used to estimate the reddening to each star, using their colour near the point of minimum brightness. The method was first proposed by \cite{sturch66}, and further developed in several studies, most recently by \cite{guldenschuh05} and \cite{kunder10}. \cite{guldenschuh05} find that the intrinsic $(V-I)_0$ colour of RR0 stars between phases 0.5 and 0.8 is $(V-I)_0^{\phi[0.5-0.8]}=0.58 \pm 0.02$ mag. By calculating the $(V-I)$ colour for each of our light curves between those phases and comparing it to the value of \cite{guldenschuh05}, we can therefore calculate values of $E(V-I)$, which we can then convert through $E(B-V)=E(V-I)/1.616$ \citep[e.g.][]{arellano13}. Stars showing Blazhko modulation were excluded from this calculation, as were stars with poor $I-$band light curves, leading to reddening estimates for V2, V12, V14, V22, V23, V35 and V46. We found a mean reddening of $0.05 \pm 0.05$ mag, within the range of values published (see \Sec{sec:rrl_dist}).

\subsection{Distance}

\subsubsection{Using the RRL stars}\label{sec:rrl_dist}

The Fourier decomposition of RRL stars can also be used to derive a distance modulus to the cluster. The Fourier fit parameter $A_0$ corresponds to the intensity-weighted mean $V$ magnitude of the light curve, and since we also derived absolute magnitudes for each star with a good Fourier decomposition, this can be used to calculate the distance modulus to each star.

For RR0 stars, the mean value of $A_0$ is $15.64 \pm 0.08$ mag, while the mean absolute value is $\langle M_{V} \rangle=0.49 \pm 0.07$ mag. From this, we find a distance modulus $\mu=15.15 \pm 0.11$ mag. For RR1 stars, we find $\langle A_0 \rangle=15.68 \pm 0.03$ mag and $\langle M_{V} \rangle=0.53 \pm 0.01$ mag, yielding $\mu=15.15 \pm 0.03$ mag.

Many values of $E(B-V)$ have been published for M68, ranging from 0.01 mag \citep[e.g.][]{racine73} to 0.07 mag \citep[e.g.][]{bica83}. Recent studies have tended to use $E(B-V)=0.05$ mag, which is the value listed in the catalogue of \cite{harris96}. \cite{piersimoni02} derived a value of $E(B-V)=0.05 \pm 0.01$ from $BVI$ photometry of RRL stars and empirical relations, and this coincides with the value we derived in \Sec{sec:rrl_red}, although uncertainties on our estimate are larger. We therefore adopt the value of \cite{piersimoni02}, and a value of $R_V=3.1$ for the Milky Way. From this we find true distance moduli of $\mu_0=15.00 \pm 0.11$ mag and $\mu_0=15.00 \pm 0.05$ mag from RR0 and RR1 stars, respectively. These values correspond to physical distances of $10.00 \pm 0.49$ kpc and $9.99 \pm 0.21$ kpc, respectively, and are consistent with values reported in the literature, examples of which we list in \Tab{tab:lit_dist}.

\subsubsection{Using SX Phoenicis stars}\label{sec:sxp_dist}

We can derive distances to SX Phe stars thanks to empirical period-luminosity ($P-L$) relations \citep[e.g.][]{jeon03}. Here we calculate distances to the SX Phe stars in our images (see Table \ref{tab:variables}) using the $P-L$ relation of \cite{cohen12},

\begin{equation}\label{eq:sxpl}
M_V	= -(1.640 \pm 0.110) - (3.389 \pm 0.090) \log P_f\, ,
\end{equation}

\noindent
where $P_f$ is the fundamental-mode pulsation period. This provides us with another independent estimate of the distance to M68. \cite{arellano11} derived their own $P-L$ relation, using only SX Phe stars in metal-poor globular clusters, and have shown a good match of their relation to the locus of SX Phe in other clusters \citep[e.g.][]{arellano14b, arellano14}. 

\begin{figure}
  \centering
  \includegraphics[width=9cm, angle=0]{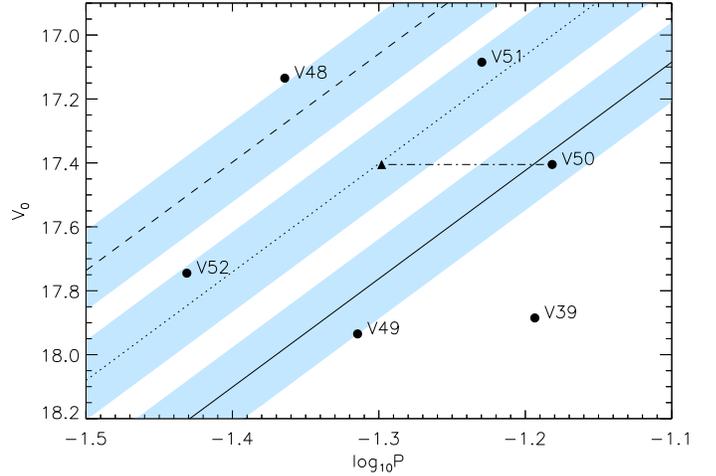}
  \caption{$\log P - V_0$ plot of the SX Phe stars in M68, with the $P-L$ relation of \cite{cohen12} for fundamental, first-overtone, and second-overtone pulsators as solid, dotted, and dashed lines respectively. Shaded areas show 1-$\sigma$ boundaries for each case. The relations have been shifted to a distance of $\drrc$ kpc (see \Sec{sec:rrl_dist}), and the SX Phe magnitudes have been dereddened using our adopted value of $E(B-V)=0.05 \pm 0.01$ mag. The location of the black filled circles in this plot correspond to the periods listed in \Tab{tab:variables}, and for V50 the first-overtone period is shown as a triangle, with a dash-dot line connecting the two periods. \label{fig:sxpl}}
\end{figure}

Assuming that the values given in \Tab{tab:variables} are the fundamental periods and using them with \Eq{eq:sxpl} as well as the relation of \cite{arellano11}, the distances derived for V39, V48, V51 and V52 disagree with the values we found using RRL stars, and with most values in the literature (see  \Tab{tab:lit_dist}). In \Fig{fig:sxpl}, we plot $\log P$ against dereddened magnitudes $V_0$ (using our adopted value $E(B-V)=0.05 \pm 0.01$ mag), as well as the relations of \cite{cohen12} for fundamental, first and second overtone SX Phe pulsators, shifted to a distance of $\drrc$ kpc, the cluster distance we derived in \Sec{sec:rrl_dist} using RR1 stars. From this, it is clear that V49 pulsates in the fundamental mode, and that the periods of the double-mode pulsator V50 (see \Sec{sec:sxp_double}) do correspond to fundamental and first-overtone periods. \Fig{fig:sxpl} also suggests that V51 and V52 are first-overtone pulsators, and that V48 is a second-overtone pulsator; W94 had suggested that V48 might be a first-overtone pulsator, noting however that it would be bright for its period. \Fig{fig:sxpl} also suggest that V39 might not be a cluster member, with the large distance hinting that it may be a background object. 

For the overtone pulsators, we can ``fundamentalise" the periods to then use \Eq{eq:sxpl} in order to derive empirical values of $M_V$, $\mu_0$, and the distance. To do this, we use the canonical first- and second-overtone to fundamental  period ratios, respectively $P_1/P_f \sim 0.783$ \citep[e.g.][]{jeon04b, santolamazza01, gilliland98} and $P_2/P_f \sim 0.62$ \citep{santolamazza01}. Using these values to calculate $P_f$, we find the corrected absolute magnitudes and distances listed in \Tab{tab:sx_prop_correct}. 

We note that the discrepant distance estimates could alternatively be due to V48 being a foreground object, while V51 clearly suffers from blending from two nearby stars of comparable brightness, meaning that its reference flux could be over-estimated by as much as a factor of 2. If this is the case, V51 would fall on the track for fundamental pulsators on \Fig{fig:sxpl}. Higher-resolution data could help determine which explanation is correct for V51, while radial velocity measurements would help address the cluster membership of V39 and V48.

Using the corrected values, and excluding V39, we find an average distance modulus of $\mu_0=14.97 \pm 0.11$ mag, corresponding to a physical distance of $9.84 \pm 0.50$ kpc, in good agreement with the values we found using RRL stars in the previous section.

\begin{table*}
\begin{center}
  \begin{tabular}{cccc}

\hline
    Reference		&$\mu_0$ [mag]		&Distance [kpc]		&Method	\\
\hline

This work			&$15.00 \pm 0.11$ 	& $10.00 \pm 0.49$ 	&Fourier decomposition of RR0 light curves \\
This work			&$15.00 \pm 0.05$	&$9.99 \pm 0.21$	&Fourier decomposition of RR1 light curves \\
This work			& $14.97 \pm 0.11$ 	& $9.84 \pm 0.50$ 	&SX Phe $P-L$ relation \\
This work			&$15.00 \pm 0.07$ 	& $10.00 \pm 0.30$ 	&$M_V-$[Fe/H] relation for cluster RRL stars \\
\cite{rosenberg99}	&15.02		&10.07  	&CMD analysis \\
\cite{caloi97}		&15.15		&10.72  	&CMD isochrone fitting \\
\cite{brocato97}		&15.16 $\pm$ 0.10	& $10.76 \pm 0.50$   	&CMD isochrone fitting \\
\cite{gratton97}		&15.24 $\pm$ 0.08	& $11.17 \pm 0.41$  	&CMD analysis \\
\cite{harris96}		&15.06	&10.3  	&Globular cluster catalogue \\
\cite{straniero91}	& 14.99 $\pm$ 0.03	&9.94 $\pm$ 0.14 	&CMD isochrone fitting \\
\cite{mcclure87}	& 15.03 $\pm$ 0.15	&10.14 $\pm$ 0.70	&CMD analysis  \\
\cite{alcaino77}		& 14.97	& 9.8 	&Magnitude of the HB  \\
\cite{harris75}		& 14.91	& 9.6 	&Mean magnitude of RRL stars  \\

\hline \hline
\end{tabular}

  \caption{Distance modulus and physical distance estimates for M68 in the literature. \label{tab:lit_dist}}
  \end{center}
\end{table*}

\begin{table}
\begin{center}
  \begin{tabular}{cccc}

\hline
    Reference		&$M_V [mag]$	&$\mu_0$ [mag]		&Distance [kpc]\\
\hline

V39			&2.40$\pm$0.11 	& 15.48$\pm$0.25 	&12.48$\pm$1.42  \\
V48			&2.28$\pm$0.11 	& 14.86$\pm$0.13 	&9.35$\pm$0.55  \\
V49			&2.81$\pm$0.11	& 15.12$\pm$0.21 	&10.57$\pm$1.03  \\
V50			&2.36$\pm$0.11 	& 14.04$\pm$0.15 	&10.19$\pm$0.73  \\
V51			&2.17$\pm$0.11 	& 14.92$\pm$0.14 	&9.63$\pm$0.64  \\
V52			&2.85$\pm$0.11 	& 14.89$\pm$0.14 	&9.53$\pm$0.60  \\

\hline \hline
\end{tabular}

  \caption{Absolute magnitudes, true distance moduli and physical distance for the SX Phe stars in M68, derived using the $P-L$ relation of \cite{cohen12} after ``fundamentalisation" of the pulsation period. \label{tab:sx_prop_correct}}
  \end{center}
\end{table}

\subsection{Distance and metallicity from the $M_V-${\rm[Fe/H]} relation}

Many studies in the literature have compiled magnitude and metallicity measurements from globular cluster observations to fit a linear relation between the metallicity of a cluster and the mean magnitude of its RRL stars, $M_V=\alpha{\rm [Fe/H]}\, +\beta$. Here we use the relation derived by \cite{kains12b}, with coefficients $\alpha=0.16 \pm 0.01$ and $\beta=0.85 \pm 0.02$, for a metallicity given on the ZW scale. Using the metallicity value of $\fehzw=\clustermet$ derived in \Sec{sec:clustermet}, we find a mean absolute magnitude of $0.52 \pm 0.03$ mag, in good agreement with the value found in \Sec{sec:rrl_dist}. We can now use this to estimate the distance to M68, in a similar way to what was done in that section. Since the $M_V-$[Fe/H] does not distinguish between RR0 and RR1, we use average value over all RRL types in this calculation.

We find a mean magnitude for all RRL stars of $\langle A_0 \rangle = 15.68 \pm 0.05$ mag, yielding a distance modulus of $\mu = 15.16 \pm 0.06$ mag. With $E(B-V)=0.05 \pm 0.01$ mag and $R_V=3.1$ as in \Sec{sec:rrl_dist}, this gives a true distance modulus of $15.00 \pm 0.07$ mag, corresponding to a physical distance of $10.00 \pm 0.30$ kpc.

Conversely, we can also estimate the mean metallicity of the cluster using the $M_V-$[Fe/H] relation, using the values found from Eqs. (\ref{eq:mvrr0}) and (\ref{eq:mvrr1}). We find a mean value for all RRL stars with a good Fourier decomposition of $M_V=0.52 \pm 0.04$, yielding $\fehzw=\metmvfeh$, in agreement with our values in \Sec{sec:clustermet}, but with a much larger error bar.

\subsection{Age}

Although our CMD does not allow us to derive a precise age, we use our data to check consistency of our CMD with age estimates of M68 from the literature (\Tab{tab:lit_age}), by overplotting the isochrones of \cite{dotter08}, using our estimate of the cluster metallicity of $\fehzw=\clustermet$. We interpolated those isochrones to match the alpha-element abundance for this cluster of [$\alpha$/Fe] of $\sim 0.3$ \citep[e.g.][]{carney96}. We found that our CMD is consistent with an age of $13.0 \pm 0.50$ Gyr, in agreement with the most recent values in the literature.

\begin{table*}
\begin{center}
  \begin{tabular}{cccc}

     \hline
    Reference	&Age [Gyr]		&Method	\\
 \hline
\cite{vandenberg13}	&12.00$\pm$0.25   	& CMD isochrone fitting \\
\cite{rakos05}		&11.2   	& Str\"{o}mgren photometry \\
\cite{salaris02}		&11.2 $\pm$ 0.9   	&CMD analysis  \\
\cite{caloi97}		&12 $\pm$ 2   	&CMD analysis  \\
\cite{gratton97}		&10.1 $\pm$ 1.2   	&CMD isochrone fitting  \\
\cite{gratton97}		&11.4 $\pm$ 1.4   	&CMD isochrone fitting  \\
\cite{brocato97}		&10   	&CMD analysis  \\
\cite{jimenez96}	&12.6 $\pm$ 2   	&CMD analysis  \\
\cite{chaboyer96}	&12.8 $\pm$ 0.3  	&CMD analysis  \\
\cite{sandage93}	&12.1 $\pm$ 1.2  	&CMD isochrone fitting  \\
\cite{straniero91}	&19 $\pm$ 1  	&CMD isochrone fitting  \\
\cite{alcaino90}		&13 $\pm$ 3  	&CMD isochrone fitting  \\
\cite{chieffi89}		&16  	 & CMD isochrone fitting \\
\cite{mcclure87}	&14 $\pm$ 1  	&CMD isochrone fitting  \\
\cite{peterson87}	&15.5  	&CMD analysis  \\
\cite{vandenberg86}	&18  	&CMD analysis \\

	\hline \hline
  \end{tabular}

  \caption{Age estimates for M68 in the literature. \label{tab:lit_age}}
  \end{center}
\end{table*}

\section{Conclusions}\label{sec:conclusions}

We carried out a detailed survey of variability in M68 using CCD observations as well as some EMCCD images. We showed that data from identical telescopes in a telescope network could be reduced using a reference image from a single telescope, i.e. that the network could be considered to be a single telescope for the purposes of data handling. This facilitates the analysis of time-series data taken with such networks significantly. 

Using our observations we were able to recover all known variables within our field of view, as well as to detect four new SX Phe variables near the cluster core. We used the light curves of RRL stars in our data to derive estimates for their metallicity, effective temperature, luminosity and mean mass. Those were then used to infer values for the cluster metallicity, reddening and distance. Furthermore, the light curves of the SX Phe stars were also used to obtain an additional independent estimate of distance. 

We found a metallicity [Fe/H]=$\clustermet$ on the ZW scale and $\clustermetuves$ on the UVES scale. We derived distance moduli of $\mu_0=\murrab$ mag (using RR0 stars), $\mu_0=\murrc$ mag (using RR1 stars), $\mu_0=\musx$ mag (using SX Phe stars), and $\mu_0=\mumvfeh$ mag (using the $M_V-$[Fe/H] relation for RRL stars), corresponding to physical distances of $\drrab$, $\drrc$, $\dsx$, and $\dmvfeh$ kpc, using RR0, RR1, SX Phe stars, and the $M_V-$[Fe/H] relation respectively. Finally, we used our CMD to check the consistency of age estimates in the literature for M68. We also used archival data to refine period estimates, and calculate period changes where appropriate, for the RRL stars.  The grid search we used here is particularly useful to estimate period changes for stars for which not many maxima were observed; in those cases, the traditional \textit{O-C} method struggles to yield a period-change value that phases all data sets well, while the grid search performs better. We also examined in detail the light curves of double-mode pulsators in this cluster. The ratios of the RR01 periods enabled us to estimate their masses with the new mass-period-metallicity empirical relation of Marconi et al. (2015, submitted).

Thanks to the latest DIA methods, we can now be confident that all of the RRL stars in M68 are known. Carrying out such studies for more Milky Way globular clusters will help us strengthen observational evidence for the Oosterhoff dichotomy, which in turn can be used to shed light on the origin of the Galactic halo.

\section*{Acknowledgements}
The research leading to these results has received funding from the European Community's Seventh Framework Programme 
(/FP7/2007-2013/) under grant agreements No. 229517 and 268421. AAF acknowledges the support of DGAPA-UNAM through project IN104612. 
This publication was made possible by NPRP grant \# X-019-1-006 from the Qatar National Research Fund (a member of Qatar Foundation).
OW and JS acknowledge support from the Communaut\'{e} fran\c{c}aise de Belgique -- Actions de recherche concert\'{e}es -- Acad\'{e}mie universitaire Wallonie-Europe. 
TCH gratefully acknowledges financial support from the Korea Research Council for Fundamental Science and Technology (KRCF) through the Young Research Scientist Fellowship Program. TCH acknowledges financial support from KASI (Korea Astronomy and Space Science Institute) grant number 2012-1-410-02.
KA, MD, MH, and CL are supported by NPRP grant NPRP-09-476-1-78 from the Qatar National Research Fund (a member of Qatar Foundation).
The Danish 1.54m telescope is operated based on a grant from the Danish Natural Science Foundation (FNU). 
SHG and XBW acknowledge support from National Natural Science Foundation of China (grants Nos. 10373023 and 10773027).
HK acknowledges support from a Marie Curie Intra-European Fellowship. 
M.R. acknowledges support from FONDECYT postdoctoral fellowship Nr. 3120097.
This work has made extensive use of the ADS and SIMBAD services, for which we are thankful.

\bibliographystyle{aa}
\bibliography{../thesisbib}

\begin{thebibliography}{131}
\expandafter\ifx\csname natexlab\endcsname\relax\def\natexlab#1{#1}\fi

\bibitem[{{Alard}(1999)}]{alard99}
{Alard}, C. 1999, \aap, 343, 10

\bibitem[{{Alard}(2000)}]{alard00}
{Alard}, C. 2000, \aaps, 144, 363

\bibitem[{{Albrow} {et~al.}(2009){Albrow}, {Horne}, {Bramich}, {Fouqu{\'e}},
  {Miller}, {Beaulieu}, {Coutures}, {Menzies}, {Williams}, {Batista},
  {Bennett}, {Brillant}, {Cassan}, {Dieters}, {Prester}, {Donatowicz},
  {Greenhill}, {Kains}, {Kane}, {Kubas}, {Marquette}, {Pollard}, {Sahu},
  {Tsapras}, {Wambsganss}, \& {Zub}}]{albrow09}
{Albrow}, M.~D., {Horne}, K., {Bramich}, D.~M., {et~al.} 2009, \mnras, 397,
  2099

\bibitem[{{Alcaino}(1977)}]{alcaino77}
{Alcaino}, G. 1977, \aaps, 29, 9

\bibitem[{{Alcaino} {et~al.}(1990){Alcaino}, {Liller}, {Alvarado}, \&
  {Wenderoth}}]{alcaino90}
{Alcaino}, G., {Liller}, W., {Alvarado}, F., \& {Wenderoth}, E. 1990, \aj, 99,
  1831

\bibitem[{{Arellano Ferro} {et~al.}(2014){Arellano Ferro}, {Ahumada},
  {Calder{\'o}n}, \& {Kains}}]{arellano14b}
{Arellano Ferro}, A., {Ahumada}, J.~A., {Calder{\'o}n}, J.~H., \& {Kains}, N.
  2014, \rmxaa, 50, 307

\bibitem[{{Arellano Ferro} {et~al.}(2013{\natexlab{a}}){Arellano Ferro},
  {Bramich}, {Figuera Jaimes}, {Giridhar}, {Kains}, {Kuppuswamy},
  {J{\o}rgensen}, {Alsubai}, {Andersen}, {Bozza}, {Browne}, {Calchi Novati},
  {Damerdji}, {Diehl}, {Dominik}, {Dreizler}, {Elyiv}, {Giannini},
  {Harps{\o}e}, {Hessman}, {Hinse}, {Hundertmark}, {Juncher}, {Kerins},
  {Korhonen}, {Liebig}, {Mancini}, {Mathiasen}, {Penny}, {Rabus}, {Rahvar},
  {Ricci}, {Scarpetta}, {Skottfelt}, {Snodgrass}, {Southworth}, {Surdej},
  {Tregloan-Reed}, {Vilela}, {Wertz}, \& {Mindstep Consortium}}]{arellano13}
{Arellano Ferro}, A., {Bramich}, D.~M., {Figuera Jaimes}, R., {et~al.}
  2013{\natexlab{a}}, \mnras, 434, 1220

\bibitem[{{Arellano Ferro} {et~al.}(2013{\natexlab{b}}){Arellano Ferro},
  {Bramich}, {Giridhar}, {Figuera Jaimes}, {Kains}, \&
  {Kuppuswamy}}]{arellano14}
{Arellano Ferro}, A., {Bramich}, D.~M., {Giridhar}, S., {et~al.}
  2013{\natexlab{b}}, \actaa, 63, 429

\bibitem[{{Arellano Ferro} {et~al.}(2011){Arellano Ferro}, {Figuera Jaimes},
  {Giridhar}, {Bramich}, {Hern{\'a}ndez Santisteban}, \&
  {Kuppuswamy}}]{arellano11}
{Arellano Ferro}, A., {Figuera Jaimes}, R., {Giridhar}, S., {et~al.} 2011,
  \mnras, 416, 2265

\bibitem[{{Arellano Ferro} {et~al.}(2010){Arellano Ferro}, {Giridhar}, \&
  {Bramich}}]{arellano10}
{Arellano Ferro}, A., {Giridhar}, S., \& {Bramich}, D.~M. 2010, \mnras, 402,
  226

\bibitem[{{Arellano Ferro} {et~al.}(2008){Arellano Ferro}, {Rojas L{\'o}pez},
  {Giridhar}, \& {Bramich}}]{arellano08b}
{Arellano Ferro}, A., {Rojas L{\'o}pez}, V., {Giridhar}, S., \& {Bramich},
  D.~M. 2008, \mnras, 384, 1444

\bibitem[{{Bellini} {et~al.}(2011){Bellini}, {Anderson}, \&
  {Bedin}}]{bellini11}
{Bellini}, A., {Anderson}, J., \& {Bedin}, L.~R. 2011, \pasp, 123, 622

\bibitem[{{Belserene}(1964)}]{belserene64}
{Belserene}, E.~P. 1964, \aj, 69, 475

\bibitem[{{Benedict} {et~al.}(2002){Benedict}, {McArthur}, {Fredrick},
  {Harrison}, {Lee}, {Slesnick}, {Rhee}, {Patterson}, {Nelan}, {Jefferys}, {van
  Altena}, {Shelus}, {Franz}, {Wasserman}, {Hemenway}, {Duncombe}, {Story},
  {Whipple}, \& {Bradley}}]{benedict02}
{Benedict}, G.~F., {McArthur}, B.~E., {Fredrick}, L.~W., {et~al.} 2002, \aj,
  123, 473

\bibitem[{{Bessell}(2005)}]{bessell05}
{Bessell}, M.~S. 2005, \araa, 43, 293

\bibitem[{{Bica} \& {Pastoriza}(1983)}]{bica83}
{Bica}, E.~L.~D. \& {Pastoriza}, M.~G. 1983, \apss, 91, 99

\bibitem[{{Bla{\v z}ko}(1907)}]{blazhko1907}
{Bla{\v z}ko}, S. 1907, Astronomische Nachrichten, 175, 325

\bibitem[{{Bono} {et~al.}(2003){Bono}, {Caputo}, {Castellani}, {Marconi},
  {Storm}, \& {Degl'Innocenti}}]{bono03}
{Bono}, G., {Caputo}, F., {Castellani}, V., {et~al.} 2003, \mnras, 344, 1097

\bibitem[{{Bragaglia} {et~al.}(2001){Bragaglia}, {Gratton}, {Carretta},
  {Clementini}, {Di Fabrizio}, \& {Marconi}}]{bragaglia01}
{Bragaglia}, A., {Gratton}, R.~G., {Carretta}, E., {et~al.} 2001, \aj, 122, 207

\bibitem[{{Bramich}(2008)}]{bramich08}
{Bramich}, D.~M. 2008, \mnras, 386, L77

\bibitem[{{Bramich} {et~al.}(2011){Bramich}, {Figuera Jaimes}, {Giridhar}, \&
  {Arellano Ferro}}]{bramich11}
{Bramich}, D.~M., {Figuera Jaimes}, R., {Giridhar}, S., \& {Arellano Ferro}, A.
  2011, \mnras, 413, 1275

\bibitem[{{Bramich} \& {Freudling}(2012)}]{bramich12b}
{Bramich}, D.~M. \& {Freudling}, W. 2012, \mnras, 424, 1584

\bibitem[{{Bramich} {et~al.}(2013){Bramich}, {Horne}, {Albrow}, {Tsapras},
  {Snodgrass}, {Street}, {Hundertmark}, {Kains}, {Arellano}, {Figuera}, \&
  {Giridhar}}]{bramich13}
{Bramich}, D.~M., {Horne}, K., {Albrow}, M.~D., {et~al.} 2013, \mnras, 428,
  2275

\bibitem[{{Brocato} {et~al.}(1997){Brocato}, {Castellani}, \&
  {Piersimoni}}]{brocato97}
{Brocato}, E., {Castellani}, V., \& {Piersimoni}, A. 1997, \apj, 491, 789

\bibitem[{{Brocato} {et~al.}(1994){Brocato}, {Castellani}, \&
  {Ripepi}}]{brocato94}
{Brocato}, E., {Castellani}, V., \& {Ripepi}, V. 1994, \aj, 107, 622

\bibitem[{{Cacciari} {et~al.}(2005){Cacciari}, {Corwin}, \&
  {Carney}}]{cacciari05}
{Cacciari}, C., {Corwin}, T.~M., \& {Carney}, B.~W. 2005, \aj, 129, 267

\bibitem[{{Caloi} {et~al.}(1997){Caloi}, {D'Antona}, \& {Mazzitelli}}]{caloi97}
{Caloi}, V., {D'Antona}, F., \& {Mazzitelli}, I. 1997, \aap, 320, 823

\bibitem[{{Carney}(1996)}]{carney96}
{Carney}, B.~W. 1996, \pasp, 108, 900

\bibitem[{{Carretta} {et~al.}(2009{\natexlab{a}}){Carretta}, {Bragaglia},
  {Gratton}, {D'Orazi}, \& {Lucatello}}]{carretta09}
{Carretta}, E., {Bragaglia}, A., {Gratton}, R., {D'Orazi}, V., \& {Lucatello},
  S. 2009{\natexlab{a}}, \aap, 508, 695

\bibitem[{{Carretta} {et~al.}(2009{\natexlab{b}}){Carretta}, {Bragaglia},
  {Gratton}, \& {Lucatello}}]{carretta09b}
{Carretta}, E., {Bragaglia}, A., {Gratton}, R., \& {Lucatello}, S.
  2009{\natexlab{b}}, \aap, 505, 139

\bibitem[{{Carretta} {et~al.}(2009{\natexlab{c}}){Carretta}, {Bragaglia},
  {Gratton}, {Lucatello}, {Catanzaro}, {Leone}, {Bellazzini}, {Claudi},
  {D'Orazi}, {Momany}, {Ortolani}, {Pancino}, {Piotto}, {Recio-Blanco}, \&
  {Sabbi}}]{carretta09a}
{Carretta}, E., {Bragaglia}, A., {Gratton}, R.~G., {et~al.} 2009{\natexlab{c}},
  \aap, 505, 117

\bibitem[{{Castelli}(1999)}]{castelli99}
{Castelli}, F. 1999, \aap, 346, 564

\bibitem[{{Catelan}(2009)}]{catelan09}
{Catelan}, M. 2009, \apss, 320, 261

\bibitem[{{Chaboyer} {et~al.}(1996){Chaboyer}, {Demarque}, {Kernan}, {Krauss},
  \& {Sarajedini}}]{chaboyer96}
{Chaboyer}, B., {Demarque}, P., {Kernan}, P.~J., {Krauss}, L.~M., \&
  {Sarajedini}, A. 1996, \mnras, 283, 683

\bibitem[{{Chieffi} \& {Straniero}(1989)}]{chieffi89}
{Chieffi}, A. \& {Straniero}, O. 1989, \apjs, 71, 47

\bibitem[{{Clement}(1990)}]{clement90}
{Clement}, C.~M. 1990, \aj, 99, 240

\bibitem[{{Clement} {et~al.}(1993){Clement}, {Ferance}, \& {Simon}}]{clement93}
{Clement}, C.~M., {Ferance}, S., \& {Simon}, N.~R. 1993, \apj, 412, 183

\bibitem[{{Clement} {et~al.}(2001){Clement}, {Muzzin}, {Dufton}, {Ponnampalam},
  {Wang}, {Burford}, {Richardson}, {Rosebery}, {Rowe}, \& {Hogg}}]{clement01}
{Clement}, C.~M., {Muzzin}, A., {Dufton}, Q., {et~al.} 2001, \aj, 122, 2587

\bibitem[{{Cohen} \& {Sarajedini}(2012)}]{cohen12}
{Cohen}, R.~E. \& {Sarajedini}, A. 2012, \mnras, 419, 342

\bibitem[{{Cox}(1991)}]{cox91}
{Cox}, A.~N. 1991, \apjl, 381, L71

\bibitem[{{Di Fabrizio} {et~al.}(2005){Di Fabrizio}, {Clementini}, {Maio},
  {Bragaglia}, {Carretta}, {Gratton}, {Montegriffo}, \&
  {Zoccali}}]{difabrizio05}
{Di Fabrizio}, L., {Clementini}, G., {Maio}, M., {et~al.} 2005, \aap, 430, 603

\bibitem[{{Dotter} {et~al.}(2007){Dotter}, {Chaboyer}, {Jevremovi{\'c}},
  {Baron}, {Ferguson}, {Sarajedini}, \& {Anderson}}]{dotter07}
{Dotter}, A., {Chaboyer}, B., {Jevremovi{\'c}}, D., {et~al.} 2007, \aj, 134,
  376

\bibitem[{{Dotter} {et~al.}(2008){Dotter}, {Chaboyer}, {Jevremovi{\'c}},
  {Kostov}, {Baron}, \& {Ferguson}}]{dotter08}
{Dotter}, A., {Chaboyer}, B., {Jevremovi{\'c}}, D., {et~al.} 2008, \apjs, 178,
  89

\bibitem[{{Dworetsky}(1983)}]{dworetsky83}
{Dworetsky}, M.~M. 1983, \mnras, 203, 917

\bibitem[{{Feuchtinger}(1998)}]{feuchtinger98}
{Feuchtinger}, M.~U. 1998, \aap, 337, L29

\bibitem[{{Figuera Jaimes} {et~al.}(2013){Figuera Jaimes}, {Arellano Ferro},
  {Bramich}, {Giridhar}, \& {Kuppuswamy}}]{figuera13}
{Figuera Jaimes}, R., {Arellano Ferro}, A., {Bramich}, D.~M., {Giridhar}, S.,
  \& {Kuppuswamy}, K. 2013, \aap, 556, A20

\bibitem[{{Freedman} {et~al.}(2001){Freedman}, {Madore}, {Gibson}, {Ferrarese},
  {Kelson}, {Sakai}, {Mould}, {Kennicutt}, {Ford}, {Graham}, {Huchra},
  {Hughes}, {Illingworth}, {Macri}, \& {Stetson}}]{freedman01}
{Freedman}, W.~L., {Madore}, B.~F., {Gibson}, B.~K., {et~al.} 2001, \apj, 553,
  47

\bibitem[{{Gilliland} {et~al.}(1998){Gilliland}, {Bono}, {Edmonds}, {Caputo},
  {Cassisi}, {Petro}, {Saha}, \& {Shara}}]{gilliland98}
{Gilliland}, R.~L., {Bono}, G., {Edmonds}, P.~D., {et~al.} 1998, \apj, 507, 818

\bibitem[{{Gratton} {et~al.}(2004){Gratton}, {Bragaglia}, {Clementini},
  {Carretta}, {Di Fabrizio}, {Maio}, \& {Taribello}}]{gratton04}
{Gratton}, R.~G., {Bragaglia}, A., {Clementini}, G., {et~al.} 2004, \aap, 421,
  937

\bibitem[{{Gratton} {et~al.}(1997){Gratton}, {Fusi Pecci}, {Carretta},
  {Clementini}, {Corsi}, \& {Lattanzi}}]{gratton97}
{Gratton}, R.~G., {Fusi Pecci}, F., {Carretta}, E., {et~al.} 1997, \apj, 491,
  749

\bibitem[{{Gratton} \& {Ortolani}(1989)}]{gratton89}
{Gratton}, R.~G. \& {Ortolani}, S. 1989, \aap, 211, 41

\bibitem[{{Greenstein} {et~al.}(1947){Greenstein}, {Bidelman}, \&
  {Popper}}]{greenstein47}
{Greenstein}, J.~L., {Bidelman}, W.~P., \& {Popper}, D.~M. 1947, \pasp, 59, 143

\bibitem[{{Guldenschuh} {et~al.}(2005){Guldenschuh}, {Layden}, {Wan},
  {Whiting}, {van der Bliek}, {Baca}, {Carlin}, {Freismuth}, {Mora}, {Salyk},
  {Vera}, {Verdugo}, \& {Young}}]{guldenschuh05}
{Guldenschuh}, K.~A., {Layden}, A.~C., {Wan}, Y., {et~al.} 2005, \pasp, 117,
  721

\bibitem[{{Harps{\o}e} {et~al.}(2012){Harps{\o}e}, {J{\o}rgensen}, {Andersen},
  \& {Grundahl}}]{harpsoe12}
{Harps{\o}e}, K.~B.~W., {J{\o}rgensen}, U.~G., {Andersen}, M.~I., \&
  {Grundahl}, F. 2012, \aap, 542, A23

\bibitem[{{Harris}(1975)}]{harris75}
{Harris}, W.~E. 1975, \apjs, 29, 397

\bibitem[{{Harris}(1996)}]{harris96}
{Harris}, W.~E. 1996, \aj, 112, 1487

\bibitem[{{Jeon} {et~al.}(2003){Jeon}, {Lee}, {Kim}, \& {Lee}}]{jeon03}
{Jeon}, Y.-B., {Lee}, M.~G., {Kim}, S.-L., \& {Lee}, H. 2003, \aj, 125, 3165

\bibitem[{{Jeon} {et~al.}(2004){Jeon}, {Lee}, {Kim}, \& {Lee}}]{jeon04b}
{Jeon}, Y.-B., {Lee}, M.~G., {Kim}, S.-L., \& {Lee}, H. 2004, \aj, 128, 287

\bibitem[{{Jimenez} {et~al.}(1996){Jimenez}, {Thejll}, {Jorgensen},
  {MacDonald}, \& {Pagel}}]{jimenez96}
{Jimenez}, R., {Thejll}, P., {Jorgensen}, U.~G., {MacDonald}, J., \& {Pagel},
  B. 1996, \mnras, 282, 926

\bibitem[{{Jurcsik}(1995)}]{jurcsik95}
{Jurcsik}, J. 1995, \actaa, 45, 653

\bibitem[{{Jurcsik}(1998)}]{jurcsik98}
{Jurcsik}, J. 1998, \aap, 333, 571

\bibitem[{{Jurcsik} {et~al.}(2012){Jurcsik}, {Hajdu}, {Szeidl}, {Ol{\'a}h},
  {Kelemen}, {S{\'o}dor}, {Saha}, {Mallick}, \& {Claver}}]{jurcsik12}
{Jurcsik}, J., {Hajdu}, G., {Szeidl}, B., {et~al.} 2012, \mnras, 419, 2173

\bibitem[{{Jurcsik} \& {Kov\'{a}cs}(1996)}]{jurcsik96}
{Jurcsik}, J. \& {Kov\'{a}cs}, G. 1996, \aap, 312, 111

\bibitem[{{Kains} {et~al.}(2013){Kains}, {Bramich}, {Arellano Ferro}, {Figuera
  Jaimes}, {J{\o}rgensen}, {Giridhar}, {Penny}, {Alsubai}, {Andersen}, {Bozza},
  {Browne}, {Burgdorf}, {Calchi Novati}, {Damerdji}, {Diehl}, {Dodds},
  {Dominik}, {Elyiv}, {Fang}, {Giannini}, {Gu}, {Hardis}, {Harps{\o}e},
  {Hinse}, {Hornstrup}, {Hundertmark}, {Jessen-Hansen}, {Juncher}, {Kerins},
  {Kjeldsen}, {Korhonen}, {Liebig}, {Lund}, {Lundkvist}, {Mancini}, {Martin},
  {Mathiasen}, {Rabus}, {Rahvar}, {Ricci}, {Sahu}, {Scarpetta}, {Skottfelt},
  {Snodgrass}, {Southworth}, {Surdej}, {Tregloan-Reed}, {Vilela}, {Wertz}, \&
  {Williams}}]{kains13b}
{Kains}, N., {Bramich}, D.~M., {Arellano Ferro}, A., {et~al.} 2013, \aap, 555,
  A36

\bibitem[{{Kains} {et~al.}(2012){Kains}, {Bramich}, {Figuera Jaimes}, {Arellano
  Ferro}, {Giridhar}, \& {Kuppuswamy}}]{kains12b}
{Kains}, N., {Bramich}, D.~M., {Figuera Jaimes}, R., {et~al.} 2012, \aap, 548,
  A92

\bibitem[{{Kinman}(2002)}]{kinman02}
{Kinman}, T.~D. 2002, Information Bulletin on Variable Stars, 5354, 1

\bibitem[{{Kov{\'a}cs}(1998)}]{kovacs98}
{Kov{\'a}cs}, G. 1998, \memsai, 69, 49

\bibitem[{{Kov{\'a}cs}(2002)}]{kovacs02}
{Kov{\'a}cs}, G. 2002, in Astronomical Society of the Pacific Conference
  Series, Vol. 265, Omega Centauri, A Unique Window into Astrophysics, ed.
  F.~{van Leeuwen}, J.~D. {Hughes}, \& G.~{Piotto}, 163

\bibitem[{{Kovacs} {et~al.}(1991){Kovacs}, {Buchler}, \& {Marom}}]{kovacs91}
{Kovacs}, G., {Buchler}, J.~R., \& {Marom}, A. 1991, \aap, 252, L27

\bibitem[{{Kov{\'a}cs} \& {Walker}(2001)}]{kovacs01}
{Kov{\'a}cs}, G. \& {Walker}, A.~R. 2001, \aap, 371, 579

\bibitem[{{Kunder} {et~al.}(2010){Kunder}, {Chaboyer}, \& {Layden}}]{kunder10}
{Kunder}, A., {Chaboyer}, B., \& {Layden}, A. 2010, \aj, 139, 415

\bibitem[{{Kunder} {et~al.}(2013{\natexlab{a}}){Kunder}, {Stetson}, {Cassisi},
  {Layden}, {Bono}, {Catelan}, {Walker}, {Paredes Alvarez}, {Clem},
  {Matsunaga}, {Salaris}, {Lee}, \& {Chaboyer}}]{kunder13b}
{Kunder}, A., {Stetson}, P.~B., {Cassisi}, S., {et~al.} 2013{\natexlab{a}},
  \aj, 146, 119

\bibitem[{{Kunder} {et~al.}(2013{\natexlab{b}}){Kunder}, {Stetson}, {Catelan},
  {Walker}, \& {Amigo}}]{kunder13}
{Kunder}, A., {Stetson}, P.~B., {Catelan}, M., {Walker}, A.~R., \& {Amigo}, P.
  2013{\natexlab{b}}, \aj, 145, 33

\bibitem[{{Lane} {et~al.}(2009){Lane}, {Kiss}, {Lewis}, {Ibata}, {Siebert},
  {Bedding}, \& {Sz{\'e}kely}}]{lane09}
{Lane}, R.~R., {Kiss}, L.~L., {Lewis}, G.~F., {et~al.} 2009, \mnras, 400, 917

\bibitem[{{Lee} \& {Carney}(1999)}]{lee99}
{Lee}, J.-W. \& {Carney}, B.~W. 1999, \aj, 118, 1373

\bibitem[{{Lee}(1991)}]{lee91}
{Lee}, Y.-W. 1991, \apj, 367, 524

\bibitem[{{Lee} {et~al.}(1990){Lee}, {Demarque}, \& {Zinn}}]{lee90}
{Lee}, Y.-W., {Demarque}, P., \& {Zinn}, R. 1990, \apj, 350, 155

\bibitem[{{Lenz} \& {Breger}(2005)}]{lenz05}
{Lenz}, P. \& {Breger}, M. 2005, Communications in Asteroseismology, 146, 53

\bibitem[{{McClure} {et~al.}(1987){McClure}, {Hesser}, {Stetson}, {Vandenberg},
  \& {Bell}}]{mcclure87}
{McClure}, R.~D., {Hesser}, J.~E., {Stetson}, P.~B., {Vandenberg}, D.~A., \&
  {Bell}, R.~A. 1987, \aj, 93, 1144

\bibitem[{{Minniti} {et~al.}(1993){Minniti}, {Geisler}, {Peterson}, \&
  {Claria}}]{minniti93}
{Minniti}, D., {Geisler}, D., {Peterson}, R.~C., \& {Claria}, J.~J. 1993, \apj,
  413, 548

\bibitem[{{Montegriffo} {et~al.}(1998){Montegriffo}, {Ferraro}, {Origlia}, \&
  {Fusi Pecci}}]{montegriffo98}
{Montegriffo}, P., {Ferraro}, F.~R., {Origlia}, L., \& {Fusi Pecci}, F. 1998,
  \mnras, 297, 872

\bibitem[{{Morgan} {et~al.}(2007){Morgan}, {Wahl}, \& {Wieckhorst}}]{morgan07}
{Morgan}, S.~M., {Wahl}, J.~N., \& {Wieckhorst}, R.~M. 2007, \mnras, 374, 1421

\bibitem[{{Nemec} {et~al.}(2013){Nemec}, {Cohen}, {Ripepi}, {Derekas},
  {Moskalik}, {Sesar}, {Chadid}, \& {Bruntt}}]{nemec13}
{Nemec}, J.~M., {Cohen}, J.~G., {Ripepi}, V., {et~al.} 2013, \apj, 773, 181

\bibitem[{{Nemec} {et~al.}(1985){Nemec}, {Hazen-Liller}, \&
  {Hesser}}]{nemec85b}
{Nemec}, J.~M., {Hazen-Liller}, M.~L., \& {Hesser}, J.~E. 1985, \apjs, 57, 329

\bibitem[{{Oosterhoff}(1939)}]{oosterhoff39}
{Oosterhoff}, P.~T. 1939, The Observatory, 62, 104

\bibitem[{{Petersen}(1973)}]{petersen73}
{Petersen}, J.~O. 1973, \aap, 27, 89

\bibitem[{{Peterson}(1987)}]{peterson87}
{Peterson}, C.~J. 1987, \pasp, 99, 1153

\bibitem[{{Piersimoni} {et~al.}(2002){Piersimoni}, {Bono}, \&
  {Ripepi}}]{piersimoni02}
{Piersimoni}, A.~M., {Bono}, G., \& {Ripepi}, V. 2002, \aj, 124, 1528

\bibitem[{{Pojmanski}(2002)}]{pojmanski02}
{Pojmanski}, G. 2002, \actaa, 52, 397

\bibitem[{{Poretti} {et~al.}(2010){Poretti}, {Papar{\'o}}, {Deleuil}, {Chadid},
  {Kolenberg}, {Szab{\'o}}, {Benk{\H o}}, {Chapellier}, {Guggenberger}, {Le
  Borgne}, {Rostagni}, {Trinquet}, {Auvergne}, {Baglin}, {Sarro}, \&
  {Weiss}}]{poretti10}
{Poretti}, E., {Papar{\'o}}, M., {Deleuil}, M., {et~al.} 2010, \aap, 520, A108

\bibitem[{{Pritzl} {et~al.}(2001){Pritzl}, {Smith}, {Catelan}, \&
  {Sweigart}}]{pritzl01}
{Pritzl}, B.~J., {Smith}, H.~A., {Catelan}, M., \& {Sweigart}, A.~V. 2001, \aj,
  122, 2600

\bibitem[{{Pritzl} {et~al.}(2002){Pritzl}, {Smith}, {Catelan}, \&
  {Sweigart}}]{pritzl02}
{Pritzl}, B.~J., {Smith}, H.~A., {Catelan}, M., \& {Sweigart}, A.~V. 2002, \aj,
  124, 949

\bibitem[{{Racine}(1973)}]{racine73}
{Racine}, R. 1973, \aj, 78, 180

\bibitem[{{Rakos} \& {Schombert}(2005)}]{rakos05}
{Rakos}, K. \& {Schombert}, J. 2005, \pasp, 117, 245

\bibitem[{{Rathbun} \& {Smith}(1997)}]{rathbun97}
{Rathbun}, P. \& {Smith}, H. 1997, \pasp, 109, 1128

\bibitem[{{Rosenberg} {et~al.}(1999){Rosenberg}, {Saviane}, {Piotto}, \&
  {Aparicio}}]{rosenberg99}
{Rosenberg}, A., {Saviane}, I., {Piotto}, G., \& {Aparicio}, A. 1999, \aj, 118,
  2306

\bibitem[{{Rosino} \& {Pietra}(1953)}]{rosino53}
{Rosino}, L. \& {Pietra}, S. 1953, \memsai, 24, 331

\bibitem[{{Rosino} \& {Pietra}(1954)}]{rosino54}
{Rosino}, L. \& {Pietra}, S. 1954, \memsai, 25, 227

\bibitem[{{Rutledge} {et~al.}(1997){Rutledge}, {Hesser}, {Stetson}, {Mateo},
  {Simard}, {Bolte}, {Friel}, \& {Copin}}]{rutledge97}
{Rutledge}, G.~A., {Hesser}, J.~E., {Stetson}, P.~B., {et~al.} 1997, \pasp,
  109, 883

\bibitem[{{Salaris} {et~al.}(1993){Salaris}, {Chieffi}, \&
  {Straniero}}]{salaris93}
{Salaris}, M., {Chieffi}, A., \& {Straniero}, O. 1993, \apj, 414, 580

\bibitem[{{Salaris} \& {Weiss}(2002)}]{salaris02}
{Salaris}, M. \& {Weiss}, A. 2002, \aap, 388, 492

\bibitem[{{Sandage}(1993)}]{sandage93}
{Sandage}, A. 1993, \aj, 106, 687

\bibitem[{{Santolamazza} {et~al.}(2001){Santolamazza}, {Marconi}, {Bono},
  {Caputo}, {Cassisi}, \& {Gilliland}}]{santolamazza01}
{Santolamazza}, P., {Marconi}, M., {Bono}, G., {et~al.} 2001, \apj, 554, 1124

\bibitem[{{Sariya} {et~al.}(2014){Sariya}, {Lata}, \& {Yadav}}]{sariya14}
{Sariya}, D.~P., {Lata}, S., \& {Yadav}, R.~K.~S. 2014, \na, 27, 56

\bibitem[{{Sekiguchi} \& {Fukugita}(2000)}]{sekiguchi00}
{Sekiguchi}, M. \& {Fukugita}, M. 2000, \aj, 120, 1072

\bibitem[{{Shapley}(1919)}]{shapley19}
{Shapley}, H. 1919, \pasp, 31, 226

\bibitem[{{Shapley}(1920)}]{shapley20}
{Shapley}, H. 1920, \apj, 51, 49

\bibitem[{{Simon} \& {Clement}(1993)}]{simon93}
{Simon}, N.~R. \& {Clement}, C.~M. 1993, \apj, 410, 526

\bibitem[{{Skottfelt} {et~al.}(2013){Skottfelt}, {Bramich}, {Figuera Jaimes},
  {J{\o}rgensen}, {Kains}, {Harps{\o}e}, {Liebig}, {Penny}, {Alsubai},
  {Andersen}, {Bozza}, {Browne}, {Calchi Novati}, {Damerdji}, {Diehl},
  {Dominik}, {Elyiv}, {Giannini}, {Hessman}, {Hinse}, {Hundertmark}, {Juncher},
  {Kerins}, {Korhonen}, {Mancini}, {Martin}, {Rabus}, {Rahvar}, {Scarpetta},
  {Southworth}, {Snodgrass}, {Street}, {Surdej}, {Tregloan-Reed}, {Vilela}, \&
  {Williams}}]{skottfelt13}
{Skottfelt}, J., {Bramich}, D.~M., {Figuera Jaimes}, R., {et~al.} 2013, \aap,
  553, A111

\bibitem[{{Smith} \& {Manduca}(1983)}]{smith83}
{Smith}, H.~A. \& {Manduca}, A. 1983, \aj, 88, 982

\bibitem[{{Smith} \& {Wesselink}(1977)}]{smith77}
{Smith}, H.~A. \& {Wesselink}, A.~J. 1977, \aap, 56, 135

\bibitem[{{Sollima} {et~al.}(2014){Sollima}, {Cassisi}, {Fiorentino}, \&
  {Gratton}}]{sollima14}
{Sollima}, A., {Cassisi}, S., {Fiorentino}, G., \& {Gratton}, R.~G. 2014,
  \mnras, 444, 1862

\bibitem[{{Soszy{\'n}ski} {et~al.}(2009){Soszy{\'n}ski}, {Udalski},
  {Szyma{\'n}ski}, {Kubiak}, {Pietrzy{\'n}ski}, {Wyrzykowski}, {Szewczyk},
  {Ulaczyk}, \& {Poleski}}]{soszynski09}
{Soszy{\'n}ski}, I., {Udalski}, A., {Szyma{\'n}ski}, M.~K., {et~al.} 2009,
  \actaa, 59, 1

\bibitem[{{Stagg} \& {Wehlau}(1980)}]{stagg80}
{Stagg}, C. \& {Wehlau}, A. 1980, \aj, 85, 1182

\bibitem[{{Stetson}(2000)}]{stetson00}
{Stetson}, P.~B. 2000, \pasp, 112, 925

\bibitem[{{Straniero} \& {Chieffi}(1991)}]{straniero91}
{Straniero}, O. \& {Chieffi}, A. 1991, \apjs, 76, 525

\bibitem[{{Sturch}(1966)}]{sturch66}
{Sturch}, C. 1966, \apj, 143, 774

\bibitem[{{Suntzeff} {et~al.}(1991){Suntzeff}, {Kinman}, \&
  {Kraft}}]{suntzeff91}
{Suntzeff}, N.~B., {Kinman}, T.~D., \& {Kraft}, R.~P. 1991, \apj, 367, 528

\bibitem[{{Terzan} {et~al.}(1973){Terzan}, {Rutily}, \& {Ounnas}}]{terzan73}
{Terzan}, A., {Rutily}, B., \& {Ounnas}, C. 1973, in Astrophysics and Space
  Science Library, Vol.~36, IAU Colloq. 21: Variable Stars in Globular Clusters
  and in Related Systems, ed. J.~D. {Fernie}, 76

\bibitem[{{van Agt} \& {Oosterhoff}(1959)}]{vanagt59}
{van Agt}, S.~L.~T.~J. \& {Oosterhoff}, P.~T. 1959, Annalen van de Sterrewacht
  te Leiden, 21, 253

\bibitem[{{van Albada} \& {Baker}(1971)}]{vanalbada71}
{van Albada}, T.~S. \& {Baker}, N. 1971, \apj, 169, 311

\bibitem[{{Vandenberg}(1986)}]{vandenberg86}
{Vandenberg}, D.~A. 1986, \memsai, 57, 373

\bibitem[{{VandenBerg} {et~al.}(2013){VandenBerg}, {Brogaard}, {Leaman}, \&
  {Casagrande}}]{vandenberg13}
{VandenBerg}, D.~A., {Brogaard}, K., {Leaman}, R., \& {Casagrande}, L. 2013,
  \apj, 775, 134

\bibitem[{{Walker}(1994)}]{walker94}
{Walker}, A.~R. 1994, \aj, 108, 555

\bibitem[{{Yoon} \& {Lee}(2002)}]{yoon02}
{Yoon}, S.-J. \& {Lee}, Y.-W. 2002, Science, 297, 578

\bibitem[{{Zacharias} {et~al.}(2010){Zacharias}, {Finch}, {Girard}, {Hambly},
  {Wycoff}, {Zacharias}, {Castillo}, {Corbin}, {DiVittorio}, {Dutta}, {Gaume},
  {Gauss}, {Germain}, {Hall}, {Hartkopf}, {Hsu}, {Holdenried}, {Makarov},
  {Martinez}, {Mason}, {Monet}, {Rafferty}, {Rhodes}, {Siemers}, {Smith},
  {Tilleman}, {Urban}, {Wieder}, {Winter}, \& {Young}}]{zacharias10}
{Zacharias}, N., {Finch}, C., {Girard}, T., {et~al.} 2010, \aj, 139, 2184

\bibitem[{{Zinn}(1980)}]{zinn80}
{Zinn}, R. 1980, \apjs, 42, 19

\bibitem[{{Zinn}(1993{\natexlab{a}})}]{zinn93a}
{Zinn}, R. 1993{\natexlab{a}}, in Astronomical Society of the Pacific
  Conference Series, Vol.~48, The Globular Cluster-Galaxy Connection, ed. G.~H.
  {Smith} \& J.~P. {Brodie}, 302

\bibitem[{{Zinn}(1993{\natexlab{b}})}]{zinn93b}
{Zinn}, R. 1993{\natexlab{b}}, in Astronomical Society of the Pacific
  Conference Series, Vol.~48, The Globular Cluster-Galaxy Connection, ed. G.~H.
  {Smith} \& J.~P. {Brodie}, 38

\bibitem[{{Zinn} \& {West}(1984)}]{zinn84}
{Zinn}, R. \& {West}, M.~J. 1984, \apjs, 55, 45

\bibitem[{{Zorotovic} {et~al.}(2010){Zorotovic}, {Catelan}, {Smith}, {Pritzl},
  {Aguirre}, {Angulo}, {Aravena}, {Assef}, {Contreras}, {Cort{\'e}s}, {De
  Martini}, {Escobar}, {Gonz{\'a}lez}, {Jofr{\'e}}, {Lacerna}, {Navarro},
  {Palma}, {Prieto}, {Recabarren}, {Trivi{\~n}o}, \& {Vidal}}]{zorotovic10}
{Zorotovic}, M., {Catelan}, M., {Smith}, H.~A., {et~al.} 2010, \aj, 139, 357

\end{thebibliography}

\label{lastpage}

\end{document}